\documentclass[letterpaper, 12pt]{JHEP3}
\def\slash#1{\setbox0=\hbox{$#1$}#1\hskip-\wd0\hbox to\wd0{\hss\sl/\/\hss}}
\usepackage{epsf, cite, amsmath, amssymb}
\usepackage{epsfig}
\usepackage[all]{xy}
\let\non\nonumber

\newcommand{\bea}{\begin{eqnarray}}
\newcommand{\eea}{\end{eqnarray}}
\newcommand{\be}{\begin{equation}}
\newcommand{\ee}{\end{equation}}

\newcommand{\Z}{{\mathbb Z}}

\newcommand{\m}{\mu}
\newcommand{\n}{\nu}
\newcommand{\p}{\partial}

\newcommand{\al}{\alpha}
\newcommand{\bt}{\beta}
\newcommand{\g}{\gamma}
\newcommand{\s}{\sigma}
\newcommand{\del}{\delta}
\newcommand{\Ot}{{\mathcal{O}}_\tau}
\newcommand{\Obt}{{\mathcal{O}}_{\bar\tau}}
\newcommand{\OT}{{\mathcal{O}}_2}
\newcommand{\Om}{\mathcal{O}}
\newcommand{\dt}{\dot\theta}
\newcommand{\dal}{\dot\alpha}
\newcommand{\dbt}{\dot\beta}
\newcommand{\dg}{\dot\gamma}
\newcommand{\ddel}{\dot\delta}
\newcommand{\pmh}{\pm \frac{1}{2}}

\newcommand{\Oc}{\mathcal{C}}

\newcommand{\C}[1]{$(\ref{#1})$}

\title{Some Systematics of the Coupling Constant Dependence of $N=4$ Yang--Mills }
\author{ Anirban Basu$^{\dagger}$,
Michael B. Green$^{\dagger\dagger}$ and
Savdeep Sethi$^{\dagger}$
\\
{$^{\dagger}$   Enrico Fermi Institute, University of Chicago,
Chicago, IL 60637, USA}
\\
\email{basu@theory.uchicago.edu; sethi@theory.uchicago.edu}
\\
{$^{\dagger\dagger}$ \it   Department of Applied Mathematics and
Theoretical Physics,\\ Wilberforce Road, Cambridge CB3 0WA,
UK}
\\
\email{M.B.Green@damtp.cam.ac.uk}}

\abstract{
The operator, $\Ot$, that generates infinitesimal changes of the coupling
constant in $N=4$ Yang--Mills sits in the same supermultiplet as the
superconformal currents.  We show how superconformal current
Ward identities determine a class of terms in
the operator product expansion of $\Ot$
with any other operator. In certain cases, this
leads to constraints on the coupling dependence of
 correlation functions in $N=4$ Yang-Mills. As an
application, we demonstrate the exact non-renormalization of
two and certain three-point correlation functions of BPS operators. }
\preprint{DAMTP-2004-66, EFI-04-20 }

\begin{document}

\baselineskip=18pt

\section{Introduction}
\label{intro}

Maximally supersymmetric Yang-Mills in $4$ dimensions has played a central role in the
development of duality in both field theory and string theory. At loci
in the Coulomb branch where some non-abelian gauge symmetry is
unbroken, there exists an interacting superconformal field theory parametrized by a
coupling constant, $\tau$, given in terms of the Yang-Mills coupling constant
and theta-angle by
\be \tau = \frac{\theta_{\rm YM}}{2\pi} + \frac{4\pi i}{g_{\rm YM}^2}.\ee
The superconformal field theory is defined by correlation functions of
local operators. There
is, by now, convincing evidence that $N=4$ Yang-Mills is
invariant under an $SL(2,\Z)$ strong-weak coupling duality
group which acts on $\tau$~\cite{Sen:1994yi, Montonen:1977sn}. For
simply-laced gauged groups, this duality can be naturally understood
by viewing $N=4$ Yang-Mills as a torus reduction of the $6$-dimensional
$(2,0)$ chiral tensor theory~\cite{Witten:1995zh}. The $SL(2,\Z)$
duality group then corresponds to the symmetry group of the
compactification torus.

What this picture intuitively suggests is that the coupling constant
dependence of correlation functions should be controllable to roughly
the same degree as the space-time dependence. Both dependences originate from
diffeomorphisms in $6$ dimensions. The goal of this work is
to explore the extent to which this statement can be made
precise. Consider a correlator of local operators,
\be \langle \Om_1(x_1) \cdots \Om_n(x_n) \rangle.  \ee
To determine the coupling dependence, we want to evaluate
\be \frac{\partial}{\partial \tau} \langle \Om_1(x_1) \cdots \Om_n(x_n) \rangle,  \ee
but this has two distinct contributions. The first comes from the
explicit derivative acting on each operator, while the second
corresponds to the insertion of the holomorphic part of the
action~\cite{Intriligator:1998ig},
\be \label{dertau}
\frac{\partial}{\partial \tau} \langle \prod_i {\mathcal{O}}_i (x_i) \rangle
= \langle \frac{\partial}{\partial \tau} \left(\prod_i {\mathcal{O}}_i
(x_i)  \right)\rangle +
\frac{i}{4\tau_2} \int d^4 z \langle \Ot (z)
\prod_i {\mathcal{O}}_i (x_i) \rangle,
\ee
where the full action is given by
\be\label{action} S = \frac{i}{4\tau_2} \int d^4 z \left\{  \tau \Ot (z) - \bar\tau
{\bar\Om}_{\tau}(z)\right\}. \ee
What is special about $\Ot$ is that it sits in the current multiplet
of $N=4$ Yang-Mills, together with the stress-energy tensor, the
supercurrents and the $SU(4)$ R-symmetry currents~\cite{Howe:1981qj,
  Intriligator:1998ig}.  This is in
agreement with our higher-dimensional intuition and will play a
crucial role in our analysis.

Although $\Ot$ is
not a current, we will show that its OPE with any other operator is
special in the same way that a current OPE is special. For example, for
a scalar operator $\Om(y)$, this OPE takes the
schematic form (suppressing all indices and details)
\bea
\Ot(x) \Om(y)
&\sim &  \frac{1}{|x-y|^2} \{Q ,[Q ,\{{\bar{Q}}, [{\bar{Q}} ,\Om
    (y)]\}]\}  \non \\
&&  +\frac{1}{|x-y|} \{Q ,[Q,\{{{Q}}, [{\bar{Q}},\{{\bar{Q}},[\bar{Q},\Om
     (y)]\}]\}]\} + \ldots. \non
\eea
This expression captures all local operators in the same
supermultiplet as $\Om(y)$. It omits operators that live in different
supermultiplets.  The
precise form of the OPE is derived in section~\ref{OtOPE}. The OPE
coefficients are, a priori, arbitrary functions of the coupling. It is
important to note that the most singular term we might have expected,
$1/|x-y|^4$, does not appear on the right hand side.

The determination of this OPE leads to a number of results. In this
paper, we will show that the structure of the OPE together with
superconformal Ward identities leads to a non-renormalization theorem
for two-point functions of $1/2$, $1/4$ and $1/8$ BPS operators. We
also show the exact non-renormalization of three-point correlators
of $1/2$, $1/4$ and $1/8$ BPS superconformal primary operators. This
non-renormalization result extends to three-point functions of
superconformal descendents of the current multiplet via the results
of~\cite{Dolan:2001tt}. However, it is difficult to extend our
argument to three-point functions of generic BPS operators. In cases
where there are no instanton corrections to a given correlator, we
give an argument in section~\ref{descendentcomments}\ for exact
non-renormalization. It is an interesting open question to prove (or
disprove) exact non-renormalization for generic BPS three-point
correlators.

This non-renormalization result,
originally conjectured for $1/2$ BPS superconformal primary operators
in~\cite{Lee:1998bx}\ and verified at
$1$-loop in~\cite{D'Hoker:1998tz, Gonzalez-Rey:1999ih},  has
been argued using an on-shell superspace
formalism~\cite{Howe:1997rk, Eden:1999gh, Howe:1999hz, Howe:2001je, Eden:2001ec,
  Heslop:2001gp, Heslop:2003xu}. The non-renormalization of two and
three-point functions involving $1/4$ BPS operators has been
conjectured  and verified at
$1$-loop in~\cite{Ryzhov:2001bp, D'Hoker:2001bq}.

In section~\ref{nice}, we show that the OPE
coefficients between $\Ot$ and any BPS operator are not
renormalized. This leads to a pretty formula for the integrated OPE
between $\Ot$ and a BPS operator $\Om$,
\be
\int d^4z \Ot(z) \Om(x) \sim \sum_{i} \Om_i'(x) + \ldots,
\ee
where $\Om_i'$ is also BPS and sits in the same supermultiplet as
$\Om$. The omitted terms involve long and semi-short operators.
A much stranger renormalization result will be described in a
sequel~\cite{toappear}. It
might also be possible to use our results to study the bonus $U(1)_Y$ predictions
described in~\cite{Intriligator:1998ig, Intriligator:1999ff} and
perhaps study extremal correlators~\cite{D'Hoker:1999ea}.

\section{Short and Current Multiplets}
\subsection{The structure of the multiplets}
The superconformal symmetry group of $N=4$ Yang-Mills  is generated by $16$ real
supersymmetry generators which we denote by $Q_i{}_{\! \al},
{\bar{Q}}^{i}_{\dal}$ where $i=1,\ldots,4$ and $\al=1,2$. Our notation
closely follows that of~\cite{Dolan:2002zh}. In addition,
there are $16$ superconformal charges $S^i{}^{\! \al},
{\bar{S}}_i^{\dal}$ and an $SU(4)_R$ symmetry with generators
$R^i_j$. The structure of the superconformal algebra is given in
Appendix~\ref{superconformal}.

There are superconformal multiplets of different sizes. A
multiplet contains a state of lowest conformal dimension which is
annihilated by both the superconformal charges and by the generators
of special conformal transformations, $K_{\mu}$. These properties
define the (unique)
superconformal primary state, which we characterize by its  $SU(4)_R$
Dynkin labels $[k,p,q]$ $(k,p,q \geq 0)$ and its spin quantum numbers
$(j,\bar\jmath)$ under $Spin(3,1) \approx SU(2)_L \otimes
SU(2)_R$. This state cannot be obtained by acting on any other state
with combinations of  $ Q_i{}_{\! \al}$ or
${\bar{Q}}^i_{\dal}$. We denote this state by
$|k,p,q;j,\bar\jmath\rangle^{\rm hw}$.

By acting on this state with  $Q_i{}_{\! \al},
{\bar{Q}}^{i}_{\dal}$, we generate the remaining states in the
multiplet. These states (including the superconformal primary) are
conformal primaries. Acting further on these states with $P_{\mu}$
generates conformal descendents. In this way, we construct the entire
multiplet. If the  superconformal primary is annihilated by some of
the supersymmetry generators, the multiplet is reduced in size. It is
akin to a BPS particle. For example, if all the supercharges kill the
superconformal primary then it must be the unique vacuum
state. Standard short representations are annihilated by 8 supersymmetry
charges (hence $\frac{1}{2}$ BPS)~\cite{Gunaydin:1985fk}. If the
superconformal primary is not annihilated by any supercharges then the
multiplet is long.

{}For a
short representation,
following~\cite{Dolan:2002zh},  we choose a basis where
\be \label{basisq}
Q_{i\al} |k,p,q;j,\bar\jmath\rangle^{\rm hw} = {\bar{Q}}^j_{\dal}
|k,p,q;j,\bar\jmath\rangle^{\rm hw} =0,\ee
for $i=1,2$ and $j=3,4$.
We are free to act on the superconformal primary state by any of the remaining $8$
supersymmetry charges. As derived in Appendix~\ref{shortstructure}, a
superconformal primary  has quantum numbers $[0,p,0]_{(0,0)}$ in the
terser notation $[k,p,q]_{(j,\bar\jmath)}$ and conformal dimension
$\Delta =p$.

The $Q, \bar{Q}$ operators can be used to build a multiplet `up' by
acting on the superconformal primary. On the other hand, the
$S,\bar{S}$ operators move us `down' the multiplet. We summarize
the structure of the multiplet pictorially in diagram~\C{BigM}; the
details of the construction appear in Appendix~\ref{shortstructure}.
The $\swarrow$ denotes the action of
$Q$ while $\searrow$
denotes the action of $\bar{Q}$)~\cite{Dolan:2001tt},
\be 
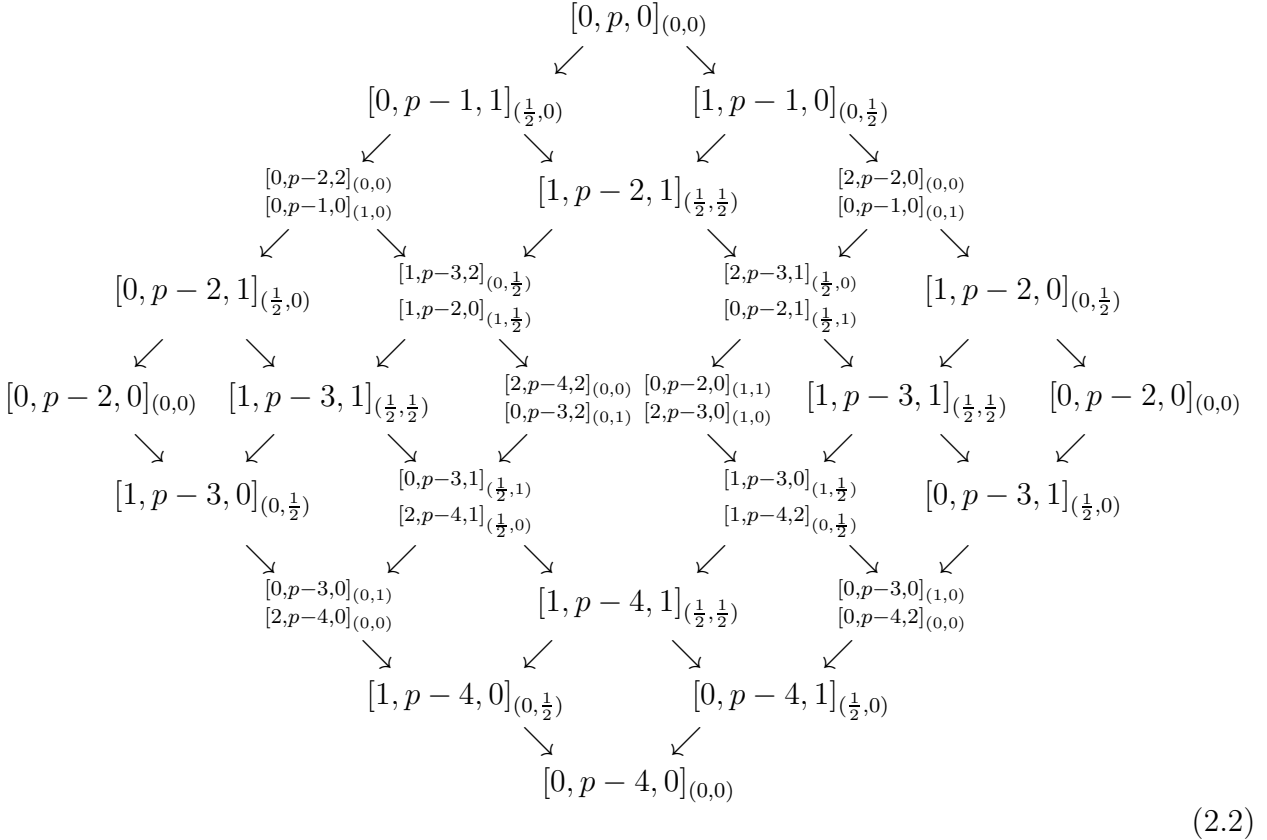
\label{BigM}
\begin{matrix}
&  &  &  &\!\!\!\!\!\!\!\!\!\!\!\!\!\!\!\!\!\! [0,p,0]_{(0,0)} & & & &  \\

&  &  &~~~~~~~~~~\swarrow  &~~~~ \!\!\!\!\searrow & & & &  \\

& & &\!\!\!\!\!\!\!\!\!\!\!\!\!\!\!\!\!\![0,p-1,1]_{(\frac{1}{2},0)} &
&\!\!\!\!\!\!\!\!\!\!\!\!\!\!\!\!\!\![1,p-1,0]_{(0,\frac{1}{2})}&&&  \\

& & \swarrow & ~~~~\searrow &~~~~\swarrow &~~~~\searrow & & &  \\

&  &\!\!\!\!\!\!\!\!\!\!\!\!\!\!\!\!\!\!{[0,p-2,2]_{(0,0)} \atop
[0,p-1,0]_{(1,0)} } & &\!\!\!\!\!\!\!\!\!\!\!\!\!\!\!\!\!\!
[1,p-2,1]_{(\frac{1}{2},\frac{1}{2})}  & &
\!\!\!\!\!\!\!\!\!\!\!\!\!\!\!\!\!\!\!\!{ [2,p-2,0]_{(0,0)} \atop
[0,p-1,0]_{(0,1)} }  & &  \\

& ~~\swarrow& ~~\searrow &~~~~\swarrow & ~~~~~~\searrow &~~\swarrow &
\searrow  & &  \\

&\!\!\!\!\!\!\!\!\!\!\!\!\!\!\!\!\!\![0,p-2,1]_{(\frac{1}{2},0)} &
&\!\!\!\!\!\!\!\!\!\!\!\!\!\!\!\!\!\!{[1,p-3,2]_{(0,\frac{1}{2})}
\atop [1,p-2,0]_{(1,\frac{1}{2})} }&  &
\!\!\!\!\!\!\!\!\!\!\!\!\!\!\!\!\!\!{ [2,p-3,1]_{(\frac{1}{2},0)}
\atop [0,p-2,1]_{(\frac{1}{2},1)} }& &\!\!\!\!\!\!\!\!\!\!\!\!\!\!\!\!\!\!
[1,p-2,0]_{(0,\frac{1}{2})}&  \\

\swarrow&\searrow & ~~\swarrow & \searrow &~~~~~~~\swarrow &\searrow &
\swarrow & \searrow &\\

\!\!\!\!\!\!\!\!\!\!\!\!\!\!\!\!\!\![0,p-2,0]_{(0,0)}& &
\!\!\!\!\!\!\!\!\!\!\!\!\!\!\!\!\!\![1,p-3,1]_{(\frac{1}{2},\frac{1}{2})}&
&\!\!\!\!\!\!\!\!\!\!\!\!\!\!\!\!\!\!{{[2,p-4,2]_{(0,0)}~
[0,p-2,0]_{(1,1)}} \atop {[0,p-3,2]_{(0,1)}~[2,p-3,0]_{(1,0)}}}& &
\!\!\!\!\!\!\!\!\!\!\!\!\!\!\!\!\!\![1,p-3,1]_{(\frac{1}{2},\frac{1}{2})}&
&\!\!\!\!\!\!\!\!\!\!\!\!\!\!\!\![0,p-2,0]_{(0,0)}  \\

\searrow & \swarrow & ~~~~\searrow &\swarrow& ~~~~~~\searrow&\swarrow
&\searrow &\swarrow  &\\

&\!\!\!\!\!\!\!\!\!\!\!\!\!\!\!\!\!\![1,p-3,0]_{(0,\frac{1}{2})} & &
\!\!\!\!\!\!\!\!\!\!\!\!\!\!\!\!\!\!{[0,p-3,1]_{(\frac{1}{2},1)} \atop
[2,p-4,1]_{(\frac{1}{2},0)} }&  &\!\!\!\!\!\!\!\!\!\!\!\!\!\!\!\!\!\!
{ [1,p-3,0]_{(1,\frac{1}{2})} \atop [1,p-4,2]_{(0,\frac{1}{2})} }& &
\!\!\!\!\!\!\!\!\!\!\!\!\!\!\!\!\!\![0,p-3,1]_{(\frac{1}{2},0)}&  \\

& \searrow&  ~~~~\swarrow& ~~~~\searrow&~~~~\swarrow&~~~~\searrow&\swarrow
& &\\
 &  &\!\!\!\!\!\!\!\!\!\!\!\!\!\!\!\!\!\!{[0,p-3,0]_{(0,1)} \atop
[2,p-4,0]_{(0,0)} } & &\!\!\!\!\!\!\!\!\!\!\!\!\!\!\!\!\!\!
[1,p-4,1]_{(\frac{1}{2},\frac{1}{2})}  & &
\!\!\!\!\!\!\!\!\!\!\!\!\!\!\!\!\!\!\!\!{ [0,p-3,0]_{(1,0)} \atop
 [0,p-4,2]_{(0,0)} }  & &  \\

 &  & \searrow &~~~~\swarrow&\searrow&\swarrow&&    &\\

& & &\!\!\!\!\!\!\!\!\!\!\!\!\!\!\!\!\!\![1,p-4,0]_{(0,\frac{1}{2})} &
 &\!\!\!\!\!\!\!\!\!\!\!\!\!\!\!\!\!\![0,p-4,1]_{(\frac{1}{2},0)}&&&  \\

 &  & &~~~~\searrow &\swarrow&& &  &\\

&  &  &  &\!\!\!\!\!\!\!\!\!\!\!\!\!\!\!\!\!\! [0,p-4,0]_{(0,0)} & & & &  \\
\end{matrix}
\ee
\vskip 1in
\noindent
The dimension of this short representation is equal to ${\frac{64}{3}}
p^2 (p^2 -1)$. The conformal dimensions range from $p$ to $p+4$.

Let us now discuss the particular case of $p=2$ which is central to
our later discussion. We shall also see how multiplet shortening occurs
{}for this multiplet. This is a $256$-dimensional representation
consisting of $128$  bosonic and $128$ fermionic degrees of freedom.
The multiplet is given by the following diagram~\cite{Dolan:2002zh}
\be \label{LitM}
\begin{matrix}
&  &  &  &\!\!\!\!\!\!\!\!\!\!\!\!\!\!\!\!\!\! [0,2,0]_{(0,0)} & & & &  \\

&  &  &\swarrow  &~~~~ \!\!\!\!\searrow & & & &  \\

& & &\!\!\!\!\!\!\!\!\!\!\!\!\!\!\!\!\!\![0,1,1]_{(\frac{1}{2},0)} &
&\!\!\!\!\!\!\!\!\!\!\!\!\!\!\!\!\!\![1,1,0]_{(0,\frac{1}{2})}&&&  \\

& & \swarrow & ~~~~\searrow &~~~~\swarrow &~~~~\searrow & & &  \\

&  &\!\!\!\!\!\!\!\!\!\!\!\!\!\!\!\!{[0,0,2]_{(0,0)} \atop
[0,1,0]_{(1,0)} } & &\!\!\!\!\!\!\!\!\!\!\!\!\!\!\!\!\!\!
[1,0,1]_{(\frac{1}{2},\frac{1}{2})}  & &
\!\!\!\!\!\!\!\!\!\!\!\!\!\!\!\!\!\!\!\!\!\!{ [2,0,0]_{(0,0)} \atop
[0,1,0]_{(0,1)} }  & &  \\

& ~~\swarrow& ~~\searrow &~~~~\swarrow & ~~~~~~\searrow &~~\swarrow &
\searrow  & &  \\

&\!\!\!\!\!\!\!\!\!\!\!\!\!\!\!\!\!\![0,0,1]_{(\frac{1}{2},0)} &
&\!\!\!\!\!\!\!\!\!\!\!\!\!\!\!\!\!\!
[1,0,0]_{(1,\frac{1}{2})} &  &
\!\!\!\!\!\!\!\!\!\!\!\!\!\!\!\!\!\!
[0,0,1]_{(\frac{1}{2},1)} & &\!\!\!\!\!\!\!\!\!\!\!\!\!\!\!\!\!\!
~~~~~~[1,0,0]_{(0,\frac{1}{2})}&  \\

\swarrow& & & \searrow &~~~~~~~\swarrow & &
& \searrow &\\

\!\!\!\!\!\!\!\!\!\!\!\!\!\!\!\!\!\![0,0,0]_{(0,0)}& &
\!\!\!\!\!\!\!\!\!\!\!\!\!\!\!\!\!\!&
&\!\!\!\!\!\!\!\!\!\!\!\!\!\!\!\!\!\![0,0,0]_{(1,1)}& &
\!\!\!\!\!\!\!\!\!\!\!\!\!\!\!\!\!\!&
&\!\!\!\!\!\!\!\!\!\!\!\!\!\!\!\![0,0,0]_{(0,0)}
\end{matrix}
\ee
Note that there is a difference between diagrams \C{BigM}\ (for $p=2$)
and \C{LitM},  which we now explain.
First, there are 12 representations in \C{BigM}\ which involve $p-3$ as
a Dynkin label and they are absent for $p=2$. This actually follows
from the Racah-Speiser algorithm discussed in
Appendix~\ref{shortstructure}. One can show that all $SU(4)$
representations characterized by a highest weight state with $-1$ as a Dynkin label
vanish using \C{RSalg2} in the tensor product decomposition \C{RSalg}.

Second, there are $9$ representations with $p-4$ as a Dynkin label.
For $p=2$, using the Racah-Speiser algorithm, it can be shown that $5$ of
them vanish in \C{BigM}\ and so do not appear in \C{LitM}. The $4$ surviving
ones are: $[2,-2,2]_{(0,0)}$ with $\Delta =4$, $[2,-2,1]_{(\frac{1}{2},0)}$
and $[1,-2,2]_{(0,\frac{1}{2})}$ with $\Delta =\frac{9}{2}$ and
$[1,-2,1]_{(\frac{1}{2},\frac{1}{2})}$ with $\Delta =5$ and we have removed
these representations from \C{LitM}\ leading to multiplet shortening.

The reason we remove these representations is that they vanish after
we impose current conservation. In constructing the on-shell short
multiplet, we will impose current conservation. Without the on-shell condition, the
dimension of the multiplet would be incorrect.
This will become explicit when we discuss the operators corresponding
to this representation. As an aside, note that for $p>2$, there is no
multiplet shortening. For
$p=3$, the representations in \C{BigM}\ which have
$p-4$ as a Dynkin label vanish directly by the Racah-Speiser
algorithm; we do not have to impose the equations of motion.
For higher powers of $p$, all the Dynkin labels which appear in
\C{BigM}\ are non-negative and again there is no multiplet
shortening.

\subsection{More on the current multiplet}

The $p=2$ multiplet is the current multiplet~\cite{Bergshoeff:1981is}.
We will need to identify
operators  with the states of \C{LitM}. The
current multiplet contains
the energy momentum tensor $T_{\mu \nu} \sim [0,0,0]_{(1,1)}$, the
supersymmetry currents $J^{\mu}_{i\al} \sim [1,0,0]_{(1,\frac{1}{2})}$
and their conjugates ${\bar{J}}^{\mu i}_{\dal} \sim
[0,0,1]_{(\frac{1}{2},1)}$ and the $R$-symmetry currents $R^{\mu i}{}_j
\sim [1,0,1]_{(\frac{1}{2},\frac{1}{2})}$.

Apart from the currents mentioned above,
the bosonic operators in the multiplet are the real scalars $Q^{[ij]}{}_{[kl]}
\sim [0,2,0]_{(0,0)}$, the complex scalars ${\mathcal{E}}^{(ij)} \sim
[0,0,2]_{(0,0)}$,  $\Ot \sim [0,0,0]_{(0,0)}$, their conjugates
${\bar{\mathcal{E}}}_{(ij)} \sim [2,0,0]_{(0,0)}$ and $\bar\Om_{\tau}
\sim [0,0,0]_{(0,0)}$.
Lastly, there is an antisymmetric $2$-form $B^{[ij]}_{\mu\nu}
\sim [0,1,0]_{(1,0)}$ and its conjugate ${\bar{B}}_{[ij] \mu\nu}
\sim [0,1,0]_{(0,1)}$. It is critical that $\Ot$ sits in this
multiplet.

The remaining fermionic operators in the multiplet are
 the spin-$\frac{1}{2}$ fermions $\chi^i{}_{jk} \sim
[0,1,1]_{(\frac{1}{2},0)}$, $\Lambda^i \sim [0,0,1]_{(\frac{1}{2},0)}$
and their conjugates  ${\bar\chi}_i{}^{jk} \sim [1,1,0]_{(0,\frac{1}{2})}$
and ${\bar\Lambda}_i \sim [1,0,0]_{(0,\frac{1}{2})}$. These composite operators
can be constructed in terms of the fundamental fields in the abelian
theory~\cite{Bergshoeff:1981is}.
However, for most of our discussion, we will not need the classical
 expressions for the operators (note that some of the operators for the
 non-abelian theory have been written down in~\cite{Green:2002vf}).

Current conservation leads to multiplet shortening as discussed
before. To see this  note that
$$\partial^\mu T_{\mu \nu} \sim
[1,-2,1]_{(\frac{1}{2},\frac{1}{2})}, \qquad \partial_\mu J^{\mu}_{i\al} \sim
[2,-2,1]_{(\frac{1}{2},0)}, $$ $$ \partial_\mu {\bar{J}}^{\mu i}_{\dal} \sim
[1,-2,2]_{(0,\frac{1}{2})}, \qquad \partial_\mu R^{\mu i}{}_j
\sim [2,-2,2]_{(0,0)}.$$
The conservation of these currents
leads to the vanishing of these four representations.

There are certain aspects of the current multiplet which will be
useful later. We restrict ourselves to the free abelian theory for
this part of the discussion. This restriction imposes no loss of
generality since the operators in the non-abelian theory must satisfy
the same algebra.
The supersymmetry transformations in this case (with gauge
covariant derivatives going over to ordinary derivatives and with all
commutators set to zero) are given by
\bea \label{susy}
 \hat\del \varphi^{ij} &=& \frac{1}{2} (\lambda^i \eta^j -\lambda^j \eta^i)
+ \frac{1}{2} \epsilon^{ijkl} {\bar\eta}_k {\bar\lambda}_l, \cr
 \hat\del \lambda^i{}_{\al} &=& -\frac{1}{2}
(\s^{\mu \nu})_{\al}{}^{\! \beta} F_{\mu \nu} \eta^{\! ~i}{}_{\beta} +4i
\partial_{\al \dal} \varphi^{ij} {\bar\eta}_j{}^{\dal}, \cr
 \hat\del A_\mu &=& -i (\lambda^i \s_\mu {\bar\eta}_i +\eta^i \s_\mu
{\bar\lambda}_i),
\eea
where $\eta$ is a Grassmann parameter.
It will be important to remember that the supersymmetry transformations
are independent of the coupling with our choice of
action~\C{action}.
We can construct the supersymmetry
transformations of the current multiplet using \C{susy}, which we
display~\cite{Bergshoeff:1981is}
\bea \label{supervar}
\hat\del Q^{ij}{}_{kl} & =  & \frac{4}{3} i {\bar\eta}_{[k}
{\bar\chi}_{l]}{}^{ij} -\frac{2}{3} i \del^{[i}{}_{[k} {\bar\eta}_m
{\bar\chi}_{l]}{}^{mj]} + {\rm h.c.}, \cr
\hat\del \chi^k{}_{ij} & = & \frac{3}{4}\Big\{ i \epsilon_{ijmn} \s^{\mu \nu}
B^{km}_{\mu\nu} \eta^n  + i \epsilon_{ijmn} {\mathcal{E}}^{mk}
\eta^n  \cr &&
- i \s^\mu  R^{k}{}_{\mu [i} {\bar\eta}_{j]}
+2 i \s^\mu \partial_\mu Q^{kl}{}_{ij}
{\bar\eta}_{l} \Big\} -{\rm trace},\cr
\hat\del B^{ij}_{\mu\nu} & = & -\frac{1}{2} \epsilon^{ijkl} {\bar\eta}_{k}
{\bar\s}^\rho \s_{\mu \nu} J_{\rho l} - \eta^{[i} \s_{\mu \nu} \Lambda^{j]}
+\frac{2}{3} i \epsilon^{ijkl} {\bar\eta}_n  {\bar\s}^{\rho} \s_{\mu \nu}
\partial_\rho \chi^n{}_{kl},\cr
\hat\del {\mathcal{E}}^{ij} & =& \eta^{(i} \Lambda^{j)} +\frac{2}{3} i
\epsilon^{mnk(i}  {\bar\eta}_{k} {\bar\s}^\mu \partial_\mu
\chi^{j)}{}_{mn}, \cr
\hat\del R^{\mu i}{}_j & = & -\eta^{i} J^\mu_j +\frac{1}{4} \del^{i}{}_{j}
\eta^{k} J^\mu_k + \frac{8}{3} i\eta^{k} \s^{\mu \nu} \partial_\nu
\chi^i{}_{jk} -{\rm h.c.},\cr
\hat\del J_\mu{}_i & = & -\s^{\nu} T_{\mu \nu} {\bar\eta}_{i}
-2(\s_\rho {\bar\s}_{\mu \nu} -\frac{1}{3} \s_{\mu \nu} \s_\rho)
\partial^\nu R^{\rho k}{}_i {\bar\eta}_{k} \cr
&& -(\s_{\rho \s} \s_{\mu \nu}
+\frac{1}{3} \s_{\mu \nu} \s_{\rho \s}) \epsilon_{iklm} \partial^\nu
B^{kl \rho\s} \eta^m , \cr
\hat\del \Lambda^i & = & -i \eta^i \Ot  +\s_\mu \partial^\mu {\mathcal{E}}^{ik}
{\bar\eta}_{k} -\s^{\mu \nu} \partial_\rho B^{ij}_{\mu\nu} \s^\rho
{\bar\eta}_{j} ,\cr
\hat\del T_{\mu \nu} & = & \eta^i \s_{(\mu \vert \rho \vert}
\partial^\rho J_{\nu)i}
 + {\rm h.c.} , \cr
\hat\del \Ot & = & 2 i{\bar\eta}_i {\bar\s}^\mu \partial_\mu \Lambda^i +i\eta^i
(\partial^\mu J_{\mu i} +2
\s^\mu \bar\s^\nu \partial_\mu J_{\nu i}),
\eea
There are similar expressions for the conjugate operators.
Note that the terms $\eta^i \partial^\mu J_{\mu i}$ and $\eta^i \s^\mu
\bar\s^\nu \partial_\mu J_{\nu i}$ in the variation
of $\Ot$ do not appear in~\cite{Bergshoeff:1981is}.
This is because both these terms vanish classically by current
conservation and the spin-$1/2$ anomaly cancellation condition.
However, both these conditions are violated in the quantum theory;
this can be seen by studying
Ward identities where the violation is caused by contact terms. We
will analyze this violation in considerable detail in the following
section where it will become clearer why we require these additional
terms in the
supervariation of $\Ot$.

There are a few points worth highlighting in \C{supervar}.
First, as expected, the current multiplet varies into itself.
Also, note that  $B^{ij}_{\mu\nu}$ transforms into $J^\mu_i$, while
${\mathcal{E}}^{ij}$ does not; this will also be important later.
The variation of $Q^{ij}{}_{kl}$ contains no conformal descendent. On
the other hand,  $T_{\mu \nu}, \Ot$ and $\bar{\Om}_\tau$ vary only
into conformal descendents. The remaining operators vary into
combinations of primaries and descendents.

{}From the
supersymmetry transformations, we see that it is easy to recover \C{LitM}.
{}For example, from the variation $\hat\del \chi^i{}_{jk}$, we see
that $Q$ acting on $\chi^i{}_{jk}$ gives $B^{ij}_{\mu\nu}$ and ${\mathcal{E}}^{ij}$,
while acting with $\bar{Q}$ gives $R^{\mu i}{}_j$. The other term in
$\hat\del \chi^i{}_{jk}$ is a conformal descendent.

\subsection{The structure of $\del \Ot$}
\label{desc}

As a matter of notation, we will
denote the superconformal primary in the representation $[0,p,0]_{(0,0)}$
 for arbitrary $p$ by
${\mathcal{O}}_p$. We will also
denote the action of $Q$ by $\del$ and the action of $\bar{Q}$ by
$\bar\del$ as in~\cite{Intriligator:1998ig}. Acting on operators,
$\del^r {\bar\del}^s$ stands
for a sequence of graded commutators; for example,
$$\del {\bar\del}^2 {\mathcal{O}} \,\leftrightarrow\, [Q,[ \bar{Q},[\bar{Q},
{\mathcal{O}}]_\pm]_\mp]_\pm. $$
 Note that for all values of $p$, $\{Q,\bar{Q}\}$ is always zero
(and never proportional to $P_{\mu}$) for the particular supercharges
 used in diagram~\C{BigM}. Therefore, starting from ${\mathcal{O}}_p$,
 we can reach any conformal primary in
the multiplet by acting suitably with $\del$ and $\bar\del$ operators in
an arbitrary way (up to an overall sign), without worrying about the
ordering of the operators.

{}For the current multiplet, it might appear from diagram~\C{LitM}\
that $\del^5 =0$. This, however, is not the case as we will now
demonstrate. Consider the two-point correlator $\langle \Ot (x)
\Ot (y) \rangle$. We can express this correlator in the form
\be \label{otot}
\langle \Ot (x) \Ot (y) \rangle = \langle \del (\del^3 \OT (x)
\Ot (y)) \rangle +\langle \del^3 \OT (x) \del^5 \OT (y) \rangle\ee
where the first term on the right hand side vanishes because $\del$
kills the vacuum. By computing both sides in the abelian theory, we
can determine the structure of $\del^5 \OT = \del \Ot$.

In the abelian theory, the operator $\Ot$ is given by
\be \label{act} \Ot = \frac{\tau_2}{8\pi} \left(F_{\mu \nu} F^{\mu \nu} +i
F_{\mu \nu} {\widetilde{F}}^{\mu \nu} -4i \bar\lambda_i \bar\sigma_\mu
\partial^\mu \lambda^i +
4 \bar\varphi_{ij}\partial^\mu \partial_\mu\varphi^{ij} \right)\ee
where $ {\widetilde{F}} = \ast F$.
{}For clarity, let us first restrict to the terms in $\Ot$ that depend
only on the field strength. We will show that the supervariation of
these terms gives rise to a conformal
descendent (the analysis for the other terms is similar).
Using the supervariations \C{susy}, we see that
\be \label{onemore}
\hat\del \Ot = -\frac{i\tau_2}{2\pi}
\left[(F_{\mu \nu} +i {\widetilde{F}}_{\mu \nu}) (\partial^\mu
\lambda^i \s^\nu {\bar\eta}_i +\eta^i \s^\nu
\partial^\mu {\bar\lambda}_i)\right]+\ldots \ee
where the omitted terms are generated from varying the scalars and
fermions of~\C{act}.
After some lengthy algebra, we can write this as
\be \label{cftdes}
\hat\del \Ot = 2i ({\bar\eta}_i \bar\s^\mu \partial_\mu \Lambda^i +
\eta^i \partial^\mu J_{\mu i}) + \ldots,
\ee
which includes one of the terms of $\del \Ot$ appearing in \C{supervar} and where
\bea \label{vardef}
& \Lambda^i{}_{\al} =  \left(\del^3 \OT\right)^i{}_{\! \al} =
-\frac{\tau_2}{4\pi} (\s^{\mu \nu})_{\al}{}^{\beta}
\left(F_{\mu \nu} \lambda^i{}_{\beta}\right),  \cr
&  J^\mu_{i\al} = \left(\del^2 \bar\del \OT \right)^\mu_{i\al}
=-\frac{\tau_2}{4\pi} \left(F_{\rho \s} (\s^{\rho \s} \s^\mu
{\bar\lambda}_i)_\al +\ldots\right).
\eea
The omitted terms again involve $(\bar\varphi, \lambda)$.
{}From the expression for $J_i$ in \C{vardef}, we see that
the spin-$1/2$ anomaly cancellation condition,
\be \label{spin12}\bar\s_\mu J^\mu_i =0, \ee
is trivially satisfied without
using the equations of motion and so plays no role in \C{cftdes}. We also
see that
\be \label{spinanom} \partial_\mu J^\mu_i =-\frac{\tau_2}{4\pi} \partial^\mu
\left[(F_{\mu \nu} +i {\widetilde{F}}_{\mu \nu}) \s^\nu
{\bar\lambda}_i \right]+ \ldots,\ee
which vanishes on-shell. After using the Bianchi identity and the
equations of motion~\C{spinanom}\ agrees
with~\C{onemore}.

Let us compute the left hand side of~\C{otot}\ in free field theory.
We use the gauge field propagator in Feynman gauge given by
\be \label{prop} \Delta_{\mu \nu}^{ab} (x-y) =
\frac{\del^{ab} \eta_{\mu \nu}}{\pi\tau_2 (x-y)^2}.\ee
This satisfies the usual relation
$\partial^2 \Delta_{\mu \nu}^{ab} (x-y) =-\frac{4\pi}{\tau_2}
\del^4 (x-y) \del^{ab} \eta_{\mu \nu}$. Using the contraction
\be
\langle {F_{\mu \nu}^a (x) F_{\rho \s}^b (y)} \rangle =
\frac{\del^{ab}}{\pi \tau_2}
[\eta_{\nu \s} \partial_{\mu}^x \partial_{\rho}^y -\eta_{\nu \rho}
\partial_{\mu}^x \partial_{\s}^y -\eta_{\mu \s} \partial_{\nu}^x
\partial_{\rho}^y +\eta_{\mu \rho} \partial_{\nu}^x
\partial_{\s}^y] \frac{1}{(x-y)^2}, \ee
we deduce the relations
\bea \label{eqns}
\langle (F_{\mu \nu}^a F^{\mu \nu a}) (x)
(F_{\rho \s}^b F^{\rho \s b}) (y) \rangle =\frac{16}{(\pi \tau_2)^2}
\left[(\partial_\mu \partial_\rho \frac{1}{(x-y)^2})^2 +8\pi^4
(\del^4 (x-y))^2 \right],\cr
\langle (F_{\mu \nu}^a {\widetilde{F}}^{\mu \nu a}) (x)
(F_{\rho \s}^b {\widetilde{F}}^{\rho \s b}) (y) \rangle
=\frac{16}{(\pi \tau_2)^2}
\left[(\partial_\mu \partial_\rho \frac{1}{(x-y)^2})^2 -16\pi^4
(\del^4 (x-y))^2 \right],
\eea
while
\be \langle (F_{\mu \nu}^a F^{\mu \nu a}) (x) (F_{\rho \s}^b
{\widetilde{F}}^{\rho \s b}) (y)\rangle =0.\ee
Note that up to total derivatives, the second equation in~\C{eqns}\ is
zero using
\be \left(\partial_\mu \partial_\rho \frac{1}{(x-y)^2} \right)^2
=\left(\partial^2 \frac{1}{(x-y)^2} \right)^2
= 16\pi^4 \left\{ \del^4 (x-y) \right\}^2. \ee
After summing these various contributions, the left hand side of \C{otot}\ yields
\be \label{value} \langle \Ot (x) \Ot (y) \rangle = \frac{3}{2}
\left\{ \del^4 (x-y) \right\}^2.\ee
The left hand side of \C{otot} is non-vanishing -- in fact, it is
a contact term. What about the right-hand side of \C{otot}? Only the
second term can be non-vanishing and indeed, $\del^5 \OT$ is a
conformal descendent as we see from~\C{cftdes}.

{}From \C{onemore}, note that
\be
\left( \del \Ot \right)_{i \al} = \frac{i\tau_2}{2\pi} \partial^\mu
 \left[(F_{\mu \nu} +i {\widetilde{F}}_{\mu \nu})\s^\nu
{\bar\lambda}_i \right]_\al. \ee
We now compute
$\langle \Lambda_{\al}{}^{\! i} (x) \left( \del \Ot \right)_{i}{}^{\! \al} (y) \rangle$.
There is no sum over $i$ or $\al$, but it is easier to evaluate the
correlator by summing and dividing by $8$.
We use the fermion propagator
\be \label{fermprop}
\langle {\lambda^i{}_{\! \al} (x) {\bar\lambda}_{j \dbt}
(y)} \rangle  = \frac{i}{\pi\tau_2} (\s^\mu)_{\al \dbt} \partial_\mu
\frac{1}{(x-y)^2} {\del}^i{}_{\! j}
\ee
and some lengthy but straightforward algebra to obtain
\be
\langle \Lambda(x) \delta \OT(y)\rangle
\equiv
\langle \del^3 \OT (x) \del^5 \OT (y) \rangle =\frac{3}{2}
\left\{\del^4 (x-y) \right\}^2.\ee
As had to be the case, this is a contact term in agreement
with~\C{value}.

So from this free field analysis, we see that $\del \Ot \sim
\eta^i \partial^\mu J_{\mu i}$.  It is similarly easy to find the $\eta^i
\s^\mu \bar\s^\nu \partial_\mu J_{\nu i}$ term in $\del \Ot$ as written in
\C{supervar}. To see this  we consider the full expression for $\Ot$
in \C{act} and consider its supervariation. Including all the
contributions and integrating by parts, the total contribution gives
\be \label{otmorevar} \del \Ot = i \eta^i
\partial^\mu J_{\mu i} +\frac{2\tau_2}{\pi}
(\eta^i \s^\nu \bar\s^\mu \partial_\mu
\lambda^j) \partial_\nu \bar\varphi_{ij} . \ee
We now consider the complete expression for $J_i$,
\be J^\mu_{i\al}
=-\frac{\tau_2}{4\pi} {\rm Tr} (F_{\rho \s} (\s^{\rho \s} \s^\mu
{\bar\lambda}_i)_\al
-2i {\bar\varphi}_{ij} {\buildrel \leftrightarrow \over
\partial}^\mu
\lambda^j{}_{\al} -\frac{4}{3} i \partial_\nu (\s^{\mu \nu}
{\bar\varphi}_{ij} \lambda^j)_\al), \ee
where we define $  \Om_i {\buildrel \leftrightarrow \over
\partial} \Om_j = \Om_i \left( \p \Om_j\right) - \left( \p \Om_i \right) \Om_j $.
The extra terms beyond those of~\C{vardef}\ do not satisfy $\bar\s^\mu
J_{\mu i} =0$ trivially; in fact, they satisfy
\be \label{notzero}
\bar\s^\mu J_{\mu i} = -\frac{i\tau_2}{\pi} \bar\varphi_{ij}
\bar\s^\mu \partial_\mu \lambda^j,\ee
which only vanishes on-shell. Using~\C{notzero}\ to compute $\eta^i \s^\mu
\bar\s^\nu \partial_\mu J_{\nu i}$, we recover the expression given
in~\C{otmorevar}. So~\C{otmorevar}\ becomes
\be \label{otfinvar} \del \Ot = i \eta^i
\partial^\mu J_{\mu i} +2i \eta^i
\s^\mu \bar\s^\nu \partial_\mu J_{\nu i},\ee
giving us the desired relation in \C{supervar}.
So while $\del \Ot$ vanishes on-shell, it is actually non-vanishing
because of contact terms. These contact terms can also be seen from a Ward
identity for condition~\C{spin12}\ which we now deduce.

\subsection{Ward identities}

Let us begin by recalling the more familiar Ward identities involving
the stress-energy tensor. We will use these identities in the
following section. Consider an operator, $\Phi^I$, with
conformal dimension $\Delta_I$ transforming
in representation $I$ of the Lorentz group with generators
$(S_{\mu\nu})^I{}_J$.
The Ward identities state that,
\bea \label{Wardo}
&\partial^\mu \langle T_{\mu\nu} (z) \Phi^{I_1} (x_1)
\cdots \Phi^{I_n} (x_n)
\rangle = \sum_{i=1}^n \partial_{\nu_i} \del^4 (z-x_i)
\langle  \Phi^{I_1} (x_1) \cdots \Phi^{I_n} (x_n) \rangle,  \non\\ &
 \langle T_{[\mu\nu]} (z) \Phi^{I_1} (x_1)
\cdots \Phi^{I_n} (x_n)
\rangle = \frac{i}{2} \sum_{i=1}^n (S_{\mu\nu})^{I_i}{}_J \del^4 (z-x_i)
\langle \Phi^{I_1} (x_1) \cdots \Phi^{J} (x_i)\cdots\Phi^{I_n} (x_n)
\rangle,  \non \\
&  \langle T^\mu_\mu (z) \Phi^{I_1} (x_1) \cdots \Phi^{I_n} (x_n)
\rangle =  \sum_{i=1}^n \Delta_{I_i} \del^4 (z-x_i)
\langle \Phi^{I_1} (x_1) \cdots \Phi^{I_n} (x_n)\rangle.
\eea
 The first and the last identities
express the breakdown of energy-momentum conservation and conformal invariance. The
second equation shows that
the stress tensor is not symmetric at the location of the operators
inserted in the correlator.

The Ward identity for the supercurrent conservation,  $\partial_\mu
J^\mu_i=0$,  is given by
\be \label{jward}
\partial^\mu \langle J^{\alpha}_{\mu i} (z) \Phi^{I_1} (x_1)
\cdots \Phi^{I_n} (x_n)
\rangle = \sum_{j=1}^n \del^4 (z-x_j)
\langle  \Phi^{I_1} (x_1) \cdots [Q_i{}^\alpha,\Phi^{I_j} (x_j)]_{\pm}
\cdots \Phi^{I_n} (x_n) \rangle. \ee
The divergence of the supercurrent is non-zero at the positions
of inserted operators in the correlator.

In order to deduce the Ward identity for $\bar\s_\mu J^\mu_i=0$, we first
consider the Ward identity for the superconformal current $I_{\mu\alpha}^i$
\be \label{iward}
\partial^\mu \langle I_{\mu\alpha}^i (z) \Phi^{I_1} (x_1)
\cdots \Phi^{I_n} (x_n)
\rangle = \sum_{j=1}^n \del^4 (z-x_j)
\langle  \Phi^{I_1} (x_1) \cdots [S_\alpha^i ,\Phi^{I_j} (x_j)]_{\pm}
\cdots \Phi^{I_n} (x_n) \rangle. \ee
We now use the relation between these two currents given
in~\cite{Ferrara:1975pz} (see~\cite{Green:1999qt}\ for a recent discussion)
\be \label{relcurr} I_{\mu\alpha}^i (x) = x_{\alpha\dal}
\bar{J}_\mu^{i\dal} (x).\ee
Inserting \C{relcurr}\ into \C{iward}, using \C{jward} and then taking the
conjugate gives
\bea \label{anward}
(\bar\s_\mu)^{\dal\alpha} \langle J^\mu_{i\alpha} (z) \Phi^{I_1} (x_1)
\cdots \Phi^{I_n} (x_n)
\rangle &=& \sum_{j=1}^n \del^4 (z-x_j)
\langle  \Phi^{I_1} (x_1)  \cdots \cr
&& \cdots  [\bar{S}_i^{\dal} -x_j^{\dal\alpha}
Q_i{}^\alpha,\Phi^{I_j} (x_j)]_{\pm} \cdots \Phi^{I_n} (x_n) \rangle. \eea
This is the desired Ward identity.

\section{Constraining the $\Ot$ OPE}
\subsection{Overview}

The strategy we will use to determine the $\Ot$ OPE
makes use of the Ward identities of the superconformal currents.
More precisely, we start with the stress-tensor
\be \label{shorteq} T \sim {\bar\del}^2 \del^2 \OT \ee
whose OPE with an operator $\Phi$
is in part determined by Ward identities.
What is not determined by
Ward identities are local operators in the $T(z)\Phi(x)$ OPE which
do not reside in the same supermultiplet as $\Phi$. These can include
long, BPS and semi-short operators. Examples of semi-short multiplets include 
multi-trace operators; for an example,  see~\cite{Arutyunov:2000ku}. For example, the
Konishi operator, ${\rm Tr} (\phi^2)$, can appear in the $T(z)T(x)$ OPE but is
not determined by Ward identities. Since this operator is long, its
dimension is renormalized. Therefore, whether this term is singular in
the OPE depends on the value of the coupling. This same caveat will
apply to all the OPEs that we determine from Ward identities.

Peeling off a $\bar\del$ in~\C{shorteq}\
results in  the OPE of ${\bar\del} \del^2 \OT$ (which
is the supercurrent, $ J^\alpha_{\mu i}$) with an arbitrary operator,
modulo that caveat mentioned above.
Since the OPE of $J^\alpha_{\mu i}$ is also determined in part
by a Ward identity, this will allow us to check that the peeling off
procedure works. We will then remove the remaining ${\bar\del}$ to
determine the OPE
of $\del^2 \OT$, which is not a current.

The next step is to construct $\del^3 \OT$ by
applying a $\del$ and recursively using the known OPE
results.  Finally, we apply $\del$ to obtain the OPE of $\Ot$
with any other operator. For a scalar operator, this OPE will depend on $15$ coefficients
$a_i(\tau,\bar\tau)$.
In section~\ref{nice}, we will show that these $15$ coefficients are
independent of the coupling for BPS operators.
It should be noted that for N=1 supersymmetric
theories, there exists a supermultiplet of
currents~\cite{Ferrara:1975pz}. The lowest component of  the multiplet
is the $R$-symmetry current, the middle component is the supercurrent, while the top
component is the stress tensor. The OPE of these currents collectively
can be determined in superspace. Our peeling off procedure will
match these known results for the first step taking us from the stress
tensor to the supercurrent.
{}For a recent discussion of the current supermultiplet and its OPE
structure, see, for example,~\cite{Anselmi:1997mq}.

\subsubsection{The stress tensor OPE}

The Ward identities~\C{Wardo}\ uniquely determine a class of terms in
the OPE of the
stress-energy tensor. The most singular terms in the OPE are given
by
\bea \label{OPE1} T_{\mu\nu} (z) \Phi^I (x) &\sim & \Phi^I (x)
\partial_\mu \partial_\nu
\frac{1}{(z-x)^2} + \Phi^I (x) \eta_{\mu \nu}
\del^4 (z-x)
+ \Big[ \frac{i}{2} (S_{\mu\nu})^I{}_J \Phi^J (x) \del^4 (z-x)
\nonumber \\ & &
+ \frac{i}{8\pi^2} \Phi^J (x) \left(S_{\lambda \mu}
\partial^\lambda \partial_\nu +S_{\lambda \nu}
\partial^\lambda \partial_\mu \right)^I_J
\frac{1}{(z-x)^2} \Big] \non\\ &&
+{\rm less ~ singular ~ terms}. \eea
These are all the most singular terms that involve operators in the
$\Phi^I$ supermultiplet.
The symbol $\sim$ in this equation indicates that we have ignored coefficients.
Some of the terms appearing on the right hand side can
have coupling dependent coefficients although the relative coefficient of the
bracketed terms is precise. In particular, those terms that
contribute to the violation of scale invariance will generally be renormalized.

Now we need to express $T_{\mu \nu}$ in terms of the supercharges and $\OT$.
{}From \C{supervar} we deduce the following
supersymmetry transformations:
\be \label{eqnone} [Q_{p \al} ,Q^{ij}{}_{kl}]= \frac{2}{3} i
\del^{[i}{}_{[k} \del^m_{\vert p \vert}
\chi^{j]}{}_{ml] \al} -\frac{4}{3} i \del^{[i}{}_p \chi^{j]}{}_{kl \al},\ee
\be \label{eqntwo} \{Q_n{}^\beta ,\chi^k{}_{ij \al}\}= -\frac{3}{4} i
\epsilon_{ijmn} \Sigma_{\alpha}{}^{\beta km} -\frac{3}{8} i \del^k{}_{[i}
\epsilon_{j]pmn} \Sigma_{\alpha}{}^{\beta pm},\ee
\be \label{eqnthree} [{\bar{Q}}^{k \dal} , B^{ij}_{\mu\nu}] =-\frac{1}{2}
\epsilon^{ijkl} ({\bar\sigma}_\rho
\sigma_{\mu \nu})^{\dal \alpha} J^\rho_{l\alpha}
+\frac{2}{3} i \epsilon^{ijnl} (\bar\sigma^\rho
\sigma_{\mu \nu})^{\dal \alpha} \partial_\rho
\chi^{k}{}_{nl \alpha},\ee
\be \label{eqnfour} [{\bar{Q}}^{k\dal} , {\mathcal{E}}^{ij}]
=\frac{2}{3} i \epsilon^{mnk(i} \partial_\mu
\chi^{j)}{}_{mn \alpha} \bar\sigma^{\mu \dal \alpha}, \ee
and
\be \label{eqnfive} \{{\bar{Q}}^j{}_{\dal}, J_\mu{}_{i\alpha}\} =
\sigma^\nu{}_{\alpha \dal} T_{\mu \nu} \del^j{}_i  +2(\sigma_\rho
\bar\sigma_{\mu \nu} -\frac{1}{3} \sigma_{\mu \nu}
\sigma_\rho)_{\alpha \dal} \partial^\nu R^{\rho j}{}_i.\ee
Here,
\be \label{sigdef}
\Sigma_{\alpha}{}^{\beta ij} =  \sigma^{\mu \nu}_{\alpha}{}^{\beta}
B^{ij}_{\mu\nu} +\del_{\alpha}{}^{\beta} {\mathcal{E}}^{ij}.\ee
Using these relations, after some tedious algebra, we find that
\be \label{stress} T_{\mu \nu} =  (\sigma_\mu){}_{\alpha \dal}
(\sigma_\nu){}_{\beta \dbt} \{{\bar{Q}}^{l\dbt},[{\bar{Q}}^{k\dal},\{
Q_j{}^\beta,[Q_i{}^\alpha,Q^{ij}{}_{kl}]\}]\}.\ee

We have set the irrelevent numerical factor on the right
hand side of
\C{stress} to one.
%
Note that the right-hand side of~\C{stress}\ is
manifestly symmetric in $\mu$ and $\nu$, using the fact that the various
$Q$ and $\bar{Q}$ operators anti-commute. Also~\C{stress}\
satisfies $T^\mu_\mu =0$ by construction. These classical constraints are
violated
quantum mechanically in accord with \C{Wardo}.
\subsubsection{The supercurrent OPE}
\label{superOPE}

Given the OPE \C{OPE1}, we now want to construct the supercurrent OPE
by peeling off a $\bar{Q}$.
We begin with the following relation:
\be \label{jdiff} [{\bar{Q}}^{k\dal},\{Q_j{}^\beta,[Q_i{}^\alpha,
Q^{ij}{}_{kl}]\}] =
\left\{ (\bar\sigma^\mu)^{\dal \beta} J^{\alpha}_{\mu l} +
(\bar\sigma^\mu)^{\dal \alpha} J^{\beta}_{\mu l}\right\},\ee
%
%
which leads to
\be \label{Jdef} J^{\alpha}_{\mu l} = (\sigma_\mu)_{\beta \dal}
[{\bar{Q}}^{k\dal},\{Q_j{}^\beta,[Q_i{}^\alpha,
Q^{ij}{}_{kl}]\}]. \ee
%
%
Note that
$\bar\sigma^\mu J_{\mu i} =0$ by construction, satisfying the
classical spin-$1/2$ anomaly cancellation condition.
In deriving these relations,
we do not pick up any descendents because the possible descendents involve
derivatives of either $R^{\mu i}{}_i$ or
$\chi^{i}{}_{ij \alpha}$, both of which vanish.

Now let us calculate the contribution of the
three-point function to the integral
\be  \int d^4 z \langle T_{\mu \nu} (z)
\Phi^I (x) \Phi^J (y) \rangle,\ee
as $z \rightarrow x$ and also as $z \rightarrow y$. Considering only
the most singular terms in the OPE \C{OPE1}, this reduces to
\bea && \int d^4 z \langle T_{\mu \nu} (z)
\Phi^I (x) \Phi^J (y) \rangle =
\langle \Phi^I (x) \Phi^J (y) \rangle \left(\int_{B_x^\epsilon} d^4 z
\partial_\mu \partial_\nu
\frac{1}{(z-x)^2} +
\int_{B_y^\epsilon} d^4 z \partial_\mu \partial_\nu
\frac{1}{(z-y)^2} \right)  \non \\
&& +2 \langle \Phi^I (x) \Phi^J (y) \rangle \eta_{\mu \nu}
+\frac{i}{2} (S_{\mu\nu})^I{}_K \langle
\Phi^K (x) \Phi^J (y) \rangle
+\frac{i}{2} (S_{\mu\nu})^J{}_K \langle
\Phi^I (x) \Phi^K (y) \rangle
\non \\
&& +\frac{i}{8\pi^2} \langle \Phi^K (x) \Phi^J (y) \rangle
\int_{B_x^\epsilon} d^4 z
\left[S_{\lambda \mu}
\partial^\lambda \partial_\nu +S_{\lambda \nu}
\partial^\lambda \partial_\mu
\right]^I{}_K \frac{1}{(z-x)^2} \qquad \qquad \nonumber
\\ &&
+\frac{i}{8\pi^2} \langle \Phi^I (x) \Phi^K (y) \rangle
\int_{B_y^\epsilon} d^4 z
\left[S_{\lambda \mu}
\partial^\lambda \partial_\nu +S_{\lambda \nu}
\partial^\lambda \partial_\mu
\right]^J{}_K \frac{1}{(z-y)^2},\qquad \qquad
\eea
where $B_x^\epsilon (B_y^\epsilon)$ is a small ball of radius $\epsilon$
centered at $x (y)$. From now on, we shall drop the integrals for
clarity and write this as
\bea \label{T3pt} \langle T_{\mu \nu} (z)
\Phi^I (x) \Phi^J (y) \rangle &=&
\langle \Phi^I (x) \Phi^J (y) \rangle
\partial_\mu \partial_\nu \left(
\frac{1}{(z-x)^2} +\frac{1}{(z-y)^2} \right)
\non\\ && +\eta_{\mu \nu} \langle \Phi^I (x) \Phi^J (y) \rangle
\left(\del^4 (z-x) +\del^4 (z-y) \right)
\nonumber\\
&& +(S_{\mu\nu})^I{}_K \langle
\Phi^K (x) \Phi^J (y) \rangle \del^4 (z-x)
+(S_{\mu\nu})^J{}_K \langle
\Phi^I (x) \Phi^K (y) \rangle \del^4 (z-y)
\nonumber\\
&& +\langle
\Phi^K (x) \Phi^J (y) \rangle \left[S_{\lambda \mu}
\partial^\lambda \partial_\nu +S_{\lambda \nu}
\partial^\lambda \partial_\mu
\right]^I{}_K
\frac{1}{(z-x)^2}
\nonumber \\
&& +\langle
\Phi^I (x) \Phi^K (y) \rangle \left[S_{\lambda \mu}
\partial^\lambda \partial_\nu +S_{\lambda \nu}
\partial^\lambda \partial_\mu
\right]^J{}_K
\frac{1}{(z-y)^2} ,
\eea
with the integrals around the various points implied.
Now the left hand side of~\C{T3pt}\ can we rewritten using~\C{stress}\
and~\C{Jdef},
\be \label{Tope}
\langle T_{\mu \nu} (z) \Phi^I (x) \Phi^J (y) \rangle
= (\sigma_\nu)_{\beta \dbt}\langle
J^{\beta}_{\mu l} (z) [{\bar{Q}}^{l\dbt} ,\Phi^I (x) \Phi^J (y)]\rangle.
\ee
%
%
We now want to write down the OPE of $J^{\alpha}_{\mu i}$ with $\Phi^I$
such that the right-hand side of~\C{Tope}\ yields all the terms in
\C{T3pt}. We make the ansatz
\bea \label{JOPEless} && J^{\alpha}_{\mu i} (z) \Phi^I (x)
 \sim
[Q_i{}^\alpha ,\Phi^I (x)]_{\pm}
\partial_\mu \frac{1}{(z-x)^2}
+[Q_i{}^\beta ,\Phi^I (x)]_{\pm} (\sigma_{\mu \nu})_\beta{}^\alpha
\partial^\nu \frac{1}{(z-x)^2} \nonumber \\
&& +
(S_{\mu\nu})^I{}_J [Q_i{}^\alpha ,\Phi^J (x)]_{\pm}
\partial^\nu \frac{1}{(z-x)^2}
+
(S_{\rho\lambda})^I{}_J  [Q_i{}^\beta ,\Phi^J (x)]_{\pm}
(\sigma^{\rho \lambda} \sigma_{\mu \nu})_\beta{}^\alpha
\partial^\nu \frac{1}{(z-x)^2}\nonumber\\
&& +
\Big\{ (S_{\mu\nu})^I{}_J [Q_i{}^\beta ,\Phi^J (x)]_{\pm}
(\sigma^{\nu \lambda})_\beta{}^\alpha
\partial_\lambda \frac{1}{(z-x)^2}
+(S^{\nu\lambda})^I{}_J [Q_i{}^\beta ,\Phi^J (x)]_{\pm}
(\sigma_{\nu\mu})_\beta{}^\alpha
\partial_\lambda \frac{1}{(z-x)^2} \Big\} \nonumber \\
&& +{\rm less ~singular ~terms}.
\eea
%
We have made this ansatz because all the terms on the right-hand side
of~\C{Tope}\ are of the form $\langle J \Phi \bar\del \Phi \rangle$ while the
right hand side of~\C{T3pt}\ behaves as
$\langle \Phi  \Phi \rangle \partial^2 (\frac{1}{z^2})$. It is therefore
necessary for the OPE to have the form
$J \Phi \sim \del \Phi$, so that terms like
$\langle \del \Phi \bar\del \Phi \rangle =\langle [Q, \Phi]_{\pm}
[{\bar{Q}}, \Phi]_{\pm} \rangle$ arise on the right hand side of~\C{Tope}.
Using $\{ {\bar{Q}},Q\} \sim P_\mu$ such terms give derivatives,  leading to
 $P_\mu \langle \Phi \Phi \rangle$. We need one more
derivative in the
OPE in order to match~\C{T3pt}\ which suggests~\C{JOPEless}.
Also, $J\Phi \sim \bar\del \Phi$ is ruled out by a
mismatch of $SU(4)$ indices.
Finally, terms with higher numbers
of $\del$ and $\bar\del$ operators have been dropped because they lead to
less singular terms.

We have used the relation  $[Q_{i\alpha},
S_{\mu\nu}] =[\bar{Q}^{j\dal}, S_{\mu\nu}]=0$ which  can be proven as follows.
the commutator $[Q_{i\alpha}, S_{\mu\nu}]$ could contain symmetry
generators in the $(1/2,0)$ or $(3/2,0)$ representations.
However, there are no $(3/2,0)$
generators and the only possible $(1/2,0)$ generator is $Q_{i\alpha}$.
The superconformal generator, $S^i{}_\alpha$, is in the
$(1/2,0)$ representation but is ruled out by its $SU(4)_R$ quantum
numbers. So the most general possibility is
\be [Q_{i\alpha}, S_{\mu\nu}] =c_i{}^j
(\s_{\mu\nu})_{\alpha}{}^{\beta} Q_{j\beta},
\ee
which, on conjugation, yields
\be [\bar{Q}^{i\dal} , S_{\mu\nu}] =c^i{}_j
(\bar\s_{\mu\nu})^{\dal}{}_{\dbt} \bar{Q}^{j\dbt},
\ee
where $(c_i{}^j)^* = c^i{}_j$. Now consider the Jacobi identity
\be \label{Jacobi}
[S_{\mu\nu}, \{Q_{i\alpha}, \bar{Q}^{j\dbt} \}] +
\{Q_{i\alpha}, [\bar{Q}^{j\dbt}, S_{\mu\nu}]\} -
\{ \bar{Q}^{j\dbt}, [S_{\mu\nu}, Q_{i\alpha} ]\} =0.
\ee
Using the relation $[S_{\mu\nu}, P_\lambda]=0$,~\C{Jacobi}\
gives
\be \label{Jacobimore}
c_i{}^j (\s_{\mu\nu} \s_\lambda \epsilon)_\alpha{}^{\dbt}
P^\lambda + c^j{}_i (\s_\lambda \epsilon)_\alpha{}^{\dal}
(\bar\s_{\mu\nu})^{\dbt}{}_{\dal} P^\lambda =0.\ee
Contracting with $(\s_{\mu\nu})_{\beta}{}^{\alpha}$,
equation~\C{Jacobimore}\  yields
\be c_i{}^j (\s_\mu \epsilon)_\beta{}^{\dbt} P^\mu =0, \ee
which implies that $c_i{}^j =0$. The crucial step is that
$[S_{\mu\nu}, P_\lambda]=0$, i.e., momentum commutes with
`intrinsic' angular momentum. On the other hand, for the
`orbital' angular momentum, $L_{\mu\nu}$, we know that
$$[L_{\mu\nu}, P_\lambda] =i(\eta_{\mu\lambda} P_\nu -
\eta_{\nu\lambda} P_\mu).$$
The Jacobi identity \C{Jacobi} for
$L_{\mu\nu}$ implies that the structure constant, $\tilde{c}$, in
 $[Q_{i\alpha}, L_{\mu\nu}]$ satisfies $\tilde{c}_i{}^j =\del_i{}^j$.
This is exactly as it should be because using $M_{\mu\nu}
= L_{\mu\nu} +iS_{\mu\nu}$, this gives
$$[Q_{i\alpha}, M_{\mu\nu}] =(\s_{\mu\nu})_\alpha{}^\beta
Q_{i\beta}. $$
In summary, inserting~\C{JOPEless}\ into~\C{Tope},
we obtain (to leading order) all terms in~\C{T3pt}. This justifies
the OPE~\C{JOPEless}. As a check, we note that from~\C{JOPEless}, we
obtain the supercurrent Ward identity~\C{jward}.
Note that the $S_{\m\n}$ terms do not contribute
to the Ward identity.

\subsubsection{The $\del^2 \OT$ OPE}

Now given \C{JOPEless}, we want to
construct the OPE of $\del^2 \OT$. So we need to peel off another
$\bar\del$.
Restricting to the most singular terms, we first calculate the
three-point function
\be \langle  J_{\alpha \mu i} (z) \Phi^I (x)
[{\bar{Q}}^j{}_{\dal}, \Phi^J (y)]_{\pm}
\rangle .\ee
Consider the contribution to the correlator as
$z \rightarrow x$ and $z \rightarrow y$. To leading
order, we see that
\bea
\label{Jcorr} && \langle  J_{\alpha \mu i} (z) \Phi^I (x)
[{\bar{Q}}^j{}_{\dal}, \Phi^J (y)]_{\pm}
\rangle =  \pm  \del^j{}_i (\sigma^\nu)_{\alpha \dal}
\langle \Phi^I (x) \Phi^J (y) \rangle
\partial_\mu \partial_\nu
\left(\frac{1}{(z-x)^2} +\frac{1}{(z-y)^2} \right) \nonumber\\
&&
\pm  \del^j{}_i (\sigma_\mu)_{\alpha \dal}
\langle \Phi^I (x) \Phi^J (y) \rangle
\left(\del^4 (z-x) + \del^4 (z-y)\right) \nonumber  \\
&&
\pm i \del^j{}_i \langle \Phi^K (x) \Phi^J (y) \rangle
\left[S_{\mu\nu}  \sigma_\lambda
\partial^\lambda \partial^\nu
+S^{\lambda\nu} \sigma_{\nu\mu}
\sigma^\rho
\partial_\lambda \partial_\rho
+S^{\nu\rho} \sigma_\nu
\partial_\rho \partial_\mu
+S^{\rho\lambda} \sigma_\nu
\bar\sigma_{\rho\lambda}
\partial_\mu \partial^\nu\right]^I_{\alpha \dal K}
\frac{1}{(z-x)^2} \nonumber \\
&& \pm i \del^j{}_i \langle \Phi^I (x) \Phi^K (y) \rangle
\left[S_{\mu\nu}  \sigma_\lambda
\partial^\lambda \partial^\nu
+S^{\lambda\nu}  \sigma_{\nu\mu}
\sigma^\rho
\partial_\lambda \partial_\rho
 +S^{\nu\rho} \sigma_\nu
\partial_\rho \partial_\mu
+S^{\rho\lambda} \sigma_\nu
\bar\sigma_{\rho\lambda}
\partial_\mu \partial^\nu\right]^J_{\alpha \dal K}
\frac{1}{(z-y)^2} \nonumber \\
&& \pm i \del^j{}_i \langle \Phi^K (x) \Phi^J (y) \rangle
[S_{\mu\nu} \sigma^\nu
+S^{\rho\lambda} \sigma_\mu  \bar\sigma_{\rho\lambda}
]^I_{\alpha \dal K}  \del^4 (z-x) \nonumber \\
&& \pm i \del^j{}_i \langle \Phi^I (x) \Phi^K (y) \rangle
[S_{\mu\nu} \sigma^\nu
+S^{\rho\lambda} \sigma_\mu  \bar\sigma_{\rho\lambda}
]^J_{\alpha \dal K} \del^4 (z-y).
\eea
We also need to make use of the relation
\be \label{qqo} \{Q_j{}^\beta,[Q_{i\alpha}, Q^{ij}{}_{kl}]\} =
\epsilon_{klmn} \Sigma_{\alpha}{}^{\beta mn},\ee
%
%
which follows from~\C{eqnone}\ and~\C{eqntwo}
and implies
\be\label{qeps}
[Q^{k \dot \beta}, \Sigma_\alpha^{\beta mn}]
\epsilon_{ikmn}\sigma^\mu_{\beta\dot \beta} = J^\mu_{\alpha i}.
\ee
Using~\C{Jdef}\ and~\C{qeps}, the left hand side of~\C{Jcorr}\
can be written as
\be \label{jmore} \langle  J^\mu_{\alpha i} (z)
\Phi^I (x)
[{\bar{Q}}^j{}_{\dal}, \Phi^J (y)]_{\pm}  \rangle =
\epsilon_{ikmn} (\sigma^\mu)_{\beta \dbt} \langle
\Sigma_{\alpha}{}^{\beta mn} (z) \{{\bar{Q}}^{k\dbt},
\Phi^I (x)
[{\bar{Q}}^j{}_{\dal}, \Phi^J (y)]_{\pm}  \}\rangle.
\ee
%
%
Exactly along the lines of section~\ref{superOPE}, we make a preliminary ansatz
of the form
\bea \label{Sans} \Sigma_{\alpha}{}^{\beta ij} (z) \Phi^I (x) &\sim &
\frac{\epsilon^{ijkl}}{(z-x)^2}
[Q_l{}^\beta, [Q_{k\alpha} ,\Phi^I (x)]_{\pm}]_{\mp}
\nonumber \\ &&
+i\frac{\epsilon^{ijkl}}{(z-x)^2} (S^{\nu\rho})^I{}_J
(\sigma_{\mu\nu})_\alpha{}^\beta
[Q_l{}^\gamma, [Q_{k\delta} ,\Phi^J (x)]_{\pm}]_{\mp}
(\sigma_\rho{}^\mu)_\gamma{}^\delta \nonumber \\
&& +{\rm less ~singular~ terms}.\eea
%
%
Using this ansatz, we can compute \C{jmore}. There are terms of the form
$$\langle [Q,[Q,\Phi]], [\bar{Q},[\bar{Q},\Phi]]\rangle$$ which give
rise to $2$ derivative terms like
$P^2 \langle \Phi \Phi \rangle$.
At the end, one obtains to leading order, the expression in~\C{Jcorr}.

If we did not have the $\Phi^I(x) \eta_{\mu\nu} \del^4 (z-x)$ contact
term in~\C{OPE1}, there would be no need for the
second term in~\C{JOPEless}. However, in this case, the contact terms
in~\C{Jcorr}\ which do not involve $S_{\m\n}$ would not appear. Then
the ansatz of~\C{Sans} would not work. So this ansatz works
consistently only when contact terms are correctly taken into acount.
However, unlike the previous OPE expressions~\C{OPE1}\
and~\C{JOPEless},~\C{Sans} is incomplete. There should be more terms in the
$\del^2 \OT$ OPE (note from~\C{eqnone}\ and~\C{eqntwo}, we
see that $\del^2 \OT \sim \Sigma$). Let us explain why this is the
case.

{}From the definition of $\Sigma$ in~\C{sigdef}, we see that it
contains  two types of terms:
\begin{itemize}
\item $\epsilon_{\gamma\beta} \sigma^{\mu \nu}_{\alpha}{}^{\gamma}
B^{ij}_{\mu\nu}$, which is antisymmetric in $ij$ and symmetric in
$\alpha\beta$.
\item $\epsilon_{\alpha\beta}
{\mathcal{E}}^{ij}$
which is symmetric in $ij$ and antisymmetric in $\alpha\beta$.
\end{itemize}
Clearly any term in the OPE of $\Sigma$ with any $\Phi^I$ should be
one of these two types. In fact, any term not of either form will not
appear in the OPE. So schematically, we can write
\be \label{general} \Sigma_{\alpha}{}^{\beta ij}\Phi \sim
\sigma^{\mu \nu}_{\alpha}{}^{\beta} {\mathcal{P}}^{[ij]}_{\mu \nu}
+ \del_\alpha{}^\beta {\mathcal{Q}}^{(ij)}. \ee
With this definition,
\be \label{BOPE} B^{ij}_{\mu\nu} \Phi \sim {\mathcal{P}}^{[ij]}_{\mu \nu}
+{\mathcal{K}}^{[ij]}_{\mu \nu},\ee
where $\sigma^{\mu \nu}_{\alpha}{}^{\beta}
{\mathcal{K}}^{[ij]}_{\mu \nu} =0$ (for example, $\bar\sigma_{\mu \nu}^{ \dal}{}_{\dbt}
{\mathcal{Z}}^{[ij]\dbt}{}_{\dal}$ is
a possible term in ${\mathcal{K}}^{[ij]}_{\mu \nu}$).
Also by definition
\be {\mathcal{E}}^{ij} \Phi \sim {
\mathcal{Q}}^{(ij)}. \ee
It is easy to see that all the terms on the
right hand side of~\C{Sans}\ take the form $(\sigma^{\mu \nu})_{\alpha}{}^{\beta}
{{\mathcal{P}}}^{[ij]I}_{\mu \nu}$,
where
\bea {{\mathcal{P}}}^{[ij]I}_{\mu \nu} &=& -\frac{1}{2}
\frac{\epsilon^{ijkl}}{(z-x)^2}
(\sigma_{\mu \nu})_{\alpha}{}^{\beta}
[Q_l{}^\alpha, [Q_{k\beta} ,\Phi^I (x)]_{\pm}]_{\mp}
\nonumber \\ &&
-i \frac{\epsilon^{ijkl}}{(z-x)^2}
(S_{\nu\rho})^I{}_J
(\sigma_\mu{}^\rho)_\alpha{}^\beta
[Q_l{}^\alpha, [Q_{k\beta} ,\Phi^J (x)]_{\pm}]_{\mp}.
\eea
So all the terms that appear
in~\C{Sans}\ are terms in the OPE of $B^{ij}_{\mu\nu}$
with $\Phi^I$.
This is what we expect because (see~\C{supervar})
under $\bar\del$, $B^{ij}_{\mu\nu}$
transforms into $J^\alpha_{\mu i}$, while
${\mathcal{E}}^{ij}$ vanishes (modulo
descendents). Schematically, $\bar\del B \sim J$
and $\bar\del \mathcal{E} \sim 0$. So on removing a
$\bar\del$ from $J^\alpha_{\mu i}$ in a correlator
to obtain the terms in the $\Sigma$ OPE, we should only obtain terms
in the $B \Phi$ OPE since $\mathcal{E}$ is in the kernel of
$\bar\del$. The descendent term plays no role because it
does not appear in our definition of the supercurrent~\C{Jdef}.

{}From our prior analysis, we do not
expect to see terms in the $\Sigma$ OPE that arise
from $\mathcal{E}$ so~\C{Sans}\ is incomplete.
In order to
obtain these terms in the OPE of ${\mathcal{E}}^{ij}$
with $\Phi^I$, we make use of the relations
\bea \label{eqnvar}
&[Q_{k\alpha} ,{\mathcal{E}}^{ij}] =
\del_k{}^{(i}\Lambda^{j)}{}_\alpha,  \quad
\{Q_j{}^\beta ,\Lambda^i{}_\alpha\} = i\del_j{}^i
\del_\alpha{}^\beta \Ot, &\non\\ &
[Q_i ,\Ot] = \partial^\mu J_{\mu i} +2\s^\mu \bar\s^\nu \partial_\mu
J_{\nu i}, &
\eea
(which follow from from \C{supervar}) to show that
\be \label{eqnE}
\partial_\mu J^\mu_{k\beta} +2 (\s^\mu \bar\s^\nu \partial_\mu
J_{\nu k})_\beta =
[Q_{k\alpha}, \{Q_j{}^\beta, [Q_{i\beta}
,{\mathcal{E}}^{ij}]\}].
\ee
We have set various numerical factors to one in this relation since their
values will not be relevant in the following analysis.
Schematically, \C{eqnE} has the form $\del^3 \mathcal{E} \sim \partial J$  so
$\mathcal{E}$ is  related to a current.

Using \C{eqnE}, we consider the equality
\bea \label{eqnE2}
\partial_\mu \langle J^\mu_{i\alpha} (z)
\Phi^I (x) [{\bar{Q}}^j{}_{\dal},\Phi^J (y)]_{\pm}
\rangle + 2 (\s_\mu \bar\s_\nu)_\alpha{}^\beta \partial^\mu \langle
J^\nu_{i\beta} (z) \Phi^I (x) [{\bar{Q}}^j{}_{\dal},\Phi^J (y)]_{\pm}
\rangle \nonumber \\
=\langle {\mathcal{E}}^{kl} (z)
\{Q_{k\beta}, [Q_l{}^\beta,
\{Q_{i\alpha}, \Phi^I (x) [{\bar{Q}}^j{}_{\dal},
\Phi^J (y)]_{\pm} \}]\}\rangle . \qquad \qquad
\eea
We will first evaluate the contribution to the correlators on the
left-hand side of \C{eqnE2}. Using the Ward identities \C{jward}
and \C{anward}, we find
\bea \label{impterm} &&
\partial_\mu \langle J^\mu_{i\alpha} (z)
\Phi^I (x) [{\bar{Q}}^j{}_{\dal},\Phi^J (y)]_{\pm}
\rangle + 2 (\s_\mu \bar\s_\nu)_\alpha{}^\beta \partial^\mu \langle
J^\nu_{i\beta} (z) \Phi^I (x) [{\bar{Q}}^j{}_{\dal},\Phi^J (y)]_{\pm}
\rangle \nonumber \\
&&= \del^4 (z-x) \langle [Q_{i\alpha} ,\Phi^I (x)]_{\pm}
[{\bar{Q}}^j{}_{\dal},\Phi^J (y)]_{\pm} \rangle  +\del^4 (z-y)
\langle \Phi^I (x) [Q_{i\alpha} ,[{\bar{Q}}^j{}_{\dal},
\Phi^J (y)]_{\pm}]_{\mp} \rangle \nonumber \\
&& \quad +\partial^z_{\alpha\dal} \del^4 (z-x) \langle
[\bar{S}_i^{\dal} -x^{\alpha\dal} Q_{i\alpha} ,\Phi^I (x)]_{\pm}
[{\bar{Q}}^j{}_{\dal},\Phi^J (y)]_{\pm} \rangle \nonumber \\
&& \quad +\partial^z_{\alpha\dal} \del^4 (z-y)
\langle \Phi^I (x) [\bar{S}_i^{\dal} -y^{\alpha\dal} Q_{i\alpha} ,
[{\bar{Q}}^j{}_{\dal},\Phi^J (y)]_{\pm}]_{\mp} \rangle .
\eea
Eventually (from~\C{dertau}) the insertion point of $\Ot$ is to be integrated over $z$.
This simplifies the possible terms that need to be considered
in the OPE. Using
\be \int d^4 z \partial^z_\mu \delta^4 (z-x) f (x) =0, \ee
we see that terms involving derivative(s) acting on delta functions can
be ignored. Hence, to leading order the terms that will be of relevance are given by
\bea \partial_\mu \langle J^\mu_{i\alpha} (z)
\Phi^I (x) [{\bar{Q}}^j{}_{\dal},\Phi^J (y)]_{\pm}
\rangle + 2 (\s_\mu \bar\s_\nu)_\alpha{}^\beta \partial^\mu \langle
J^\nu_{i\beta} (z) \Phi^I (x) [{\bar{Q}}^j{}_{\dal},\Phi^J (y)]_{\pm}
\rangle \nonumber \\
\sim [\del^4 (z-x) + \del^4 (z-y)]
\partial_{\alpha\dal} \langle \Phi^I (x) \Phi^J (y) \rangle.
\qquad \qquad \eea
Note that all the contributions come from the $\partial^\mu J_{\mu i}$
term only.

Next we need to evaluate the right hand side of
\C{eqnE2} as $z \rightarrow x$ and
$z \rightarrow y$, retaining the leading order terms. We make the ansatz
\be {\mathcal{E}}^{ij} (z)
\Phi^I (x) \sim \frac{1}{(z-x)^2}
[{\bar{Q}}^i{}_{\dal}, [{\bar{Q}}^{j\dal}
,\Phi^I (x)]_{\pm}]_{\mp}
+{\rm less ~singular ~terms},\ee
where the term on the right hand side is symmetric
in $ij$ by construction.
Inserting this ansatz into the right hand side of~\C{eqnE2}, we see
from~\C{impterm}\ that the
equality holds to this order.
This OPE does not involve any $S_{\m\n}$ terms at this order.
So we can write the complete OPE as
\bea \label{Sans2} && \Sigma_{\alpha}{}^{\beta ij} (z)
\Phi^I (x) \sim
\frac{\epsilon^{ijkl}}{(z-x)^2}
[Q_l{}^\beta, [Q_{k\alpha} ,\Phi^I (x)]_{\pm}]_{\mp}
+\frac{\del_\alpha{}^\beta}{(z-x)^2}
[{\bar{Q}}^i{}_{\dal}, [{\bar{Q}}^{j\dal}
,\Phi^I (x)]_{\pm}]_{\mp}
\nonumber \\ &&
+i\frac{\epsilon^{ijkl}}{(z-x)^2}
(S^{\nu\rho})^I{}_J
(\sigma_{\mu\nu})_\alpha{}^\beta
[Q_l{}^\gamma, [Q_{k\delta} ,\Phi^J (x)]_{\pm}]_{\mp}
(\sigma_\rho{}^\mu)_\gamma{}^\delta
+{\rm less ~singular ~terms}.\eea

It is important for our purposes to construct all the less
singular terms in~\C{Sans2}. This is because we will use the $\del^2
\OT$ OPE to recursively construct the $\Ot$ OPE which will have
singular terms of different orders. All such terms can contribute to
the integral~\C{dertau}.
Also, these terms will be needed when we check the
consistency of the $\del^2 \OT$ OPE. Since these terms involve lengthy
expressions, we present them in Appendix~\ref{appenddOT}.
The complete result for the $\del^2\OT$ OPE plays an important role in
our subsequent analysis, so we
have also provided a number of checks on the OPE in
Appendix~\ref{somechecks}.

\subsection{The OPE for $\Lambda$ and $\Ot$}
\label{OtOPE}
The final steps require us to determine the OPE structure for $\del^3
\OT \equiv \Lambda$ and then $\Ot\equiv \delta^4 \OT$ from the $\del^2 \OT$ OPE.  We will
 first express $\Ot$ in terms of the
supercurrents and $\OT$ using the relations
\be \label{eqnOne}
[Q_{k \gamma} ,\Sigma_{\alpha}{}^{\beta ij} ]=
\epsilon_{\alpha \gamma}
\del_k{}^{[i} \Lambda^{j]\beta} +\del_\gamma{}^\beta \del_k{}^{[i}
\Lambda^{j]}{}_\alpha
+\del_\alpha{}^\beta \del_k{}^{(i} \Lambda^{j)}{}_\gamma, \quad
\{Q_j{}^\beta ,\Lambda^i{}_\alpha\} = i\del_j{}^i
\del_\alpha{}^\beta
\Ot, \ee
 which can be deduced from~\C{supervar}

The supervariations of all the operators in the current multiplet
depend only on the combination $\Sigma$ and not on $B$ or $\mathcal{E}$
individually.  Very schematically, from~\C{supervar}, we see that
\be
\hat\del \chi \sim \Sigma \eta +\ldots, \qquad
\hat\del J \sim \partial\Sigma \eta +\ldots, \qquad
\hat\del \Lambda \sim \partial\Sigma \bar\eta +\ldots.
\ee
So  $\Ot$ can be expressed as $\del^2$ acting on $\Sigma$.  Explicitly,
using the supersymmetry transformations
 leads to (dropping overall numerical coefficients)
\be B^{ij}_{\mu\nu} = \frac{1}{2} (\sigma_{\mu \nu})_\alpha{}^\beta
\epsilon^{ijkl} \{ Q_n{}^\alpha, [Q_{m\beta},
Q^{mn}{}_{kl}]\}, \quad
{\mathcal{E}}^{ij} = \epsilon^{jlmn} \{ Q_l{}^\alpha, [Q_{k\alpha},
Q^{ki}{}_{mn}]\}, \quad \quad
\ee
so that
\be \Sigma_{\alpha}{}^{\beta ij} = \epsilon^{ijkl}
\{ Q_{m\alpha}, [Q_n{}^\beta, Q^{mn}{}_{kl}]\} +\del_\alpha{}^\beta
\epsilon^{jklm} \{ Q_k{}^\gamma, [Q_{n\gamma}, Q^{ni}{}_{lm}]\}.
\ee
Using \C{eqnOne}, we see that
\be \label{Leqn}  \Lambda^i_\alpha =
[Q_{j\beta},\Sigma_{\alpha}{}^{\beta ij}], \ee
and finally we arrive at the desired relation between $\Ot$ and $\Sigma$,
\be \label{Oteqn} \Ot = \{ Q_i{}^\alpha, [Q_{j\beta},
\Sigma_{\alpha}{}^{\beta ij}]\}. \ee

Now consider the leading terms~\C{Sans2}\ in the $\Sigma$ OPE.
Acting with $Q_{j\beta}$ on~\C{Sans2} and
using~\C{Leqn}, we find that
\bea \label{Lone}
\Lambda^i_\alpha (z)\Phi^I (x) + \Sigma_{\alpha}{}^{\beta ij} (z)
[Q_{j\beta} ,\Phi^I (x)]_{\pm} \sim
\frac{1}{(z-x)^2}
[Q_{j\alpha} ,[{\bar{Q}}^i{}_{\dal}, [{\bar{Q}}^{j\dal}
,\Phi^I (x)]_{\pm}]_{\mp}]_{\pm} \nonumber \\
+i\frac{\epsilon^{ijkl}}{(z-x)^2}
(S^{\nu\rho})^I{}_J
(\sigma_{\mu\nu})_\alpha{}^\beta
[Q_{j\beta} ,[Q_l{}^\gamma, [Q_{k\delta}
,\Phi^J (x)]_{\pm}]_{\mp}]_{\pm}
(\sigma_\rho{}^\mu)_\gamma{}^\delta +\ldots. \eea
Using~\C{Sans2}\ once more in the second term on the left hand
side of~\C{Lone} gives the $\del^3 \OT$ OPE
\bea \label{del3OPE} \Lambda^i_\alpha (z) \Phi^I (x) & \sim&
\frac{1}{(z-x)^2}
[Q_{j\alpha} ,[{\bar{Q}}^i{}_{\dal}, [{\bar{Q}}^{j\dal}
,\Phi^I (x)]_{\pm}]_{\mp}]_{\pm}
+\frac{1}{(z-x)^2}
[{\bar{Q}}^i{}_{\dal}, [{\bar{Q}}^{j\dal} ,
[Q_{j\alpha} ,\Phi^I (x)]_{\pm}]_{\mp}]_{\pm} \nonumber \\ &&
+i\frac{\epsilon^{ijkl}}{(z-x)^2}
(S^{\mu\nu})^I{}_J
(\sigma_{\mu\nu})_\beta{}^\gamma
[Q_{j\alpha} ,[Q_l{}^\beta, [Q_{k\gamma}
,\Phi^J (x)]_{\pm}]_{\mp}]_{\pm}
+\ldots.  \eea
Acting on \C{del3OPE} with $Q_i{}^\alpha$ and using \C{Oteqn},
gives the relation
\bea \label{leadOt}
 & & \Ot (z) \Phi^I (x) + \Lambda^i_\alpha (z)
[Q_i{}^\alpha ,\Phi^I (x)]_{\pm}
 \sim  \frac{1}{(z-x)^2}
[Q_i{}^\alpha ,[Q_{j\alpha} ,[{\bar{Q}}^i{}_{\dal}, [{\bar{Q}}^{j\dal}
,\Phi^I (x)]_{\pm}]_{\mp}]_{\pm}]_{\mp}  \non\\
&& +\frac{1}{(z-x)^2}
[Q_i{}^\alpha ,[{\bar{Q}}^i{}_{\dal}, [{\bar{Q}}^{j\dal} ,
[Q_{j\alpha} ,\Phi^I (x)]_{\pm}]_{\mp}]_{\pm}]_{\mp}
+\ldots. \eea
On using \C{del3OPE} in the second term on the left hand side of
\C{leadOt}, we arrive at the most singular terms in the $\Ot$ OPE
\bea \label{Otoneop}
\Ot (z) \Phi^I (x) & \sim &
\frac{a_1}{(z-x)^2}
[Q_i{}^\alpha ,[Q_{j\alpha} ,[{\bar{Q}}^i{}_{\dal}, [{\bar{Q}}^{j\dal}
,\Phi^I (x)]_{\pm}]_{\mp}]_{\pm}]_{\mp}\nonumber \\
&& +\frac{a_2}{(z-x)^2}
[Q_i{}^\alpha ,[{\bar{Q}}^i{}_{\dal}, [{\bar{Q}}^{j\dal}
,[Q_{j\alpha} ,\Phi^I (x)]_{\pm}]_{\mp}]_{\pm}]_{\mp}
\qquad \nonumber \\
&& +\frac{a_3}{(z-x)^2}
[{\bar{Q}}^i{}_{\dal}, [{\bar{Q}}^{j\dal}
,[Q_i{}^\alpha ,[Q_{j\alpha} ,\Phi^I (x)]_{\pm}]_{\mp}]_{\pm}]_{\mp}
+\ldots.
\eea
The coefficients, $a_i$, are generally functions of $(\tau, \bar\tau)$.
Note also that the contributions considered so far to the  $\Ot$
OPE~\C{Otoneop}\  are independent
of $S_{\m\n}$.

 Some terms in~\C{del3OPE}\ and~\C{Otoneop}\ can be
re-ordered at the expense of introducing derivatives. This only
lengthens the expressions so we will keep the displayed ordering.
To illustrate this point, note that we can re-order~\C{Otoneop}\
to give (ignoring the $a_i$ coefficients)
\bea \Ot (z) \Phi^I (x) & \sim &
\left(-4\pi^2 \del^4 (z-x) +\frac{1}{(z-x)^2} \partial^2
+2 \partial^\mu \frac{1}{(z-x)^2} \partial_\mu \right) \Phi^I (x)
\nonumber \\
&& +\partial_{\alpha\dal}  \left(\frac{1}{(z-x)^2}
[Q_i{}^\alpha ,[{\bar{Q}}^{i\dal}, \Phi^I (x)]_{\pm}]_{\mp} \right) \non\\
&& +\frac{1}{(z-x)^2}
[Q_i{}^\alpha ,[Q_{j\alpha} ,[{\bar{Q}}^i{}_{\dal}, [{\bar{Q}}^{j\dal}
,\Phi^I (x)]_{\pm}]_{\mp}]_{\pm}]_{\mp}
+\ldots.\nonumber
\eea
With this ordering, the leading singularity in the $\Ot$ OPE appears
to be a contact  term rather than a power law singularity. Indeed, if
we were to permit integration by parts (allowed, for example, when we
integrate over $z$), all the leading singular terms
can be rewritten as contact terms.

The remaining contributions to the $\Lambda$ and $\Ot$ OPE
can be deduced in a straightforward way from the expressions
in~\C{lesssing1},~\C{lesssing2},~\C{lesssing3}\
and~\C{lesssing4}.
However,
for simplicity, we will restrict to $\Phi^I$ which are Lorentz
scalars transforming in any $SU(4)_R$
representation. This simplification means that we can set all
$S_{\m\n}$ terms to zero. It is a tedious, but straightforward,
exercise to determine these terms should they be needed (we will
require these terms in~\cite{toappear}).

{}From the terms~\C{lesssing1}\ in the
$\Sigma$ OPE, it follows that the $\Lambda$ OPE gets contributions
\bea  \Lambda^i_\alpha (z) \Phi (x) & \sim &
\frac{r^\mu}{r^2}
\partial_\mu [Q_{j\alpha} ,\{{\bar{Q}}^i{}_{\dal}, [{\bar{Q}}^{j\dal}
,\Phi (x)]\}]
+\frac{r^\mu}{r^2}
\partial_\mu [{\bar{Q}}^i{}_{\dal}, \{{\bar{Q}}^{j\dal}
,[Q_{j\alpha} ,\Phi (x)]\}] \nonumber \\
&& +(\sigma_{\mu\nu})_\alpha{}^\beta
(\bar\sigma^{\mu\rho})^{\dal}{}_{\dbt}
\frac{r^\nu}{r^2} \partial_\rho
[Q_{j\beta} ,\{{\bar{Q}}^i{}_{\dal}, [{\bar{Q}}^{j\dbt}
,\Phi(x)]\}] \non\\
&& +(\sigma_{\mu\nu})_\alpha{}^\beta
(\bar\sigma^{\mu\rho})^{\dal}{}_{\dbt}
\frac{r^\nu}{r^2} \partial_\rho
[{\bar{Q}}^i{}_{\dal}, \{{\bar{Q}}^{j\dbt}
,[Q_{j\beta} ,\Phi(x)]\}],
\eea
(dropping coupling constant dependent coefficients)
while the $\Ot$ OPE gets contributions
\bea \label{Ottwoop}
\Ot (z) \Phi (x) & \sim &
a_4 \frac{r^\mu}{r^2}
\partial_\mu \{Q_i{}^\alpha ,[Q_{j\alpha} ,\{{\bar{Q}}^i{}_{\dal},
[{\bar{Q}}^{j\dal}
,\Phi (x)]\}]\}  \nonumber \\
&& +a_5 \frac{r^\mu}{r^2}
\partial_\mu \{Q_i{}^\alpha ,[{\bar{Q}}^i{}_{\dal}, \{{\bar{Q}}^{j\dal}
,[Q_{j\alpha} ,\Phi (x)]\}]\} \non\\
&& +a_6 \frac{r^\mu}{r^2}
\partial_\mu \{{\bar{Q}}^i{}_{\dal}, [{\bar{Q}}^{j\dal}
,\{Q_i{}^\alpha ,[Q_{j\alpha}
,\Phi (x)]\}]\}
\nonumber \\
&& +a_7 (\sigma_{\mu\nu})_\alpha{}^\beta
(\bar\sigma^{\mu\rho})^{\dal}{}_{\dbt}
\frac{r^\nu}{r^2} \partial_\rho
\{Q_i{}^\alpha ,[Q_{j\beta} ,\{{\bar{Q}}^i{}_{\dal}, [{\bar{Q}}^{j\dbt}
,\Phi(x)]\}]\}, \nonumber \\
&& +a_8 (\sigma_{\mu\nu})_\alpha{}^\beta
(\bar\sigma^{\mu\rho})^{\dal}{}_{\dbt}
\frac{r^\nu}{r^2} \partial_\rho
\{Q_i{}^\alpha ,[{\bar{Q}}^i{}_{\dal}, \{{\bar{Q}}^{j\dbt}
,[Q_{j\beta} ,\Phi(x)]\}]\}, \nonumber \\
&& +a_9 (\sigma_{\mu\nu})_\alpha{}^\beta
(\bar\sigma^{\mu\rho})^{\dal}{}_{\dbt}
\frac{r^\nu}{r^2} \partial_\rho
\{{\bar{Q}}^i{}_{\dal}, [{\bar{Q}}^{j\dbt}
,\{Q_i{}^\alpha ,[Q_{j\beta} ,\Phi(x)]\}]\},
\eea
where $a_4, \ldots, a_9$ are undetermined functions of the coupling.

It is easy to see that there are no terms in the OPE of
$\Lambda$ or $\Ot$ with a Lorentz scalar from~\C{lesssing2}.
However, more contributions arise by considering~\C{lesssing3}\ which gives
\bea \Lambda^i_\alpha (z) \Phi (x) && \sim
\frac{r^\mu}{r^2}
(\sigma_\mu)_{\beta\dbt} [Q_{j\alpha} ,\{{\bar{Q}}^i{}_{\dal},
[{\bar{Q}}^{j\dal}
,\{{\bar{Q}}^{k\dbt},[Q_k{}^\beta,\Phi (x)]\}]\}]
\nonumber \\
&& +\frac{r^\mu}{r^2}
(\sigma_\mu)_{\beta\dbt} [{\bar{Q}}^i{}_{\dal}, \{{\bar{Q}}^{j\dal}
,[{\bar{Q}}^{k\dbt},\{Q_k{}^\beta,[Q_{j\alpha}
,\Phi (x)]\}]\}]
  \nonumber \\
&& +(\sigma_{\mu\nu})_\alpha{}^\beta
\frac{r_\lambda}{r^2}
(\sigma^\nu)_{\gamma\dg}
(\bar\sigma^{\mu\lambda})^{\dal}{}_{\dbt}
[Q_{j\beta} ,\{{\bar{Q}}^i{}_{\dal}, [{\bar{Q}}^{j\dbt}
,\{{\bar{Q}}^{k\dg},[Q_k{}^\gamma
,\Phi (x)]\}]\}] \nonumber \\
&& +(\sigma_{\mu\nu})_\alpha{}^\beta
\frac{r_\lambda}{r^2}
(\sigma^\nu)_{\gamma\dg}
(\bar\sigma^{\mu\lambda})^{\dal}{}_{\dbt}
[{\bar{Q}}^i{}_{\dal}, \{{\bar{Q}}^{j\dbt}
,[{\bar{Q}}^{k\dg},\{Q_k{}^\gamma
,[Q_{j\beta} ,\Phi (x)]\}]\}]
\eea
and
\bea \label{Otthreeop}
&& \Ot (z) \Phi (x) \sim
a_{10} \frac{r^\mu}{r^2}
(\sigma_\mu)_{\beta\dbt} \{Q_i{}^\alpha ,[Q_{j\alpha}
,\{{\bar{Q}}^i{}_{\dal}, [{\bar{Q}}^{j\dal}
,\{{\bar{Q}}^{k\dbt},[Q_k{}^\beta,
\Phi (x)]\}]\}]\}
\nonumber \\
&& +a_{11} \frac{r^\mu}{r^2}
(\sigma_\mu)_{\beta\dbt} \{Q_i{}^\alpha ,
[{\bar{Q}}^i{}_{\dal}, \{{\bar{Q}}^{j\dal}
,[{\bar{Q}}^{k\dbt},\{Q_k{}^\beta,
[Q_{j\alpha} ,\Phi (x)]\}]\}]\}
 \nonumber \\
&& +a_{12} \frac{r^\mu}{r^2}
(\sigma_\mu)_{\beta\dbt}
\{{\bar{Q}}^i{}_{\dal}, [{\bar{Q}}^{j\dal}
,\{{\bar{Q}}^{k\dbt},[Q_k{}^\beta,
\{Q_i{}^\alpha ,[Q_{j\alpha} ,\Phi (x)]\}]\}]\}
\nonumber \\
&& + a_{13} (\sigma_{\mu\nu})_\alpha{}^\beta
\frac{r_\lambda}{r^2}
(\sigma^\nu)_{\gamma\dg}
(\bar\sigma^{\mu\lambda})^{\dal}{}_{\dbt}
\{Q_i{}^\alpha ,[Q_{j\beta} ,\{{\bar{Q}}^i{}_{\dal}, [{\bar{Q}}^{j\dbt}
,\{{\bar{Q}}^{k\dg},[Q_k{}^\gamma
,\Phi (x)]\}]\}]\}
\nonumber \\
&& + a_{14} (\sigma_{\mu\nu})_\alpha{}^\beta
\frac{r_\lambda}{r^2}
(\sigma^\nu)_{\gamma\dg}
(\bar\sigma^{\mu\lambda})^{\dal}{}_{\dbt}
\{Q_i{}^\alpha ,[{\bar{Q}}^i{}_{\dal}, \{{\bar{Q}}^{j\dbt}
,[{\bar{Q}}^{k\dg},\{Q_k{}^\gamma, [Q_{j\beta} ,
,\Phi (x)]\}]\}]\}
\nonumber \\
&& + a_{15} (\sigma_{\mu\nu})_\alpha{}^\beta
\frac{r_\lambda}{r^2}
(\sigma^\nu)_{\gamma\dg}
(\bar\sigma^{\mu\lambda})^{\dal}{}_{\dbt}
\{{\bar{Q}}^i{}_{\dal}, [{\bar{Q}}^{j\dbt}
,\{{\bar{Q}}^{k\dg}, [Q_k{}^\gamma ,\{Q_i{}^\alpha ,[Q_{j\beta} ,
\Phi (x)]\}]\}]\},
\eea
where $a_{10}, \ldots, a_{15}$ are further undetermined functions of the coupling.
{}Finally, considering the contribution from~\C{lesssing4}, we find
\be \Lambda^i_\alpha (z) \Phi (x) \sim
\epsilon^{ijkl} \frac{r^\mu}{r^2}
(\sigma_{\mu\nu})_\beta{}^\gamma \partial^\nu [Q_{j\alpha} ,\{Q_l{}^\beta,
[Q_{k\gamma} , \Phi (x)]\}],\ee
while there is no corresponding contribution to the $\Ot$ OPE.

In summary, the complete $\Ot$ OPE with a scalar $\Phi$ (modulo operators in
different supermultiplets) is given by the sum
of the expressions~\C{Otoneop},~\C{Ottwoop} and~\C{Otthreeop}.
The coefficients $(a_1,\ldots, a_{15})$ are generally undetermined
functions of $(\tau,\bar\tau)$.

\section{Some Properties of Correlation Functions}

\subsection{Operator normalization and contact terms}

Now that we have established the form of the $\Ot$ OPE, we turn to
the structure of correlation functions of local operators,
\be
\langle \Om^{I_1}(x_1) \cdots \Om^{I_n}(x_n) \rangle.
\ee
We will always consider correlators of operators at separated
points.
Each correlation function  transforms in
some representation of $SL(2,\Z)$. This
need not be a singlet representation, as shown for correlators
involving the Konishi
supermultiplet~\cite{Tseng:2002pe}. However, a correlator of BPS operators should
map back to itself since each BPS operator is uniquely specified by its
quantum numbers. Such a correlator should transform in a singlet
representation of $SL(2,\Z)$  with fixed weights  $(w, \bar{w})$.
Under an $SL(2, \Z)$ transformation,
\be
\tau \rightarrow  \left( \frac{a\tau+b}{c\tau+d}\right),
\qquad a,b,c,d\in \Z
\ee
a modular form, $ \Theta^{(w,\bar{w})}$, in a singlet representation
transforms in the following way:
\be \Theta^{(w,\bar{w})} (\tau,\bar\tau) \rightarrow (c\tau +d)^w (c\bar\tau
+d)^{\bar{w}} \Theta^{(w,\bar{w})} (\tau,\bar\tau).\ee
The weights we assign to BPS operators should be correlated with their
$U(1)_Y$ transformation properties in a way described in~\cite{Intriligator:1998ig}.

The $SL(2,\Z)$ transformation properties of a correlation function
are important because the expression,
\be \label{ddtau}\frac{\partial}{\partial \tau} \langle \prod_r
{\mathcal{O}}^{I_r} (x_r)
\rangle
= \langle \frac{\partial}{\partial \tau} \left(\prod_i {\mathcal{O}}^{I_i}
(x_i)  \right)\rangle +\frac{i}{4\tau_2} \int d^4 z \langle \Ot (z)
\prod_r {\mathcal{O}}^{I_r} (x_r) \rangle, \ee
is not modular covariant at first sight. For the moment, let us
restrict to correlators transforming in singlet
representations of $SL(2,\Z)$.
Let us examine the
explicit $\tau$ derivative acting on each $\Om^{I_i}$,
\be\label{explicit}
\frac{\p}{\p\tau}\Om^{I_i}.
\ee
The operator~\C{explicit}\ has the same quantum numbers as
$O^{I_i}$. In particular, the conformal dimensions agree.
If there is no degeneracy for operators with conformal
dimension $\Delta_i$ then~\C{explicit}\ must be proportional to
$O^{I_i}$. This is the case for short operators. If there is a
finite-dimensional degeneracy then we can again choose a basis of operators
for which this statement is true. This leaves us with the freedom to
rescale each $\Om^{I_i}$ by a function of $(\tau ,\bar\tau)$. We can
conclude that in this basis,
\be\label{explicittau}
\frac{\p}{\p\tau}\Om^{I_i} = { i \alpha_{{I_i}}(\tau,\bar\tau) \over \tau_2} \Om^{I_i}. \ee
Using our final rescaling degree of freedom, we  choose to set
$\alpha_{{I_i}}$ to a constant. For the current multiplet, the
normalization of all operators is determined by the definition of
$\Ot$ in terms of the action~\C{action}\ together with the (coupling
independent) supersymmetry transformations.

Equation~\C{ddtau}\ must be $SL(2,\Z)$ covariant if the correlator
transforms in a singlet representation. It must therefore be the case
that summing the tree-level contact
terms between $\Ot$ and $\Om^{I_i}$ together with the
$\alpha_{{I_i}}$ from~\C{explicittau}\
has the net effect of replacing
$$ i\tau_2 \frac{\p}{\p\tau} \rightarrow D_w, \qquad
 - i\tau_2 \frac{\p}{\p\bar{\tau}} \rightarrow D_{\bar{w}}.
$$
We have defined modular covariant derivatives,
\be D_w = i\left(\tau_2 \frac{\partial}{\partial \tau} -\frac{iw}{2}\right),
\qquad \bar{D}_{\bar{w}} = -i \left(\tau_2
\frac{\partial}{\partial \bar\tau} +\frac{i\bar{w}}{2}\right), \ee
where $(w,\bar{w})$ is the weight of the correlation function.
These issues have been discussed in~\cite{Petkou:1999fv}.

To see whether this actually happens requires a computation of the
tree-level contact term between $\Ot$ and each $\Om^{I_i}$. However, in
perturbation theory, precisely this piece of the contact term is
scheme-dependent~\cite{Penati:2000zv}. However, the full non-perturbative
theory must require the particular choice consistent with
$SL(2,\Z)$. We will assume that this contact term takes the value
required by $SL(2,\Z)$ for long operators. Fortunately, for
a BPS superconformal primary, we will be
able to determine the weight of the operator (normalized in a
particular way) by direct arguments
without recourse to any duality assumptions.

We now use the $\Ot$ OPE given by~\C{Otoneop},~\C{Ottwoop}\ and \C{Otthreeop}\
to analyze the coupling dependence of correlators using~\C{ddtau}.
It is difficult to control the right hand side of~\C{ddtau}\ except
for low-point functions where we know the exact space-time dependence
of the correlator. We will consider those special cases in a moment.

Let us focus on the contribution to the right hand side
of~\C{ddtau}\ from $z
\rightarrow x_i$. This contribution is dominated by the singular terms
in the OPE between $\Ot$ and $\Om^{I_i}$.
The contribution to the integral from $z$ far from
$x_i$ is also present but will not be explicitly displayed in the following equation.
In other words, we rewrite~\C{ddtau}\ in the form
\bea \label{dertau3}
\frac{\partial}{\partial \tau} \langle \prod_i {\mathcal{O}}^{I_i} (x_i)
\rangle
&=& \langle \frac{\partial}{\partial \tau} \left(\prod_i {\mathcal{O}}^{I_i}
(x_i)  \right)\rangle \non\\ && +
\frac{i}{4\tau_2} \sum_j \int_{B_{x_j}^\epsilon} d^4z{\langle \Ot
(z)
\prod_i {\mathcal{O}}^{I_i} (x_i) \rangle +\ldots},
\eea
where $B_{x_j}^\epsilon$ is a ball of radius $\epsilon$ surrounding the
point $x_j$. The omitted terms refer to the contribution to the
integral from the region outside each ball.

The right hand side of~\C{dertau3}\ is a sum of integrated $(n+1)$-point
 functions. In an approximation where we neglect the contribution
 from outside the balls,   we can use the $\Ot$ OPE to replace each  $(n+1)$-point
 function by an $n$-point function. In cases where we know something
 about the space-time dependence of the correlator, this approximation
 is sufficient to teach us something about the coupling dependence.
This is essentially what we anticipated in the
introduction. It will be interesting to see what information can be
gained by using this relation in conjunction with recent results about
the space-time dependence of $4$-point functions~\cite{Eden:2000bk,
  Arutyunov:2001mh, Dolan:2004mu}.

\subsection{Implications for two-point functions}
\label{evaluate2pt}

We consider equation~\C{dertau3}\ restricted to two operator insertions. This
analysis will constrain combinations of the $a_i$ coefficients defined
in section~\ref{OtOPE}. This is also a
good warm-up for higher point functions. To
simplify the $\Ot$ OPE, let us take the inserted operators to be
space-time scalars.
\bea \label{dertau4}
\frac{\partial}{\partial \tau} \langle {\bar\Om} (x)
\Om (y) \rangle
&=& \langle {\frac{\partial}{\partial \tau}} \left(\bar\Om (x)
\Om (y) \right)\rangle +
 \frac{i}{4\tau_2} \int_{B_{x}^\epsilon} d^4 z \langle \Ot (z)
{\bar\Om} (x) \Om (y) \rangle \non\\
&&
+\frac{i}{4\tau_2} \int_{B_{y}^\epsilon} d^4 z \langle \Ot (z)
{\bar\Om} (x) \Om (y) \rangle +\ldots.
\eea
Also, we will consider cases where the operator
$\Om$ and its superconformal descendents, $\del \Om$ and, $\del^2 \Om$,
are conformal primaries. This is automatically the case if $\Om$ is a
superconformal primary; in this case, $\bar\Om =\Om$. For more
general situations, this restriction
rules out the possibility $\Om =\bar\del \Upsilon$ because then $\del
\Om \sim \partial_\mu \Upsilon$ is a conformal descendent.
These simplifications will enable us to
calculate the correlation functions appearing on the right hand
side of~\C{dertau4}\ with more straightforward algebra.

We need to first determine the normalization of certain two-point
functions. For any conformal primary  $\Phi$,
\be\label{confinv}
\langle \bar\Phi (x) \, \bar\del \del \, \Phi (y) \rangle
= \langle \bar\Phi (x)  {\bar{Q}}^j{}_{\dal} Q_{j\alpha}
\Phi (y) \rangle = 4i  \partial^x_{\alpha\dal}
\langle \bar\Phi (x)
\Phi (y) \rangle.
\ee
This follows from conformal
invariance; if $\Phi$ has conformal dimension $\Delta$ then
$\bar\del \del \Phi$ has dimension $\Delta+1$ so the correlator
$\langle \bar\Phi (x) \, \bar\del \del \, \Phi (y) \rangle$
vanishes if $\bar\del \del \Phi$
is a conformal primary. We want to determine the exact coefficient
appearing in this relation. Using the supersymmetry algebra, we find that
\be \label{fixcoeff}
\langle [ Q_{i\alpha} ,[{\bar{Q}}^j{}_{\dal} ,\bar\Phi (x)
]_\pm ]_\mp \Phi (y) \rangle = 2 \del_i{}^j P^x_{\alpha\dal}
\langle \bar\Phi (x)
\Phi (y) \rangle + \langle \bar\Phi (x) [ Q_{i\alpha} ,
[{\bar{Q}}^j{}_{\dal} , \Phi (y)]_\pm ]_\mp \rangle. \ee
We also observe that
\bea \langle \bar\Phi (x) [ Q_{i\alpha} ,
[{\bar{Q}}^j{}_{\dal} , \Phi (y)]_\pm ]_\mp \rangle &=&
\mp  \langle [ Q_{i\alpha} ,[{\bar{Q}}^j{}_{\dal} ,\Phi (y)]_\pm ]_\mp
\bar\Phi (x) \rangle
= -\langle [ Q_{i\alpha} ,
[{\bar{Q}}^j{}_{\dal} , \Phi (x)]_\pm ]_\mp \bar\Phi (y) \rangle
\non\\ &=&
-\langle [ Q_{i\alpha} ,[{\bar{Q}}^j{}_{\dal} ,\bar\Phi
(x)]_\pm ]_\mp  \Phi (y) \rangle ,\eea
which when substituted in \C{fixcoeff} yields the relation
\be \label{fixcoeff2}
\langle [ Q_{i\alpha} ,[{\bar{Q}}^j{}_{\dal} ,\bar\Phi (x)
]_\pm ]_\mp \Phi (y) \rangle = \del_i{}^j P^x_{\alpha\dal}
\langle \bar\Phi (x)
\Phi (y) \rangle. \ee
Alternatively, one can deduce~\C{fixcoeff2}\ in the following way:
from the general arguments above, we know that
\be
\langle [ Q_{i\alpha} ,[{\bar{Q}}^j{}_{\dal} ,\bar\Phi (x)
]_\pm ]_\mp \Phi (y) \rangle = \rho \del_i{}^j P^x_{\alpha\dal}
\langle \bar\Phi (x) \Phi (y) \rangle. \ee
The coefficient, $\rho$, should be independent of the inserted
operators so we can fix it by a convenient choice of $\Phi$. We choose
$\Phi ={\bar\varphi}_{ij}$ in the abelian theory and we consider
\be \langle \{ Q_{k\alpha} ,[{\bar{Q}}^l{}_{\dal} ,\varphi^{ij}
(x) ]\} {\bar\varphi}_{ij} (y) \rangle = -i\rho
\del_k{}^l \partial^x_{\alpha\dal}
\langle \varphi^{ij} (x) {\bar\varphi}_{ij} (y) \rangle.\ee
Using the supersymmetry transformations~\C{susy}
\be [ {\bar{Q}}^k , \varphi^{ij} ]=\frac{1}{2} \epsilon^{ijkl}
{\bar\lambda}_l, \quad \{ Q_j{}^\alpha ,\bar\lambda_i{}^{\dal}
\}= 4i \partial^{\dal\al} {\bar\varphi}_{ij},
\nonumber \ee
we see that $\rho=1$ in agreement with~\C{fixcoeff2}. We can similarly
deduce
\be
\langle [ {\bar{Q}}^j{}_{\dal} , [Q_{i\alpha} ,\bar\Phi (x)
]_\pm ]_\mp \Phi (y) \rangle = \del_i{}^j P^x_{\alpha\dal}
\langle \bar\Phi (x) \Phi (y) \rangle, \ee
directly from the algebra, or by choosing $\Phi ={\bar\varphi}_{ij}$
and using the relations
\be [ Q_k , \varphi^{ij} ] = \frac{1}{2} (\del_k{}^j  \lambda^i
-\del_k{}^i  \lambda^j ), \quad \{ {\bar{Q}}^j{}_{\dal}
,\lambda^i{}_\alpha
\}= -4i \partial_{\al\dal} \varphi^{ij}. \nonumber \ee

We now proceed to evaluate the integrals over $B_x^\epsilon$ and
$B_y^\epsilon$ in~\C{dertau4}. Using the $\Ot$ OPE given
by~\C{Otoneop},~\C{Ottwoop},  and~\C{Otthreeop}, we can reduce the
right hand side of~\C{dertau4}\ to  a collection of two-point
functions, which we now evaluate.

We briefly explain the evaluation of one of the correlators, the remaining
ones can be evaluated in a similar way. Consider the $a_1$ correlator in
\C{Otoneop},
\bea \label{corrone} \langle  \{ Q_i{}^\alpha ,[Q_{j\alpha}
,\{{\bar{Q}}^i{}_{\dal},
[{\bar{Q}}^{j\dal} ,\bar\Om (x)]\}]\} \Om (y) \rangle
&=& -\langle [{\bar{Q}}^{i\dal},\bar\Om (x)] {\bar{Q}}^j{}_{\dal}
Q_{j\alpha} [Q_i{}^\alpha , \Om (y)] \rangle  \nonumber \\
&=&-4i \partial^x_{\alpha\dal} [{\bar{Q}}^{i\dal},\bar\Om (x)]
[Q_i{}^\alpha , \Om (y)] \rangle \non\\
&=& 32 \partial_x^2 \langle \bar\Om (x) \Om (y) \rangle, \eea
where we have twice made use of~\C{confinv}.  First we  chose
$\Phi = [Q_i{}^\alpha ,\Om]$ ($\del \Om$ is a
conformal primary by assumption) and then repeated the
procedure with $\Phi =\Om$.
All the remaining correlators can be evaluated in a similar way. We
list the results in Appendix~\ref{twopoints}.

Noting that derivatives obtained from $\{ Q, \bar{Q}\} \sim P_\mu$ also act
on  the $1/r^2$ and $r^\mu /r^2$ factors in the $\Ot$ OPE, we find on
substituting~\C{corrone}\ and~\C{corrthree}\ that
\bea \label{eqn2pt}
\frac{1}{16\pi}
\frac{\partial}{\partial \tau} \langle {\bar\Om} (x) {\Om}
(y)\rangle
&=& \frac{1}{16\pi}
\langle\frac{\partial}{\partial \tau} \left( {\bar\Om} (x) {\Om}
(y)\right) \rangle -\frac{2\pi i}{\tau_2} (A+2C) \langle {\bar\Om} (x) {\Om}
(y) \rangle \nonumber \\
&& + \frac{i}{4\pi \tau_2} (A+B+C) \partial^2 \langle {\bar\Om}
(x) {\Om} (y) \rangle \left(\int_{B_x^\epsilon} \frac{d^4 z}{(z-x)^2}
+ \int_{B_y^\epsilon} \frac{d^4 z}{(z-y)^2} \right) \non\\ && +\ldots,
\eea
where
\bea \label{only3} A = 2(a_1 +a_3) -3a_2, \qquad \qquad
\qquad \nonumber \\
B = -2(a_4 +a_6) +3a_5 +\frac{9}{2} (a_7 -\frac{1}{2} a_8 +a_9),
\qquad \nonumber \\
C= i\left(8 a_{10} -16a_{11} +20a_{12} -18a_{13} +24 a_{14} +3a_{15}\right).
\eea
Of the fifteen coefficients $a_1, \ldots, a_{15}$ in the $\Ot$
OPE, only the three combinations of~\C{only3}\
appear in the final expression. The integrals over $B_x^\epsilon$ and
$B_y^\epsilon$ are each equal to
$\pi^2 \epsilon^2$; however, their precise values will not be relevant
{}for us. Note that for two-point functions, the coupling
dependence is related to the space-time dependence of the same
two-point function. This is generically not the case for higher point
functions.

We can now make use of the known space-time dependence of low point
functions (fixed by conformal invariance)
to constrain the unknown coefficients in~\C{eqn2pt}.
For two non-coincident points $x$ and $y$ and a scalar
$\Om$, we use the relation
\be \label{twopt} \langle \bar{\Om}(x) \Om(y) \rangle =
\frac{\eta (\tau, \bar\tau)}{\vert x-y \vert^{2\Delta(\tau, \bar\tau)}} \ee
where $\Delta$ is the (possibly $\tau$-dependent) conformal dimension of
$\mathcal{O}$ which we substitute into~\C{ddtau}.
The left hand side of~\C{ddtau}\ is obtained
by differentiating~\C{twopt}, giving
\be \frac{\partial}{\partial \tau} \langle \bar{{\mathcal{O}}} (x) {\mathcal{O}} (y)
\rangle = \frac{1}{\vert x-y \vert^{2\Delta}} \left\{ \frac{\partial \eta}
{\partial \tau} -2 \eta \frac{\partial \Delta}{\partial \tau} {\rm ln}
\vert x-y \vert\right\}.\ee
In order to evaluate the right hand side of~\C{ddtau}, we use
the known form of the correlator of three fields (with $\Delta_x = \Delta_y = \Delta$
and $\Delta_z =4$), which is again fixed by
conformal invariance to be
\be \label{nocon} \langle \Ot (z) \bar{\mathcal{O}} (x) {\mathcal{O}} (y)\rangle =
\frac{C^\tau(\tau, \bar\tau)}{\vert z-x \vert^4 \vert z-y \vert^4
\vert x-y \vert^{2\Delta -4}} ,\ee
which is valid for non-coincident points $x,y$ and $z$.
Equating these
expressions (including the explicit derivative)  gives,
\bea \label{logterm} \frac{1}{\vert x-y \vert^{2\Delta}} \left\{\frac{\partial \eta}
{\partial \tau} -2 \eta \frac{\partial \Delta}{\partial \tau} {\rm ln}
\vert x-y \vert\right\} &=&
\frac{i}{4\tau_2} \frac{C^\tau}{\vert x-y
\vert^{2\Delta -4}} \int d^4 z \frac{1}{\vert z-x \vert^4 \vert z-y
  \vert^4}  \non\\ && +
\langle \frac{\partial}{\partial \tau}
\left( \bar{{\mathcal{O}}} (x) {\mathcal{O}} (y) \right)
\rangle
.\eea
However, this is not quite correct. In~\C{nocon}, we have neglected
contact terms that arise when $z \rightarrow x$ and $z \rightarrow
y$.
These terms play an important role in our following analysis, as we shall discuss shortly.

{}First we need to evaluate the integral on the right hand
side of~\C{logterm}. This integral is UV divergent and so needs to be
regularized. Following~\cite{Freedman:1992tk}, we use differential
regularization. Because $z \neq x$ and $z \neq y$ in~\C{nocon} we can use
the formula for non-coincident points, giving~\cite{Freedman:1992tk}
\be \int d^4 z \frac{1}{\vert z-x \vert^4 \vert z-y \vert^4} =
-\frac{\pi^2}{4} \partial^2 \left\{ \frac{{\rm ln}^2 \vert x-y \vert^2}{\vert x-y
\vert^2}\right\} = -2\pi^2 \frac{1}{\vert x-y \vert^4} \left\{ 1-2 {\rm ln}
\vert x-y \vert\right\}. \ee
The logarithmic term of~\cite{Freedman:1992tk} appears here in
the form ${\rm ln}
(M^2 x^2)$, where $M$ is a mass scale. We have set $M$ to one
to match the corresponding expression on the left
hand side of~\C{logterm}.  This equation therefore leads to
\be\label{partt}
 \left( \frac{\partial}{\partial \tau} - \frac{i(\alpha_\Om +
  \alpha_{\bar\Om})}{\tau_2}\right)\eta -2 \eta
\frac{\partial \Delta}{\partial \tau} {\rm ln}
\vert x-y \vert = -\frac{i \pi^2 C^\tau}{2 \tau_2} \left\{1-2{\rm ln} \vert x-y \vert\right\},\ee
so that
\be \label{eqnmore} \left( \frac{\partial}{\partial \tau} - \frac{i(\alpha_\Om +
  \alpha_{\bar\Om})}{\tau_2}\right)\eta =-
\frac{i\pi^2}{2 \tau_2} C^\tau, \qquad
\frac{\partial \Delta}{\partial \tau} = -\frac{i\pi^2}{2 \tau_2}
\frac{C^\tau}{\eta},
\ee
where the coefficients  $\alpha_\Om$ and $\alpha_{\bar\Om}$ arise from the last term
in~\C{logterm}.
The second relation shows that the conformal dimension depends on the
coupling only via the ratio of the correlator of $three$ operators to the
correlator of  $two$ operators.  In order to see how the contact terms modify~\C{logterm}
we now want to compare~\C{partt} with~\C{eqn2pt}.

Using~\C{twopt}, we see that
\C{eqn2pt} reduces to
\bea \label{opeconst}
&& \frac{1}{16\pi}
\frac{\partial}{\partial \tau} \langle {\bar\Om} (x) {\Om} (y)\rangle
= \left\{ \frac{i(\alpha_\Om +
  \alpha_{\bar\Om})}{16\pi \tau_2}
-\frac{2\pi i(A+2C)}{\tau_2} \right\} \frac{\eta}{(x-y)^{2\Delta}} \nonumber \\
&& + \frac{i(A+B+C)}{\pi \tau_2}
\frac{\Delta (\Delta -1) \eta}{(x-y)^{2\Delta +2}}
\left(\int_{B_x^\epsilon} \frac{d^4 z}{(z-x)^2}
+ \int_{B_y^\epsilon} \frac{d^4 z}{(z-y)^2}\right)+\ldots.
\eea
We now want  to compare~\C{logterm}\ and~\C{opeconst}.  This is facilitated by
treating the $z$ integral in~\C{logterm}\ in the same manner as
in~\C{opeconst} by dividing it up into balls, ${B_{x_i}^\epsilon}$, surrounding
each insertion
point, $x_i$, together with the smooth contribution away from the insertion points.
 The second term on the right hand side of~\C{opeconst}\ arises from a
contact term between $\Ot$ and the inserted operators. This is
precisely the contact term we neglected in~\C{logterm}\ so we will not
be able to match this term.  For the other terms, the expansion
of~\C{logterm}\ gives
\bea \label{difreg}
\frac{\partial}{\partial \tau} \langle {\bar\Om} (x) {\Om}
(y)\rangle -
\langle \frac{\partial}{\partial \tau}
\left( \bar{{\mathcal{O}}} (x) {\mathcal{O}} (y) \right)
&=&
\frac{iC^\tau}{4\tau_2 (x-y)^{2\Delta -4}} \bigg[ \int_{B_x^\epsilon}
\frac{d^4 z}{(z-x)^4 (z-y)^4} + \non\\ &&
\int_{B_y^\epsilon} \frac{d^4 z}{(z-x)^4 (z-y)^4} \bigg] +\ldots
\nonumber \\
&=& \frac{iC^\tau}{4\tau_2 (x-y)^{2\Delta}}
\left[\int_{B_x^\epsilon} \frac{d^4 z}{(z-x)^4}
+ \int_{B_y^\epsilon} \frac{d^4 z}{(z-y)^4}\right]
\nonumber \\
&& +\frac{iC^\tau}{4\tau_2 (x-y)^{2\Delta +2}}
\left[\int_{B_x^\epsilon} \frac{d^4 z}{(z-x)^2}
+\int_{B_y^\epsilon} \frac{d^4 z}{(z-y)^2} \right]
\non\\ &&+O(\frac{1}{(x-y)^{2\Delta +4}}).
\eea
The $O(1/(x-y)^{2\Delta})$ terms contain UV divergent integrals.
We can again evaluate the integrals
using differential regularization again~\cite{Freedman:1992tk},
which gives
\be \int_{B_x^\epsilon} \frac{d^4 z}{(z-x)^4} = -\frac{1}{4}
\partial_x^2 \int_{B_x^\epsilon} d^4 z \frac{{\rm ln}
(z-x)^2 M^2}{(z-x)^2} =  -\frac{1}{4} \partial_x^2 \int_{B_0^\epsilon}
d^4 z \frac{{\rm ln}z^2 M^2}{z^2} = 0, \ee
since the integral over $B_0^\epsilon$ is independent of $x$. This is
in agreement with~\C{opeconst}.  Also terms of
$O(1/(x-y)^{2\Delta +4})$ appear with integrals of the
form
\be \int_{B_x^\epsilon} d^4 z, \quad \int_{B_x^\epsilon} d^4 z (z-x)^2,
\quad \int_{B_x^\epsilon} d^4 z (z-x)^4, \quad \cdots. \ee
These integrands are regular as $z \rightarrow x$,  and so they do not
appear in the $\Ot$ OPE expansion and are absent from~\C{opeconst}.

Equating the $O(1/(x-y)^{2\Delta +2})$ terms in~\C{opeconst}\
and~\C{difreg}\ gives
\be \label{formC}
\frac{C^\tau}{64\eta} = (A+B+C)\Delta (\Delta -1).\ee
The contact term contribution in~\C{opeconst}\ together with the
explicit $\tau$ derivative (accounted for by~\C{explicittau}) of the
inserted operators modifies the
first equation of~\C{eqnmore}
\be \label{contchange}
\left(\frac{\partial}{\partial \tau} +\frac{32\pi^2 i}{\tau_2}
(A+2C) - \frac{i(\alpha_\Om + \alpha_{\bar\Om})}{\tau_2}  \right)\eta =
-\frac{16\pi}{\tau_2} (A+B+C)\Delta (\Delta -1) \eta \ee
where we have used \C{formC}. Note that $A+2C$ is generally a coupling
dependent function.

The second equation in~\C{eqnmore}\ leads to
\be \label{nocont}
\frac{\partial \Delta}{\partial \tau}
=- \frac{16\pi}{\tau_2} (A+B+C) \Delta (\Delta -1).\ee
We now argue that contact terms cannot   modify~\C{nocont}.
 From the
left hand side of~\C{logterm}, we see that any possible contact term
contribution to $\frac{\partial \Delta}{\partial \tau}$ must include
a factor of ${\rm ln} \vert x-y \vert$.
However, in finding contact term contributions, we consider
integrals over $B_x^\epsilon$ (and $B_y^\epsilon$) and pick up the
contributions when $z=x$ (and $z=y$). However, these integrals always
result in a power series expansion in $1/ \vert x-y\vert$ and not a
${\rm ln} \vert x-y \vert$ term. So no contact term is generated.

\subsection{Implications for BPS two-point functions}
\label{BPSsection}

Let us start by considering the case where $\Om$ is the superconformal primary
of a BPS multiplet; for example, $\Om$ can be ${\Om}_p$ for a $1/2$
BPS multiplet. Let $\Om$ be annihilated by $Q_i^{\al}$ so $[Q_i^{\al},\Om] =0$.
In this case, further constraints are implied by superconformal Ward identities.
In the supercurrent Ward identity~\C{jward}, take
$\Phi^{I_1} = [{\bar{Q}}^{i\dal},\Om]$ and $\Phi^{I_2} =
\Om$,\footnote{Note that if $[Q_i^{\al},\Om] =0$ then
$[{\bar{Q}}^{i\dal},\Om] \neq 0$. Otherwise, using $\{ Q,\bar{Q} \}
\sim \partial_\mu$ gives $\partial_\mu \Om =0$ and
$\Om$ must be proportional to the identity operator.}
\bea \label{jwardmore}
\partial^\mu \langle J^{\alpha}_{\mu i} (z) [{\bar{Q}}^{i\dal},\Om (x)]
\Om (y) \rangle = \Big\{ \del^4 (z-x) \langle \{Q_i^{\al}, 
[{\bar{Q}}^{i\dal},\Om (x)] \} \Om (y) \rangle
\nonumber \\
+\del^4 (z-y) \langle [{\bar{Q}}^{i\dal},\Om (x)] [Q_i^{\al},\Om 
(y)] \rangle \Big\}. \eea Using the relation, \be [Q_i ,\Ot] = 
\partial^\mu J_{\mu i} +2 \s^\mu \bar\s^\nu \partial_\mu J_{\nu i}, 
\ee the left hand side of~\C{jwardmore}\ reduces to 
\be\label{firstman} \langle [Q_i{}^\alpha ,\Ot (z)] 
[{\bar{Q}}^{i\dal},\Om (x)] \Om (y) \rangle -2 (\s_\mu 
\bar\s_\nu)_{\al}{}^\beta \partial^{z^\mu} \langle J^\nu_{i\beta} 
(z) [{\bar{Q}}^{i\dal},\Om (x)] \Om (y) \rangle . \ee The second 
term will vanish when we integrate over $z$ by essentially the same 
argument given around~\C{impterm}. The left hand side 
of~\C{jwardmore}\ is then proportional to \be \label{LHSstep}
 \partial^{\dal\al}_x \langle \Ot (z) \Om (x) \Om (y) \rangle. \ee
Finally~\C{jwardmore}\ yields the coupling independent relation
\be 
\label{px1}
\partial^{\dal\al}_x \langle \Ot (z) \Om (x) \Om
(y) \rangle \sim \left[ \del^4 (z-x) +\del^4 (z-y) \right] \partial^{\dal\al}_x
\langle \Om (x) \Om (y) \rangle.\ee
Integrating over $z$ gives
\be \label{px2}\int d^4 z \langle \Ot (z) \Om (x) \Om
(y) \rangle
\sim \langle \Om (x) \Om (y) \rangle\ee
(we need not worry about removing the $\p_x$ in going from~\C{px1}\
to~\C{px2}\ because we know the precise $x$-dependence of these
correlators). So
the only contributions from the $\Ot$ OPE are contact term
contributions which means that
\be
A+B+C=0 \quad \Rightarrow \frac{\partial\Delta}{\partial \tau} =0.
\ee
We thus recover the known result that the conformal dimensions of BPS
superconformal primary
operators are not renormalized.

To fix the metric on this subsector of BPS operators, we need to 
note that the relation~\C{jwardmore}\ is independent of the 
coupling. This is a rather important point which can be seen as 
follows. Integrate both sides of~\C{jwardmore}\ over $z$. For the 
right hand side, the integration is trivial. The left hand side 
becomes
$$ \int d^4z \, \partial^\mu \langle J^{\alpha}_{\mu i} (z)
[{\bar{Q}}^{i\dal},\Om (x)] \Om (y) \rangle =
\langle \{Q_i^{\al}, [{\bar{Q}}^{i\dal},\Om (x)] \} \Om (y) \rangle
+\langle [{\bar{Q}}^{i\dal},\Om (x)] [Q_i^{\al},\Om (y)]
\rangle $$
so the coefficient of proportionality in~\C{jwardmore}\ is just a
constant. However, we also need to note that the supersymmetry
variation of a BPS operator is not quantum corrected, unlike the case
of long operators. This follows, essentially, because the divergence of
the supercurrent sits in the same anomaly multiplet as $T^\m_\m$. For an
example of quantum corrections to the supersymmetry variation of a
long operator, see~\cite{Intriligator:1999ff}. The only way a coupling
dependence could appear is
if $x$ approaches $y$ and a long
operator emerges in the OPE of $[{\bar{Q}}^{i\dal},\Om (x)]$ and
$\Om(y)$. The quantum corrections to the supervariation of this long
operator encodes the coupling dependence of the correlation
function. However, in this case, what would remain is the $1$-point function
of a  long operator
which vanishes.

The Ward identity has therefore taught us that the only relevant
terms in the OPE between $\Ot$ and the inserted operators are the
tree-level contact terms. So we conclude that,
\be \label{BPSzero}
(\frac{\partial}{\partial \tau} +\frac{\alpha}{\tau_2})\eta
= 0, \ee
{}for some constant $\alpha$ and that $A+2C$ is a
constant independent of the coupling.

Lastly, we can set $\alpha$ to zero in the following way: we need to
examine the solutions to~\C{BPSzero}\
and its conjugate equation. A quick inspection reveals that the only
solution for $\eta$ is a single fixed power of $\tau_2$.
However, we have the freedom
to rescale $\Om$ by powers of $\tau_2$ in a way compatible
with~\C{explicittau}\ (the value of $\alpha_{{I_i}}$ shifts by such a
rescaling). Using this freedom, we set $\alpha=0$ in~\C{BPSzero}\ and
conclude that $\eta$ is a constant. With this normalization, each BPS
superconformal primary has modular weight $(0,0)$. This is consistent with our
expectations from S-duality~\cite{Intriligator:1998ig}.

The only supermultiplet with a canonical normalization is the current
multiplet. In particular, $\OT$ appears with a single factor of
$\tau_2$. In this case, we do not want to use our rescaling freedom to
change $\alpha$. However, using the canonical normalization, we see
that the tree-level $2$-point function,
$$ \langle \OT(x) \OT(y) \rangle \sim \frac{1}{|x-y|^4}, $$
is coupling independent and non-zero. Therefore, in this case  we find
$\alpha=0$ automatically with the canonical normalization. We
conclude that $2$-point functions of superconformal BPS primary operators
are not renormalized. This statement is non-perturbative and applies
to $1/2$, $1/4$ and $1/8$ BPS multiplets. We might
worry that in an instanton background, half of the supersymmetries are
broken so the Ward identities associated with the broken currents are
inapplicable. However, we are still free to choose any $J^{\alpha}_{\mu
  i}$ in~\C{jwardmore}\ which is preserved, and the rest of the
argument is unchanged.

We can extend this result to two-point functions of BPS operators
which are not superconformal primaries. This follows fairly directly
from~\cite{Intriligator:1998ig}. First note that we are free to move
all the $\del$ and $\bar\del$ operators so that they act on one of the
two inserted operators. We then observe that any two-point function of this kind
satisfies the following relation
\be \label{anotherrel}
\langle \Om(x) \del^n \bar\del^m \Om(y) \rangle \sim \delta^{(n,m)} \,
\p_y^n
\langle \Om(x)\Om(y) \rangle,
\ee
which follows for the same reasons as~\C{confinv}. This two-point
function is therefore also not renormalized.

It is worth noting that there can be contact term contributions to
two-point correlators. These contact terms are renormalized even
for BPS operators, as shown explicitly
in~\cite{Penati:2000zv}. Also, these contact terms can and do appear in
correlators like~\C{anotherrel}\ even when $n\neq m$
(in which case the correlator vanishes for separated insertion points).

\subsection{Implications for BPS three-point functions}
\label{BPS}
Let us consider the three point function $\langle {\Om}^I (x) {\Om}^J
(y) {\Om}^K (w)\rangle$ where the operators are BPS superconformal
primaries at separated points.
Using the result for conformal primaries, we define the
ring coefficients  ${\Oc}^{IJK}$,
\be \langle {\Om}^I (x) {\Om}^J (y) {\Om}^K (w) \rangle =
\frac{{\Oc}^{IJK} (\tau, \bar\tau)}{\vert x-y \vert^{\Delta_{IJK}}
\vert x-w \vert^{\Delta_{IKJ}}
\vert y-w \vert^{\Delta_{JKI}} },\ee
where $\Delta_{IJK} =\Delta_I +\Delta_J -\Delta_K$ and
$\frac{\partial\Delta_I}{\partial \tau} =0$.

The coupling dependence, which follows from~\C{ddtau}, is determined by
\bea \label{dertau3pt}
\frac{\partial}{\partial \tau} \langle {\Om}^I (x) {\Om}^J
(y) {\Om}^K (w) \rangle
&=& \langle \frac{\partial}{\partial \tau}\left( {\Om}^I (x) {\Om}^J
(y) {\Om}^K (w) \right)\rangle \non\\ &&
+ \frac{i}{4\tau_2} \int d^4 z \langle \Ot (z)
{\Om}^I (x) {\Om}^J (y) {\Om}^K (w) \rangle.  \eea
By analogy with the case of the two-point function, consider the
Ward identity~\C{jward}\ with $\Phi^{I_1} = [{\bar{Q}}^{i\dal},{\Om}^I]$,
$\Phi^{I_2} ={\Om}^J$ and $\Phi^{I_3} ={\Om}^K$ and where each ${\Om}$
is annihilated by $Q_i{}^\alpha$. Repeating our prior argument gives,
\bea \label{worry}
\partial^{\dal\al}_x \langle \Ot (z) {\Om}^I (x) {\Om}^J
(y) {\Om}^K (w) \rangle \sim \qquad \qquad \qquad  \nonumber \\
\left[\del^4 (z-x) +\del^4 (z-y)
+\del^4 (z-w) \right] \partial^{\dal\al}_x
\langle {\Om}^I (x) {\Om}^J (y) {\Om}^K (w) \rangle,\eea
with the constant of proportionality again independent of the
coupling. Indeed, the constant is just the value of the contact term
between $\Ot$ and each inserted operator. This same contact term
appeared in the determination of BPS two-point functions.

In deriving~\C{worry}, we have moved a $Q$ around the correlation
function just as in the discussion around~\C{firstman}.
If this $Q$ hits a long operator,
the relation might be renormalized. Again, a long operator can only
emerge from the OPE of two of the inserted BPS operators. However,
that would leave a two-point function of a long and a BPS operator
which always vanishes. The constant in~\C{worry}\ is therefore not
renormalized.

As an aside, we should comment that had we considered a
correlator of $4$ BPS operators, the conclusion would be different.
In this case, the Ward identity no longer guarantees
the absence of quantum corrections: a long operator can emerge
from two short operators but now the resulting three-point function need
not vanish. If all possible three-point functions of this kind were to
vanish (for other symmetry reasons) then the
$4$-point function would again be protected from renormalization.

Returning to the three-point function, we note that integrating~\C{worry}\ gives,
\be \int d^4 z \langle \Ot (z)
{\Om}^I (x) {\Om}^J (y) {\Om}^K (w) \rangle \sim
\langle {\Om}^I (x) {\Om}^J (y) {\Om}^K (w) \rangle. \ee
Thus \C{dertau3pt} implies that
\be (\frac{\partial}{\partial \tau} +\frac{\alpha'}{\tau_2})
{\Oc}^{IJK} = 0 \ee
{}for some constant $\alpha'$.
Normalizing the operators as in section~\ref{BPSsection}\ so that the
two-point functions are
independent of the coupling then trivially implies that
$\alpha'=0$; the contribution from the explicit derivative together
with the contact term with $\Ot$ vanishes separately for each
operator.
We conclude that the three-point functions of BPS superconformal primaries are
not renormalized.

\subsubsection{Comments on three-point functions of descendents}
\label{descendentcomments}
It turns out that our non-renormalization proof does not extend simply
to $3$-point correlators of descendents. For the special case of the
current multiplet, the non-renormalization result does extend to
descendents. This follows from the analysis of~\cite{Dolan:2001tt}\
where correlators of descendents were related to $\langle \OT(x)
\OT(y) \OT(w) \rangle$ using superconformal symmetry.

{}For other BPS multiplets, we can demonstrate non-renormalization
under the assumption that (anti-)instanton corrections to the $3$-point
correlator vanish. The argument uses $SL(2,\Z)$ in the following way:
any correlator of superconformal descendents is
of the form
\be
\langle \del^r \bar\del^{\bar{r}} \Om_1 (x)
\del^s \bar\del^{\bar{s}} \Om_2 (y)
\del^t \bar\del^{\bar{t}} \Om_3 (w)\rangle,
\ee
where $\Om_1,\Om_2$ and $\Om_3$ are BPS superconformal primaries. By moving
around the $\del$ and $\bar\del$ operators, these correlators can always be put
in the form
\be \label{3ptcorr}
\langle \Om_1 (x) \del^m \bar\del^{\bar{n}} \Om_2 (y)
\del^p \bar\del^{\bar{q}} \Om_3 (w) \rangle,\ee
or its space-time derivatives. So it is good enough to
analyze the coupling dependence of~\C{3ptcorr}. Note that neither
$\del^m \bar\del^{\bar{n}}$ nor $\del^p \bar\del^{\bar{q}}$ yield conformal
descendents by definition (if they do, they would merely become space-time
derivatives of correlators like~\C{3ptcorr}). To study the coupling
dependence of \C{3ptcorr}, we consider the OPE of $\Om_1 (x)$ and
$\del^m \bar\del^{\bar{n}} \Om_2 (y)$ as $x \rightarrow y$.
Assuming~\C{3ptcorr}\ is non-vanishing (otherwise there is nothing to prove),
we see that the OPE is given by
\be \label{OPE3pts}
\Om_1 (x) \del^m \bar\del^{\bar{n}} \Om_2 (y) \sim \sum_k g_k (x-y)
f_k (g^2_{YM} ,\p_\mu) \del^{u_k} \bar\del^{v_k} \Om_3 (y) +\ldots,\ee
where the omitted terms involve long operators as well as BPS operators which are
not (super)conformal descendents of $\Om_3$. We assume no instanton
corrections so that $f_k$ is independent of $\theta$.
All the (super)conformal
descendents of $\Om_3$ are included in the sum.  In the $k$-th term, $u_k$ and $v_k$
are non-negative integers and the OPE coefficients $f_k (g^2_{YM} ,\p_\mu)$
are a priori functions of the coupling. Clearly, the omitted BPS
and
long operators do not contribute to~\C{3ptcorr}.
In fact, substituting~\C{OPE3pts}\ into~\C{3ptcorr}\ gives (as
$x \rightarrow y$)
\bea & \langle \Om_1 (x) \del^m \bar\del^{\bar{n}} \Om_2 (y)
\del^p \bar\del^{\bar{q}} \Om_3 (w) \rangle &  \non \\
&=\sum_k \del^{(u_k +p,v_k +q)} g_k (x-y)
f_k (g^2_{YM} ,\p_\mu) \p_y^{u_k +p} \langle
\Om_3 (y) \Om_3 (w) \rangle. &
\eea
Since $\langle \Om_3 (y) \Om_3 (w) \rangle$ is coupling independent, in
order
to demonstrate the coupling independence of~\C{3ptcorr},  we need to show
that $f_k (g^2_{YM} ,\p_\mu)$ is independent of coupling for every $k$.

To see this,  we consider the $SL(2,\Z)$ transformation properties of
\C{OPE3pts}. The BPS superconformal primaries $\Om_1, \Om_2$ and
$\Om_3$, are $SL(2,\Z)$ invariant. Also both sides of~\C{OPE3pts}\
transform covariantly under $SL(2,\Z)$ because the left hand side is
constructed purely out of BPS operators (Every omitted term involving
a long operator is also expected to transform covariantly, though the
coupling dependent OPE coefficient and the operator need not do so
individually). Let $\del$ and $\bar\del$ transform as modular forms of
weights $(\mu, -\mu)$ and $(-\mu, \mu)$ respectively. That they have
weights of the form $(\nu, -\nu)$ follows because $\del\bar\del \sim \p_\mu$
as far as the $SL(2,\Z)$ structure is concerned and $\p_\mu$ is $SL(2,\Z)$
invariant. Here the specific value of $\mu$ (which is $1/4$) is not needed.
Now $f_k (g^2_{YM} ,\p_\mu)$ is also a modular form but, by assumption,
it is independent of
$\tau_1 = \theta/2\pi$. Hence, it can only be a power of $\tau_2$. So, let
$f_k (g^2_{YM} ,\p_\mu)= \tau_2^k f_k (\p_\mu)$. So the left hand side of
\C{OPE3pts} transforms as a modular form of weights
$\left((m-n)\mu,-(m-n)\mu \right)$, while the $k$-th term on the right hand side of
\C{OPE3pts} transforms with weights $\left(-k +(u_k -v_k)\mu,-k-(u_k -v_k)\mu\right)$.
Equating these expressions,  we find that
\be (m-n)\mu = -k +(u_k -v_k)\mu , \qquad
(m-n)\mu =k+(u_k -v_k)\mu,\ee
which leads to $k=0$. Hence, \C{3ptcorr}\ is independent of the
coupling and~\C{3ptcorr}\ must have an equal number of $\del$ and
$\bar\del$ insertions.
Thus
the $3$-point function of non-coincident BPS superconformal descendents
is not renormalized if $f_k$ in~\C{OPE3pts}\ is independent of $\tau_1$.

\subsubsection{A simplified integration formula}
\label{nice}
These results lead to a pretty formula for the integrated OPE of $\Ot$
with a BPS operator $\Om$. Consider the relation,
\be
\label{anothertau}\frac{\partial}{\partial \tau} \langle \Om(x)
\Om_i'(y)
\rangle
= \langle \frac{\partial}{\partial \tau} \left\{ \Om(x)
\Om_i'(y) \right\} \rangle +\frac{i}{4\tau_2} \int d^4 z \langle \Ot (z)
\Om(x)\Om_i'(y) \rangle,
\ee
where $\Om_i'(y)$ is any BPS operator. The left hand side vanishes by
our prior non-renormalization argument. The first term of the right
hand side is zero if $\Om_i'(y)$ is not in the same supermultiplet as
$\Om(x)$. In this case, the integrated $3$-point function vanishes.

If
$\Om_i'(y)$ is in the same supermultiplet as $\Om(x)$ then the
explicit derivative term can be non-vanishing but is always
proportional to $1/\tau_2$. This term must be cancelled by the
integrated $3$-point function which must be exact at tree-level.
It is easy to check that $\int d^4z
\Ot(z)$ is zero on-shell so the $3$-point function is purely a contact
term.\footnote{There is a caveat worth mentioning:
to see that $\int d^4z\Ot(z)$ is zero on-shell requires integrating by
parts and throwing away a boundary term. In Euclidean space, we could
consider an instanton background in which we might need to consider this
boundary term.} The BPS operators in the integrated OPE between
$\Ot$ and $\Om$ are therefore in the same supermultiplet as $\Om$ and arise only
from tree-level contact terms in the OPE (up to integration by parts). We
therefore conclude that,
\be
\int d^4z \Ot(z) \Om(x) \sim \sum_{i} \Om_i'(x) + \ldots,
\ee
where the omitted terms involve long and semi-short operators.

\subsection{A comment about generic three-point functions}

The coupling dependence of generic correlators is given by
\bea \label{threecoup}
\frac{\partial}{\partial \tau} \langle {\Om} (x) \widehat{\Om}
(y) \widetilde{\Om} (w) \rangle &=&
\langle \frac{\partial}{\partial \tau} \left({\Om} (x) \widehat{\Om}
(y) \widetilde{\Om} (w) \right)\rangle+
\frac{i}{4\tau_2} \int d^4 z \langle \Ot (z) {\Om} (x)
\widehat{\Om} (y) \widetilde{\Om} (w) \rangle \nonumber \\
&=& \langle\frac{\partial}{\partial \tau}\left( {\Om} (x) \widehat{\Om}
(y) \widetilde{\Om} (w)\right) \rangle \non\\ && +
\frac{i}{4\tau_2} \int_{B_x^\epsilon}
d^4 z \langle \Ot (z) {\Om} (x)
\widehat{\Om} (y) \widetilde{\Om} (w) \rangle +\ldots,
\eea
where $\ldots$ includes integrals over $B_y^\epsilon$ and
$B_w^\epsilon$,  and over the region outside the balls.
Again, because we know the precise form of the space-time dependence
of three-point functions,
it might be possible to learn about these correlators from the ball
approximation.

After using
the $\Ot$ OPE, we note that the correlators on the right hand side
generically involve different operators from those appearing on the
left hand side of~\C{threecoup}.
In order to see this, it is enough to consider
a particular term in the integral over $B_x^\epsilon$. Consider the
$a_2$ term in~\C{Otoneop}\ which gives a contribution
\be \label{a2term}
\frac{i a_2}{4\tau_2} \int_{B_x^\epsilon} \frac{d^4 z}{(z-x)^2}
\langle \{ Q_i{}^\alpha ,[{\bar{Q}}^i{}_{\dal}, \{ {\bar{Q}}^{j\dal}
,[Q_{j\alpha} ,\Om (x)]\}]\} \widehat{\Om} (y) \widetilde{\Om} (w)
\rangle. \ee
{}For example, take $\Om =\OT$ and take $\widehat{\Om}$ and
$\widetilde{\Om}$ to be superconformal primaries of long multiplets.
Then from~\C{supervar}, very schematically, we see that~\C{a2term}\
contributes
\be \frac{a_2}{\tau_2} \int_{B_x^\epsilon} \frac{1}{(z-x)^2}
\left[\langle T_{\mu\nu} (x)
\widehat{\Om} (y) \widetilde{\Om} (w) \rangle  +\langle
\partial_\mu R_\nu \widehat{\Om} (y) \widetilde{\Om} (w) \rangle
+ \langle
\partial_\mu \partial_\nu \OT (x) \widehat{\Om} (y)
\widetilde{\Om} (w) \rangle \right] +\ldots, \non\ee
where $\ldots$ includes terms that involve derivatives acting on
$\frac{1}{(z-x)^2}$. So we see that the operators involved differ
from those in $\langle {\Om} (x) \widehat{\Om} (y) \widetilde{\Om} (w)
\rangle$, but they are all operators in the same supermultiplet (here
$\OT$) as the original one. Clearly, this is also true for higher
point functions.

\section*{Acknowledgements}

It is our pleasure to thank D.~Z.~Freedman, J.~Harvey,
K.~Intriligator, D.~Kutasov, J.~Maldacena, H.~Osborn, K.~Skenderis 
and E.~Sokatchev for helpful discussions. The work of A.~B. is 
supported in part by  NSF Grant No. PHY-0204608.  The work of 
M.~B.~G. is supported in part by a PPARC rolling grant. The work of 
S.~S. is supported in part by NSF CAREER Grant No. PHY-0094328 and 
by the Alfred P. Sloan Foundation.

\appendix
\section{The Superconformal Algebra}
\label{superconformal}

In this Appendix, we briefly review the superconformal algebra in four dimensions.
With metric $\eta_{\mu \nu}$ =
diag($-1,1,1,1$),
the conformal algebra $SO(4,2)$ in $d=4$ is given
by the commutation relations
\bea
& [M_{\mu \nu}, P_\s] = i(\eta_{\mu \s} P_{\nu} -\eta_{\nu \s}
P_{\mu}),  \qquad [M_{\mu \nu}, K_\s] = i(\eta_{\mu \s} K_{\nu}
-\eta_{\nu \s} K_{\mu}), \non \\
& {\rm [}M_{\mu \nu}, M_{\rho \s}] = i(\eta_{\mu \rho} M_{\nu \s} -\eta_{\nu
\rho} M_{\mu \s} -\eta_{\mu \s} M_{\nu \rho} +\eta_{\nu \s}
 M_{\mu \rho}),  \\
& {\rm [} D,P_{\mu}] = iP_{\mu}, \qquad [D,K_{\mu}] = -iK_{\mu},
\qquad [K_{\mu},P_{\nu}] = -2i M_{\mu \nu} -2i \eta_{\mu \nu} D, \non
\eea
where $M_{\mu \nu}, P_{\mu}, K_{\mu}$ and $D$ are the generators of Lorentz
transformations, translations,
special conformal transformations and dilations, respectively.
In $d=4$, the conformal algebra $SO(4,2)$ can be extended to the superconformal
algebra $SU(2,2|4)$ by the inclusion of the supersymmetry charges
$Q_{i \al},
{\bar{Q}}^i_{\dal}$ and superconformal charges $S^{i \al},
{\bar{S}}_i^{\dal}$,
where $i=1,\ldots,4$. These charges transform in the (anti-)fundamental representation of
the global SU(4) $R$-symmetry group. We denote the generators of the $R$-symmetry group by
$R^i{}_{\! j}$ subject to the condition $R^i{}_{\! i} =0$. The non-zero
anti-commutation
relations between the supersymmetry charges and the superconformal
charges are given by
\bea \label{nt}
& \{ Q_{i\al}, {\bar{Q}}^j_{\dal} \} = 2 \del^j_{ i}
\s^{\mu}_{\al \dal} P_{\mu},\qquad
\{{\bar{S}}_i^{\dal}, S^{j \al} \}  = 2 \del^j_{ i}
{\bar\s}_{\mu}^{\dal \al} K^{\mu}, \non \\
& \{ Q_{i \al}, S^{j \bt} \} = 4[\del^j_i (M_\al{}^{\! \bt}
-\frac{i}{2} \del_\al{}^{\! \bt} D) -\del_\al{}^{\! \bt} R^j_i], \\ &
\{ {\bar{S}}_i^{\dal}, {\bar{Q}}^j_{\dbt} \} =
4[\del^j_{i} ({\bar{M}}^{\dal}{}_{\! \dbt} +\frac{i}{2}
\del^{\dal}{}_{\! \dbt} D) -\del^{\dal}{}_{\! \dbt} R^j_i ],\non
\eea
where the Lorentz generators are expressed in a spinorial basis,
$$ M_\al{}^{\! \bt} = -\frac{i}{4} (\s^{\mu} \bar\s^{\nu})_\al{}^{\! \bt}
M_{\mu \nu}, \qquad {\bar{M}}^{\dal}{}_{\! \dbt} = -\frac{i}{4}
(\bar\s^{\mu} \s^{\nu})^{\dal}{}_{\! \dbt} M_{\mu \nu}.$$
The non-zero
commutation relations involving $M_\al{}^{\! \bt}$,
${\bar{M}}^{\dal}{}_{\! \dbt}$, the supersymmetry charges,  and
superconformal
charges are given by
\bea
& [M_\al{}^{\! \bt},Q_{i \g}]  = \del_\g{}^{\! \bt} Q_{i \al}
-\frac{1}{2} \del_\al{}^{\! \bt} Q_{i \g}, \qquad &
{\rm [}M_\al{}^{\! \bt},S^{i \g}] = -\del_\al{}^{\! \g} S^{i \bt}
+\frac{1}{2} \del_\al{}^{\! \bt} S^{i \g}, \non \\
& {[}{\bar{M}}^{\dal}{}_{\! \dbt}, {\bar{Q}}^i_{\dg}] =
-{\del}^{\dal}{}_{\! \dg} {\bar{Q}}^i_{\dbt} +\frac{1}{2}
{\del}^{\dal}{}_{\! \dbt} {\bar{Q}}^i_{\dg}, \qquad &
{\rm [}{\bar{M}}^{\dal}{}_{\! \dbt}, {\bar{S}}_i^{\dg}]=
{\del}^{\dg}{}_{\! \dbt}
{\bar{S}}_i^{\dal} -\frac{1}{2} {\del}^{\dal}{}_{\! \dbt} {\bar{S}}_i^{\dg},
\eea
while $[M_{\mu \nu},M_{\rho \s}]$ in the spinorial basis gives
\bea \label{MMCCR}
 [M_\al{}^{\! \bt}, M_\g{}^{\! \del}] &=& {\del}_\g{}^{\! \bt} M_\al{}^{\! \del}
-\del_\al{}^{\! \del} M_\g{}^{\! \bt}, \\
 {[}{\bar{M}}^{\dal}{}_{\! \dbt},{\bar{M}}^{\dg}{}_{\! \ddel}] &=&
-{\del}^{\dal}{}_{\! \ddel} {\bar{M}}^{\dg}{}_{\! \dbt}
+{\del}^{\dg}{}_{\! \dbt} {\bar{M}}^{\dal}{}_{\! \ddel}. \non
\eea

The non-zero commutation relations involving the
dilation operator, the supersymmetry charges and the superconformal
charges are given by:
\bea
& [D,Q_{i \al}] =\frac{i}{2} Q_{i \al}, \qquad  &
{[}D,{\bar{Q}}^i_{\dal}] =\frac{i}{2} {\bar{Q}}^i_{\dal}, \\
& {[}D,S^{i \al}] =-\frac{i}{2} S^{i \al}, \qquad  &
{[}D,{\bar{S}}_i^{\dal}] =-\frac{i}{2} {\bar{S}}_i^{\dal}. \non
\eea
Those involving $P_{\mu}$ and $K_{\mu}$ are
\bea
& [K^{\mu},Q_{i \al}] =-\s^{\mu}_{\al \dal} {\bar{S}}_i^{\dal}, \qquad
{\rm [}K^{\mu},{\bar{Q}}^i_{\dal}] =S^{i \al} \s^{\mu}_{\al \dal}, \non \\
& {\rm [}P_{\mu},{\bar{S}}_i^{\dal}] =-\bar\s_{\mu}^{\dal \al} Q_{i \al},
\qquad
{\rm [}P_{\mu},S^{i \al}] = {\bar{Q}}^i_{\dal} \bar\s_{\mu}^{\dal \al}.
\eea
Finally,  we list the defining relations for the $R$-symmetry
generators
\be\label{defineR} [R^i{}_{\! j},R^k{}_{\! l}] = {\del}^k{}_{\! j} R^i{}_{\! l}
-{\del}^i{}_{\! l} R^k{}_{\! j},
\ee
and their commutation relations with the super(conformal) charges
\bea \label{RReqn}
& {[}R^i{}_{\! j},Q_{k \al}]= {\del}^i_{k} Q_{j \al}
-\frac{1}{4} {\del}^i_{ j} Q_{k \al}, \qquad &
{[}R^i{}_{\! j},{\bar{Q}}^k_{\dal}]= -{\del}^k_j {\bar{Q}}^i_{\dal}
+\frac{1}{4} {\del}^i_{ j} {\bar{Q}}^k_{\dal},\\
& {[}R^i{}_{\! j},S^{k \al}] =-{\del}^k_j S^{i \al}
+\frac{1}{4} {\del}^i_{ j} S^{k \al}, \qquad &
{[}R^i{}_{\! j}, {\bar{S}}_k^{\dal}] = {\del}^i_{ k} {\bar{S}}_j^{\dal}
-\frac{1}{4} {\del}^i_{j} {\bar{S}}_k^{\dal}.\non
\eea
{}For unitary representations of the superconformal algebra, these
operators satisfy the conditions
\be Q_{i \al}{}^\dagger ={\bar{Q}}^i_{ \dal},
\qquad {S^{i \al}}{}^\dagger
={\bar{S}}_i^{\dal}, \qquad {M_\al{}^{\! \bt}}{}^\dagger =
{\bar{M}}^{\dbt}{}_{\! \dal}, \qquad R^i{}_{\! j}{}^\dagger = R^j{}_{\! i}.\ee

\section{The Properties and Construction of Short Multiplets}
\label{shortstructure}

The short multiplet has special
properties which can be deduced from the superconformal algebra. For
example, states have conformal dimensions that are protected from
renormalization. In this Appendix, we will deduce some of these
properties while describing the explicit construction of the
multiplet. Our discussion again follows~\cite{Dolan:2002zh}. We will
need these explicit results in the main text so, for completeness, we
list them here.

{}First, we give explicit representatives for  $M_\al{}^{\! \bt},
{\bar{M}}^{\dal}{}_{\! \dbt}$ and $R^i{}_{\! j}$. Writing
\be \label{Matrix} [M_\al{}^{\! \bt}] =
\begin{pmatrix} J_3 & J_+ \cr J_- & -J_3
\end{pmatrix}, \qquad [{\bar{M}}^{\dal}{}_{\! \dbt}] =
\begin{pmatrix}  {\bar{J}}_3 & {\bar{J}}_+ \cr {\bar{J}}_- & -{\bar{J}}_3
\end{pmatrix} ,\ee
where $J_3, J_\pm$ (and ${\bar{J}}_3, {\bar{J}}_\pm$) satisfy the standard
commutation relations of $SU(2)$, we see that \C{Matrix} satisfies the
commutation
relations \C{MMCCR}.

We can also express  $R^i{}_{\! j}$ in terms of the
$SU(4)_R$ generators in the Chevalley basis. That is, for every simple root
${\al}^i$ $(i=1,2,3)$, we take ladder operators $E^{\pm i} = E^{\pm
  {\al}^i}$ and a Cartan generator $h^i$ chosen to satisfy
\be [h^i, h^j ] =0, \qquad [E^{+i},E^{-i}] = {\del}_{ij} h^j, \qquad
[h_i,E^{\pm j}] = \pm A_{ij} E^{\pm j},\ee
where
\be A_{ij} = 2\frac{({\al}^i,{\al}^j )}{\vert {\al}^j \vert^2},\ee
are elements of the Cartan matrix of $SU(4)$ given by
\be  [A_{ij}] =
\begin{pmatrix} 2 & -1 & 0 \cr -1 & 2 & -1 \cr 0 & -1 & 2
\end{pmatrix}. \ee
We can express any weight vector $\vec\lambda$ in terms of the fundamental weights with
integral coefficients specified by the Dynkin label
$[\lambda_1,\lambda_2,\lambda_3]$. Acting on the weight vector, we
note that
\be h^i \vert \vec\lambda \rangle = \lambda_i \vert\vec\lambda \rangle. \ee
Thus every representation has a highest weight state satisfying
\be h^i \vert \lambda_1,\lambda_2,\lambda_3 \rangle^{hw} =
\lambda_i \vert \lambda_1,\lambda_2,\lambda_3 \rangle^{hw}, \ee
and
\be E^{+i} \vert \lambda_1,\lambda_2,\lambda_3\rangle^{hw} = 0. \ee
We now construct the matrices  $R^i{}_{\! j}$
\be [R^i{}_{\! j}] =
\begin{pmatrix} \frac{1}{4}(3h_1 +2h_2 +h_3) & E_1^+ & [E_1^+,E_2^+] &
[E_1^+,[E_2^+,E_3^+]] \cr  E_1^- & \frac{1}{4}(-h_1 +2h_2 +h_3) & E_2^+ &
[E_2^+,E_3^+]\cr -[E_1^-,E_2^-] & E_2^- & -\frac{1}{4}(h_1 +2h_2 -h_3)& E_3^+
\cr [E_1^-,[E_2^-,E_3^-]] & -[E_2^-,E_3^-]&E_3^-&
-\frac{1}{4}(h_1 +2h_2 +3h_3)
\end{pmatrix}. \non\ee
It is not hard to check that the defining relations \C{defineR}\ are
satisfied by these representatives.

Now that we have explicit forms for the generators, we see that the
the last two equations of \C{nt}\ yield non-trivial constraints on the
conformal dimension of a short superconformal primary state.
Using the defining property \C{basisq}, we see that for $i=1,2$
\bea \frac{1}{4} \{ Q_{i \al}, S^{l \bt} \}
|k,p,q;j,\bar\jmath\rangle^{\rm hw}
&=& \begin{pmatrix} \frac{\Delta}{2} {\del}^l_i -R^l{}_{\! i}
+j {\del}^l_i &  0 \cr 0 & \frac{\Delta}{2} {\del}^l_i
-R^l{}_{\! i} -j {\del}^l_i \cr
\end{pmatrix}
|k,p,q;j,\bar\jmath\rangle^{\rm hw} \cr & &
+ {\del}^l_i \begin{pmatrix} 0 &  0 \cr \sqrt{2j} & 0\cr
\end{pmatrix} |k,p,q;j-1,\bar\jmath\rangle^{\rm hw}, \cr
&=& 0.
\eea
The conformal dimension, $\Delta$, satisfies
$$D |k,p,q;j,\bar\jmath\rangle^{\rm hw} =i\Delta
|k,p,q;j,\bar\jmath\rangle^{\rm hw}.$$
We see that $j=0$ and that
\be (\frac{\Delta}{2} {\del}^l_i -R^l{}_{\! i})
|k,p,q;0,\bar\jmath\rangle^{\rm hw} =0\ee
leading to
\be \Delta = \frac{1}{2} (2p +q), \quad k=0 .\ee
Similarly, from the ${\bar{Q}}^j_{\dal}$ constraint for $j=3,4$, we find that
$\bar\jmath =0,~q=0$ and $\Delta =p$. So the superconformal primary state
of a short multiplet is given by $[0,p,0]_{(0,0)}$ with conformal dimension
$\Delta =p$.

We now summarize the construction of the short multiplets by acting
with the superymmetry charges $Q_{i \al}$ and
${\bar{Q}}^j_{\dal}$ $(i=3,4; j=1,2)$ on the superconformal primary state
$[0,p,0]_{(0,0)}$. It will be very convenient to use the Racah-Speiser
algorithm for decomposing tensor products of representations (see Appendix
B of~\cite{Dolan:2002zh} for a review and various applications). The statement
of the algorithm goes as follows: given two representations
$R_{\underline\Lambda}$ and $R_{\underline\Lambda'}$ (where
$R_{\underline\Lambda}$ is the representation with highest weight vector
$\underline\Lambda = [\lambda_1,\cdots,\lambda_r]$, with $r$ the rank
of the group), the tensor product is given by
\be \label{RSalg} R_{\underline\Lambda} \otimes R_{\underline\Lambda'} \simeq
\sum_{\lambda \in V_{\underline\Lambda'}} R_{\underline\Lambda +
\underline\lambda},\ee
where $V_{\underline\Lambda'}$ consists of all the weight vectors for all the
states in $R_{\underline\Lambda'}$.

The algorithm further tells us that
on the right hand side of \C{RSalg},
\be \label{RSalg2} R_{\underline\lambda} = {\rm sign} (\s)
R_{{\underline\lambda}^\s}, \qquad
\lambda^\s = \s (\lambda +\rho) -\rho. \ee
Here $\s$ is an element of the Weyl group and $\rho$ is the Weyl vector.
So using this algorithm we can construct the tensor product of two
representations, where from \C{RSalg2}, we note that representations
for which $\lambda =\lambda^\s$ and $ {\rm sign(\s)} =-1$ vanish from the
right hand side of \C{RSalg}\ and so do not exist in the tensor product
decomposition.

In our applications, it tells us that when we are acting with the various
supercharges on a general representation $[k,p,q]_{(j,\bar\jmath)}$, we
naively
get all possible representations $[k',p',q']_{(j',\bar\jmath')}$ obtained
by adding the weights of the supercharges to $[k,p,q]_{(j,\bar\jmath)}$. Of
the
resulting representations, the ones where $(k',p',q',j',\bar\jmath')$ are all
non-negative are to be kept. The other representations will have negative
Dynkin labels. Some them vanish identically using \C{RSalg2}. The others
we can remove using the equations of motion or conservation laws (we will see
an example of this shortly).

So we need to know the Dynkin labels of the various supercharges. They can be
obtained by computing $[h^i,Q_{j \al}],~[h^i,S^{j \al}]$
which gives,
\bea \label{Dyn}
& Q_{1 \al} \sim [1,0,0]_{(\pmh ,0)}, \qquad  & Q_{2 \al}
\sim[-1,1,0]_{(\pmh ,0)},\cr &
Q_{3 \al} \sim [0,-1,1]_{(\pmh ,0)}, \qquad & Q_{4 \al}
\sim [0,0,-1]_{(\pmh ,0)}, \cr
& S^{1 \al} \sim [-1,0,0]_{(\pmh ,0)}, \qquad & S^{2 \al}
\sim [1,-1,0]_{(\pmh ,0)}, \cr & S^{3 \al} \sim [0,1,-1]_{(\pmh ,0)},
\qquad & S^{4 \al} \sim [0,0,1]_{(\pmh ,0)},
\eea
and for the conjugates,
\bea
& {\bar{Q}}^1_{\dal} \sim [-1,0,0]_{(0,\pmh)}, \qquad & {\bar{Q}}^2_{\dal}
\sim [1,-1,0]_{(0,\pmh)}, \cr
& {\bar{Q}}^3_{\dal} \sim [0,1,-1]_{(0,\pmh)},
\qquad & {\bar{Q}}^4_{\dal} \sim [0,0,1]_{(0,\pmh)}, \cr
& {\bar{S}}_1^{\dal} \sim  [1,0,0]_{(0,\pmh)}, \qquad & {\bar{S}}_2^{\dal}
\sim [-1,1,0]_{(0,\pmh)}, \cr & {\bar{S}}_3^{\dal} \sim [0,-1,1]_{(0,\pmh)},
\qquad & {\bar{S}}_4^{\dal} \sim [0,0,-1]_{(0,\pmh)}.
\eea
We can now go ahead with the construction of the short multiplet. Acting on the
superconformal primary state with the $Q$ operators yields

\be [0,p,0]_{(0,0)} \stackrel{Q}{\rightarrow} [0,p-1,1]_{(\frac{1}{2},0)}
\stackrel{Q^2}{\rightarrow}
 { [0,p-2,2]_{(0,0)} \atop
[0,p-1,1]_{(0,0)} }
\stackrel{Q^3}{\rightarrow}
[0,p-2,1]_{(\frac{1}{2},0)} \stackrel{Q^4}{\rightarrow} [0,p-2,0]_{(0,0)},
\ee
while acting with the $\bar{Q}$ operators on the highest weight states of the
various representations yields
\be [0,p,q]_{(j,0)} \stackrel{\bar{Q}}{\rightarrow}
[1,p-1,q]_{(j,\frac{1}{2})}
\stackrel{{\bar{Q}}^2}{\rightarrow}
 { [2,p-2,q]_{(j,0)} \atop
[0,p-1,q]_{(j,1)} }
\stackrel{{\bar{Q}}^3}{\rightarrow}
[1,p-2,q]_{(j,\frac{1}{2})} \stackrel{{\bar{Q}}^4}{\rightarrow}
[0,p-2,q]_{(j,0)}. \ee
These results lead to diagram~\C{BigM}\ displayed in the main text.

\section{The Subleading Terms in the $\del^2\OT$ OPE}
\label{appenddOT}

In this Appendix, we determine the subleading singular terms in~\C{Sans2}.
One way to obtain all the remaining terms
in~\C{Sans2}\ would be to consider all the less singular terms
in~\C{OPE1}\ and~\C{JOPEless} and repeat our prior analysis.
However, that is a rather
complicated route since there are many subleading terms in~\C{OPE1}\ and~\C{JOPEless}!

So we shall proceed by a different route. Because in \C{Sans2}, the
leading term goes like
$\sim 1/|z-x|^2$, all the possible subleading terms go like
$\sim 1/|z-x|$. So the number of these terms is far less than
the number of subleading terms in~\C{OPE1}\ or~\C{JOPEless}. Also, as
discussed before, we saw from~\C{general}\ that only terms with specific properties under
$SO(3,1)$ and $SU(4)_R$ arise in the OPE -- this drastically
reduces the possible subleading terms in~\C{Sans2}. So we shall
directly write down all possible terms that go like $\sim 1/|z-x|$
in~\C{Sans2}\ consistent with the various symmetries. This will give
the full OPE. The leading terms in~\C{Sans2}\ can be written
schematically as,
\be \frac{1}{z^2} \del^2 \Phi \qquad {\rm and} \qquad
\frac{1}{z^2} \bar\del^2 \Phi. \ee
At $O(1/(z-x))$, the following terms are the only possibilities
consistent with the $SO(3,1)$ and $SU(4)_R$ structure
of~\C{Sans2}:
\be \label{lesssing}
\frac{1}{z} \bar\del^2 \partial_\mu \Phi, \qquad
\frac{1}{z} \del^2 \partial_\mu \Phi, \qquad
\frac{1}{z} \bar\del^3 \del \Phi \quad
{\rm and} \quad  \frac{1}{z} \del^3 \bar\del \Phi.
\ee

We now list the $O(1/(z-x))$ terms in the OPE
for each of these cases (we use $r^\mu$ to denote $(z-x)^\mu$).
\vskip 0.1in
\noindent
(i) \underline{$\frac{1}{z} \, \bar\del^2 \partial_\mu \Phi$: }
\bea \label{lesssing1}
\Sigma_{\alpha}{}^{\beta ij} (z)
\Phi^I (x) \sim
\del_\alpha{}^\beta \frac{r^\mu}{r^2}
\partial_\mu [{\bar{Q}}^i{}_{\dal}, [{\bar{Q}}^{j\dal}
,\Phi^I (x)]_{\pm}]_{\mp} \nonumber
\qquad \qquad\\
+(\sigma_{\mu\nu})_\alpha{}^\beta
(\bar\sigma^{\mu\rho})^{\dal}{}_{\dbt}
\frac{r^\nu}{r^2} \partial_\rho
[{\bar{Q}}^i{}_{\dal}, [{\bar{Q}}^{j\dbt}
,\Phi^I (x)]_{\pm}]_{\mp} \nonumber
\qquad \qquad \quad \\
+i\del_\alpha{}^\beta (S^{\mu\nu})^I{}_J
\frac{r_\mu}{r^2} \partial_\nu
[{\bar{Q}}^i{}_{\dal}, [{\bar{Q}}^{j\dal}
,\Phi^J (x)]_{\pm}]_{\mp} \nonumber
\qquad \qquad \quad \quad \\
+i{\del}_\alpha{}^\beta \epsilon_{\mu\nu\rho\lambda}
(S^{\mu\nu})^I{}_J
\frac{r^\rho}{r^2} \partial^\lambda
[{\bar{Q}}^i{}_{\dal}, [{\bar{Q}}^{j\dal}
,\Phi^J (x)]_{\pm}]_{\mp} \nonumber
\qquad \qquad \quad \\
+i(\sigma_{\mu\nu})_\alpha{}^\beta
(S^{\mu\lambda})^I{}_J
(\bar\sigma_{\rho\lambda})^{\dal}{}_{\dbt}
\frac{r^\nu}{r^2} \partial^\rho
[{\bar{Q}}^i{}_{\dal}, [{\bar{Q}}^{j\dbt}
,\Phi^J (x)]_{\pm}]_{\mp} \quad \quad \quad \quad
\nonumber \\
+i\epsilon^{\mu\nu\rho\s} (\sigma_{\mu\lambda})_\alpha{}^\beta
(S_{\nu\omega})^I{}_J
(\bar\sigma_{\rho\tau})^{\dal}{}_{\dbt}
\frac{r^\lambda r^\tau r^\omega}{r^4}
\partial_\s
[{\bar{Q}}^i{}_{\dal}, [{\bar{Q}}^{j\dbt}
,\Phi^J (x)]_{\pm}]_{\mp}\quad \quad \nonumber \\
+i(\sigma_{\mu\nu})_\alpha{}^\beta
(S^{\mu\lambda})^I{}_J
(\bar\sigma_{\rho\tau})^{\dal}{}_{\dbt}
\frac{r^\nu r^\rho r_\lambda}{r^4}
\partial^\tau
[{\bar{Q}}^i{}_{\dal}, [{\bar{Q}}^{j\dbt}
,\Phi^J (x)]_{\pm}]_{\mp}
+{\rm permutations}.
\eea
Unlike leading order where there are no $\mathcal{E}$ type $S_{\m\n}$
terms, there are $S_{\m\n}$ terms of both $\mathcal{E}$ and $B$ type
at this order. There are several $B$ type $S_{\m\n}$ terms -- for
brevity,
we have written one for each distinct structure, with the remaining independent terms
generated by permutations. For example, by permutation, we should include
terms like
\be
i(\sigma_{\mu\nu})_\alpha{}^\beta
(S^{\mu\lambda})^I{}_J
(\bar\sigma^{\nu\rho})^{\dal}{}_{\dbt}
\frac{r_\lambda}{r^2} \partial_\rho
[{\bar{Q}}^i{}_{\dal}, [{\bar{Q}}^{j\dbt}
,\Phi^J (x)]_{\pm}]_{\mp} ,
\ee
which is a permutation of the fifth term of~\C{lesssing1}. Also among
the permutations is
\be
i\epsilon^{\mu\nu\rho\s} (\sigma_{\mu\lambda})_\alpha{}^\beta
(S_{\nu\omega})^I{}_J
(\bar\sigma_{\rho\tau})^{\dal}{}_{\dbt}
\frac{r^\lambda r^\tau r_\s}{r^4}
\partial^\omega
[{\bar{Q}}^i{}_{\dal}, [{\bar{Q}}^{j\dbt}
,\Phi^J (x)]_{\pm}]_{\mp}
\ee
which is related to the sixth term of~\C{lesssing1}.

\vskip 0.1in
\noindent
(ii) \underline{$ \frac{1}{z} \, \del^2 \partial_\mu \Phi$: }
\bea \label{lesssing2}
\Sigma_{\alpha}{}^{\beta ij} (z)
\Phi^I (x) \sim
\epsilon^{ijkl} \frac{r^\mu}{r^2}
\partial_\mu [Q_l{}^\beta, [Q_{k\alpha} ,
\Phi^I (x)]_{\pm}]_{\mp} \qquad \qquad \nonumber \\
+i\epsilon^{ijkl}
(S^{\mu\nu})^I{}_J
\frac{r_\mu}{r^2}
\partial_\nu [Q_l{}^\beta, [Q_{k\alpha} ,
\Phi^J (x)]_{\pm}]_{\mp} \qquad \qquad
\qquad \nonumber\\
+i\epsilon^{ijkl} \epsilon_{\mu\nu\rho\lambda}
(S^{\mu\nu})^I{}_J
\frac{r^\rho}{r^2}
\partial^\lambda [Q_l{}^\beta, [Q_{k\alpha} ,
\Phi^J (x)]_{\pm}]_{\mp} \qquad \qquad
\qquad \nonumber\\
+i\epsilon^{ijkl} (\sigma_{\mu\nu})_\alpha{}^\beta
(S^{\mu\lambda})^I{}_J
(\sigma^\nu{}_\lambda)_\gamma{}^\delta
\frac{r^\rho}{r^2}
\partial_\rho
[Q_l{}^\gamma, [Q_{k\delta} ,\Phi^J (x)]_{\pm}]_{\mp}
\qquad \qquad \nonumber \\
+i\epsilon^{ijkl} \epsilon^{\mu\nu\rho\s}
(\sigma_{\mu\omega})_\alpha{}^\beta
(S_{\nu\lambda})^I{}_J
(\sigma_{\rho\tau})_\gamma{}^\delta
\frac{r^\omega r^\tau r^\lambda}{r^4}
\partial_\s
[Q_l{}^\gamma, [Q_{k\delta} ,\Phi^J (x)]_{\pm}]_{\mp}
\qquad  \nonumber \\
+i\epsilon^{ijkl}
(\sigma_{\mu\nu})_\alpha{}^\beta
(S^{\mu\lambda})^I{}_J
(\sigma_{\rho\tau})_\gamma{}^\delta
\frac{r^\nu r^\tau r_\lambda}{r^4}
\partial^\rho
[Q_l{}^\gamma, [Q_{k\delta} ,\Phi^J (x)]_{\pm}]_{\mp}
+{\rm permutations},
\eea
where the permutations include, for example,
\be
i\epsilon^{ijkl} (\sigma_{\mu\nu})_\alpha{}^\beta
(S^{\mu\rho})^I{}_J
(\sigma_{\rho\lambda})_\gamma{}^\delta
\frac{r^\nu}{r^2}
\partial^\lambda
[Q_l{}^\gamma, [Q_{k\delta} ,\Phi^J (x)]_{\pm}]_{\mp}.
\ee
In this case all the terms are $B$ type.

\vskip 0.1in
\noindent
(iii) \underline{$ \frac{1}{z} \, \bar\del^3 \del \Phi$: }
\bea \label{lesssing3}
\Sigma_{\alpha}{}^{\beta ij} (z)
\Phi^I (x) \sim
\del_\alpha{}^\beta \frac{r^\mu}{r^2}
(\sigma_\mu)_{\gamma\dg} [{\bar{Q}}^i{}_{\dal}, [{\bar{Q}}^{j\dal}
,[{\bar{Q}}^{k\dg},[Q_k{}^\gamma,\Phi^I (x)]_{\pm}]_{\mp}]_{\pm}]_{\mp}
\nonumber \\
+(\sigma_{\mu\nu})_\alpha{}^\beta
\frac{r_\lambda}{r^2}
(\sigma^\nu)_{\gamma\dg}
(\bar\sigma^{\mu\lambda})^{\dal}{}_{\dbt}
[{\bar{Q}}^i{}_{\dal}, [{\bar{Q}}^{j\dbt}
,[{\bar{Q}}^{k\dg},[Q_k{}^\gamma
,\Phi^I (x)]_{\pm}]_{\mp}]_{\pm}]_{\mp} \nonumber
\quad \\
+i\del_\alpha{}^\beta (S^{\mu\nu})^I{}_J
\frac{r_\mu}{r^2} (\sigma_\nu)_{\gamma\dg}
[{\bar{Q}}^i{}_{\dal}, [{\bar{Q}}^{j\dal}
,[{\bar{Q}}^{k\dg},[Q_k{}^\gamma
,\Phi^J (x)]_{\pm}]_{\mp}]_{\pm}]_{\mp} \nonumber
\qquad \\
+i\del_\alpha{}^\beta
\epsilon_{\mu\nu\rho\lambda}
(S^{\mu\nu})^I{}_J
\frac{r^\rho}{r^2} (\sigma^\lambda)_{\gamma\dg}
[{\bar{Q}}^i{}_{\dal}, [{\bar{Q}}^{j\dal}
,[{\bar{Q}}^{k\dg},[Q_k{}^\gamma
,\Phi^J (x)]_{\pm}]_{\mp}]_{\pm}]_{\mp} \nonumber \\
+i\del_\alpha{}^\beta (S^{\mu\nu})^I{}_J
\frac{r^\lambda}{r^2} (\sigma_{\mu\nu}
\sigma_\lambda)_{\gamma\dg}
[{\bar{Q}}^i{}_{\dal}, [{\bar{Q}}^{j\dal}
,[{\bar{Q}}^{k\dg},[Q_k{}^\gamma
,\Phi^J (x)]_{\pm}]_{\mp}]_{\pm}]_{\mp} \nonumber
\qquad \\
+i(\sigma_{\mu\nu})_\alpha{}^\beta
(S^{\mu\lambda})^I{}_J
(\bar\sigma_{\rho\lambda})^{\dal}{}_{\dbt}
\frac{r^\nu}{r^2} (\sigma^\rho)_{\gamma\dg}
[{\bar{Q}}^i{}_{\dal}, [{\bar{Q}}^{j\dbt}
,[{\bar{Q}}^{k\dg},[Q_k{}^\gamma
,\Phi^J (x)]_{\pm}]_{\mp}]_{\pm}]_{\mp} \nonumber \\
+i\epsilon^{\mu\nu\rho\s}
(\sigma_{\mu\omega})_\alpha{}^\beta
(S_{\nu\tau})^I{}_J
(\bar\sigma_{\rho\lambda})^{\dal}{}_{\dbt}
\frac{r^\tau r^\omega r_\s}{r^4} (\sigma^\lambda)_{\gamma\dg}
[{\bar{Q}}^i{}_{\dal}, [{\bar{Q}}^{j\dbt}
,[{\bar{Q}}^{k\dg},[Q_k{}^\gamma
,\Phi^J (x)]_{\pm}]_{\mp}]_{\pm}]_{\mp}
\nonumber\\
+i(\sigma_{\mu\rho})_\alpha{}^\beta
(S^{\mu\tau})^I{}_J
(\bar\sigma_{\tau\nu})^{\dal}{}_{\dbt}
\frac{r^\rho r^\lambda r^\nu}{r^4} (\sigma_\lambda)_{\gamma\dg}
[{\bar{Q}}^i{}_{\dal}, [{\bar{Q}}^{j\dbt}
,[{\bar{Q}}^{k\dg},[Q_k{}^\gamma
,\Phi^J (x)]_{\pm}]_{\mp}]_{\pm}]_{\mp}
\nonumber \\
+{\rm permutations}. \qquad \qquad
\qquad\qquad\qquad\qquad \eea
In this case, there are only  permutations of $B$ type terms.

\vskip 0.1in
\noindent
(iv) \underline{$ \frac{1}{z} \, \del^3 \bar\del \Phi$:}
\bea \label{lesssing4}
\Sigma_{\alpha}{}^{\beta ij} (z)
\Phi^I (x) \sim
\epsilon^{ijkl} \frac{r^\mu}{r^2}
(\sigma_\mu)_{\gamma\dg} [Q_l{}^\beta, [Q_{k\alpha} ,
[Q_m{}^\gamma,[{\bar{Q}}^{m\dg},
\Phi^I (x)]_{\pm}]_{\mp}]_{\pm}]_{\mp}
\qquad \qquad \nonumber \\
+\epsilon^{klm[i} \frac{r^\mu}{r^2}
(\sigma_\mu)_{\gamma\dg}
[Q_l{}^\beta, [Q_{k\alpha} ,
[Q_m{}^\gamma,[{\bar{Q}}^{j]\dg},
\Phi^I (x)]_{\pm}]_{\mp}]_{\pm}]_{\mp}
\qquad \qquad \qquad \qquad \nonumber \\
+i\epsilon^{ijkl} \frac{r_\mu}{r^2}
(\sigma_\nu)_{\gamma\dg} (S^{\mu\nu})^I{}_J
[Q_l{}^\beta, [Q_{k\alpha} ,
[Q_m{}^\gamma,[{\bar{Q}}^{m\dg},
\Phi^J (x)]_{\pm}]_{\mp}]_{\pm}]_{\mp}
\qquad \qquad \qquad \nonumber \\
+i\epsilon^{ijkl}
\epsilon_{\mu\nu\rho\lambda}
\frac{r^\rho}{r^2}
(\sigma^\lambda)_{\gamma\dg} (S^{\mu\nu})^I{}_J
[Q_l{}^\beta, [Q_{k\alpha} ,
[Q_m{}^\gamma,[{\bar{Q}}^{m\dg},
\Phi^J (x)]_{\pm}]_{\mp}]_{\pm}]_{\mp}
\qquad \qquad \nonumber \\
+i\epsilon^{ijkl} \frac{r^\mu}{r^2}
(\sigma_{\rho\lambda} \sigma_\mu)_{\gamma\dg}
(S^{\rho\lambda})^I{}_J
[Q_l{}^\beta, [Q_{k\alpha} ,
[Q_m{}^\gamma,[{\bar{Q}}^{m\dg},
\Phi^J (x)]_{\pm}]_{\mp}]_{\pm}]_{\mp}
\qquad \qquad \qquad \nonumber \\
+i\epsilon^{ijkl} (\sigma_{\mu\nu})_\alpha{}^\beta
(S^{\mu\lambda})^I{}_J
(\sigma^\nu{}_\lambda)_\gamma{}^\delta
\frac{r_\rho}{r^2}
(\sigma^\rho)_{\theta\dt}
[Q_l{}^\gamma, [Q_{k\delta} ,
[Q_m{}^\theta,[{\bar{Q}}^{m\dt},
\Phi^J (x)]_{\pm}]_{\mp}]_{\pm}]_{\mp}
\qquad \qquad \nonumber \\
+i\epsilon^{ijkl} \epsilon^{\mu\nu\rho\s}
(\sigma_{\mu\lambda})_\alpha{}^\beta
(S_{\nu\tau})^I{}_J
(\sigma_{\rho\omega})_\gamma{}^\delta
(\sigma_\s)_{\theta\dt}
\frac{r^\omega r^\lambda r^\tau}{r^4}
[Q_l{}^\gamma, [Q_{k\delta} ,
[Q_m{}^\theta,[{\bar{Q}}^{m\dt},
\Phi^J (x)]_{\pm}]_{\mp}]_{\pm}]_{\mp}
\qquad \qquad \nonumber \\
+i\epsilon^{ijkl}
(\sigma_{\mu\lambda})_\alpha{}^\beta
(S^{\mu\tau})^I{}_J
(\sigma_{\nu\tau})_\gamma{}^\delta
(\sigma_\omega)_{\theta\dt}
\frac{r^\omega r^\lambda r^\nu}{r^4}
[Q_l{}^\gamma, [Q_{k\delta} ,
[Q_m{}^\theta,[{\bar{Q}}^{m\dt},
\Phi^J (x)]_{\pm}]_{\mp}]_{\pm}]_{\mp}
\qquad \qquad \nonumber \\
+i \del_\alpha{}^\beta \epsilon^{klm(i}
(\sigma_{\mu\nu})_\gamma{}^\delta
(S^{\mu\nu})^I{}_J
\frac{r_\rho}{r^2}
(\sigma^\rho)_{\theta\dt}
[Q_l{}^\gamma, [Q_{k\delta} ,
[Q_m{}^\theta,[{\bar{Q}}^{j)\dt},
\Phi^J (x)]_{\pm}]_{\mp}]_{\pm}]_{\mp}
\qquad \qquad  \nonumber\\
+i \del_\alpha{}^\beta \epsilon^{\mu\rho\tau\s}
\epsilon^{klm(i}
(S_{\mu\nu})^I{}_J
(\sigma_{\rho\lambda})_\gamma{}^\delta
(\sigma_\tau)_{\theta\dt}
\frac{r^\nu r^\lambda r_\s}{r^4}
[Q_l{}^\gamma, [Q_{k\delta} ,
[Q_m{}^\theta,[{\bar{Q}}^{j)\dt},
\Phi^J (x)]_{\pm}]_{\mp}]_{\pm}]_{\mp}
\qquad \qquad  \nonumber\\
+i \del_\alpha{}^\beta
\epsilon^{klm(i}
(S_{\mu\nu})^I{}_J
(\sigma^{\rho\nu})_\gamma{}^\delta
(\sigma^\lambda)_{\theta\dt}
\frac{r^\mu r_\lambda r_\rho}{r^4}
[Q_l{}^\gamma, [Q_{k\delta} ,
[Q_m{}^\theta,[{\bar{Q}}^{j)\dt},
\Phi^J (x)]_{\pm}]_{\mp}]_{\pm}]_{\mp}
\qquad \qquad  \nonumber\\
+{\rm permutations},\qquad \qquad
\qquad \qquad\qquad \qquad \qquad \qquad \eea
Note that every $\mathcal{E}$
term is an $S_{\m\n}$ term. Also this is the only
case where the permutations include both $\mathcal{E}$ and $B$ type
terms. For example, the  $\mathcal{E}$ type term
\be
i \del_\alpha{}^\beta \epsilon^{klm(i}
(\sigma_{\mu\nu})_\gamma{}^\delta
(S^{\mu\lambda})^I{}_J
\frac{r^\nu}{r^2}
(\sigma_\lambda)_{\theta\dt}
[Q_l{}^\gamma, [Q_{k\delta} ,
[Q_m{}^\theta,[{\bar{Q}}^{j)\dt},
\Phi^J (x)]_{\pm}]_{\mp}]_{\pm}]_{\mp}.
\ee
So including all these additional terms, we have the complete OPE of
$\del^2 \OT$. The leading terms are given in~\C{Sans2}, while
the subleading terms are given
in~\C{lesssing1},~\C{lesssing2},~\C{lesssing3} and~\C{lesssing4}.

\section{Some Consistency Checks of the $\del^2 \OT$ OPE}
\label{somechecks}

Since the $\del^2 \OT$ OPE plays an important role in our analysis,
we will perform some consistency checks of this OPE here. The
checks will involve computing the OPE of $\mathcal{E}$ with selected
operators.

\subsection{ $\mathcal{E}(z) \bar\Lambda(x)$ in free field
theory}

We consider the $\bar\Lambda = \bar\del^3 \OT$ operator which is given by
\be
\bar\Lambda_i{}^{\dal} =
-\frac{1}{g_{YM}^2} (\bar\s^{\mu \nu})^{\dal}{}_{\dbt}
F_{\mu \nu} \bar\lambda_i{}^{\dbt}, \ee
and we want to compute
\be {\mathcal{E}}^{kl} (z) \bar\Lambda_i{}^{\dal} (x),
\ee
in the free field theory,
where $\mathcal{E}$ is given by
\be {\mathcal{E}}^{ij} = \frac{1}{g_{YM}^2} \lambda^i \lambda^j. \ee
A straightforward calculation yields that
\be \label{checkterm}
{\mathcal{E}}^{kl} (z) \bar\Lambda_i{}^{\dal} (x)
\sim -\frac{i}{8\pi^2 g_{YM}^2} \del_i{}^l
(\bar\s^{\mu \nu} \epsilon)^{\dal \dbt} \partial_{\gamma \dbt}
\frac{1}{(z-x)^2}
\left( F_{\mu \nu} \lambda^{k\gamma} \right)(x) -(k \leftrightarrow l),  \ee
to leading order, where we have used the free fermion
propagator \C{fermprop}.
We now show that the leading term in \C{Sans2} given by
\be \label{leadE} {\mathcal{E}}^{kl} (z) \bar\Lambda_i{}^{\dal} (x)
\sim \frac{1}{(z-x)^2}
[{\bar{Q}}^k{}_{\dbt}, \{ {\bar{Q}}^{l\dbt}
,\bar\Lambda_i{}^{\dal} (x)\}] ,
\ee
gives us \C{checkterm} on integrating by parts. Since we will
eventually integrate over $z$, equivalence up to total derivatives is
sufficient for us.

Using the relations
\be \{{\bar{Q}}^{i\dal} ,\bar\Lambda_j{}^{\dbt}\} =
\del_j{}^i \epsilon^{\dal \dbt} \Obt ,\qquad
[{\bar{Q}}^i,\bar\Ot] =
\partial^\mu \bar{J}_\mu^i +2 \bar\s^\mu \s^\nu \partial_\mu
\bar{J}_\nu^i
,\ee
we see that \C{leadE} gives
\be {\mathcal{E}}^{kl} (z) \bar\Lambda_i{}^{\dal} (x)
\sim -\frac{2\del_i{}^l}{(z-x)^2}
(\bar\s^{\mu \nu})^{\dal}{}_{\dbt} \partial_\mu {\bar{J}}_\nu^{k\dbt} (x)
-(k \leftrightarrow l).\ee
Now ${\bar{J}}_\mu^{i\dal}$ is given by
\be \label{jbar} {\bar{J}}_\mu^{i\dal}
=-\frac{1}{g_{YM}^2} \left(F_{\rho \s} (\bar\sigma^{\rho \s}
\bar\s_\mu
\lambda^i)^{\dal} +2i \varphi^{ij} {\buildrel \leftrightarrow \over
\partial}^\mu
\bar\lambda_j{}^{\dal} +\frac{4}{3} i \partial^\nu
(\varphi^{ij} \bar\s_{\mu \nu}\bar\lambda_j)^{\dal} \right).  \ee
Again, a straightforward calculation leads to
\be \label{checkterm2}
{\mathcal{E}}^{kl} (z) \bar\Lambda_i{}^{\dal} (x)
\sim \frac{1}{2g_{YM}^2} \del_i{}^l
(\bar\s^{\mu \nu} \epsilon)^{\dal \dbt} \partial_{\gamma \dbt}
\frac{1}{(z-x)^2}
\left(F_{\mu \nu} \lambda^{k\gamma}\right) (x) +(k \leftrightarrow l), \ee
where we have integrated by parts. The total contribution
from the second and the third terms in~\C{jbar}\ cancels.
So we see that up to an overall numerical factor,~\C{checkterm}\
and~\C{checkterm2}\  match
exactly -- demonstrating the existence of the leading term in the
$\mathcal{E}$ OPE in~\C{Sans2}.

\subsection{$\mathcal{E}(z) \bar\del \OT(x)$ in free field
theory}

We consider this example because it is a case where the subleading
terms in the $\mathcal{E}$ OPE are needed for agreement with direct free
field computations.
Start with the $\bar\del \OT$ operator given by
\be \bar\chi^{ij}{}_k =\frac{1}{2g_{YM}^2}\epsilon^{ijmn}
\left(\bar\varphi_{mn} \bar\lambda_k +
\bar\varphi_{kn} \bar\lambda_m \right), \ee
to leading order and compute in free field theory
\be {\mathcal{E}}^{kl} (z) \bar\chi^{ij\dal}{}_m (x) \sim
-\frac{i}{4\pi^2 g_{YM}^2} \epsilon^{ijpq} \partial^{\dal\al}
\frac{1}{(z-x)^2} \left(\del_m{}^{(k} \lambda^{l)} \bar\varphi_{pq}
+\del_p{}^{(k} \lambda^{l)} \bar\varphi_{mq}\right)_\al (x), \ee
This expression can also be written as
\be \label{sublead}
{\mathcal{E}}^{kl} (z) \bar\chi^{ij\dal}{}_m (x) \sim
-\frac{i}{4\pi^2 g_{YM}^2} \partial^{\dal\al} \frac{1}{(z-x)^2}
\left(3 \del_m{}^{(k} \lambda^{l)} \varphi^{ij}
+2\del_m{}^{[i} \varphi^{j](k} \lambda^{l)}\right)_\al (x)
\ee
using $\bar\varphi_{ij} =\frac{1}{2} \epsilon_{ijkl} \varphi^{kl}$.
Now the leading term in the OPE \C{Sans2} is schematically $ \frac{1}{z^2}
\bar\del^2 \bar\chi \sim \frac{1}{z^2} \bar\Lambda$, which is not
the correct term; hence we can conclude its coefficient vanishes. Similarly, all the
coefficients of all terms in~\C{lesssing1},~\C{lesssing2} and~\C{lesssing3}\  must vanish as well.

{}Finally, we
consider the terms in \C{lesssing4}, whose contribution $
\frac{1}{z^2} \del^3 \bar\del \bar\chi$ contains $J$ and $\chi$.
So we need to find the terms in \C{lesssing4}\ that reproduce
\C{sublead}\ -- these terms must have non-vanishing coefficients.
{}For this example, note that
\be (S_{\mu\nu})^I{}_J \Phi^J = (\bar\s_{\mu\nu})^{\dal}{}_{\dbt}
\bar\chi^{ij\dbt}{}_m.\ee
Many of the $\mathcal{E}$ type $S_{\m\n}$ term contributions in~\C{lesssing4}\
vanish identically for this choice of $S_{\mu\nu}$.
For simplicity, we consider a particular non-vanishing term in
\C{lesssing4}. We shall see that this term is sufficient to
reproduce the structures appearing in~\C{sublead}. However, we will
not calculate any precise
coefficients since other terms in~\C{lesssing4}\ can also give rise
to the terms of~\C{sublead}\ (also, the algebra is quite cumbersome).

Consider the term in~\C{lesssing4}\ given by
\bea \label{echi}
{\mathcal{E}}^{kl} (z) \bar\chi^{ij\dal}{}_m (x)
&\sim & i \epsilon^{\mu\rho\tau\omega}
\epsilon^{pqs(k}
(\bar\s_{\mu\nu})^{\dal}{}_{\dbt}
(\sigma_{\rho\s})_\gamma{}^\delta
(\sigma_\tau)_{\theta\dt} \nonumber \\
&& \times \frac{r^\nu r^\s r_\omega}{r^4}
[Q_q{}^\gamma, \{Q_{p\delta} ,
[Q_s{}^\theta,\{{\bar{Q}}^{l)\dt},
\bar\chi^{\dbt ij}{}_m (x)\}]\}].
\eea
In order to proceed, we need the supersymmetry transformations
from~\C{supervar}
\bea \{Q_l{}^\alpha ,\bar\chi^{ij\dal}{}_{k}\} &=& -\frac{3}{4} i
(\bar\s_\mu)^{\dal\alpha} R^{\mu[i}{}_k \del^{j]}{}_l
+\frac{3}{2} i (\bar\s^\mu)^{\dal\alpha} \partial_\mu Q^{ij}{}_{kl}
\nonumber \\ &&
-\frac{3}{16} i (\bar\s_\mu)^{\dal\alpha} \del^{[i}{}_k
R^{\mu j]}{}_l +\frac{3}{4} i (\bar\s^\mu)^{\dal\alpha}
\del^{[i}{}_k \partial_\mu Q^{j]m}{}_{ml},
\eea
\be [Q_i,T_{\mu\nu}] = \s_{(\mu \vert \rho \vert}
\partial^\rho J_{\nu)i},
\ee
\be [Q_{l\alpha},R^{\mu i}{}_j] = -\del_l{}^i J^\mu_{j\alpha}
+\frac{1}{4} \del_j{}^i J^\mu_{l\alpha} +\frac{8}{3} i
(\s^{\mu\nu} \partial_\nu \chi^i{}_{jl})_\alpha,
\ee
as well as \C{eqnone},~\C{eqntwo},~\C{eqnthree},~\C{eqnfour} and~\C{eqnfive}.
We then calculate~\C{echi}\ with a liberal use of integration by
parts.  After some tedious calculations,
we see that there are only three distinct structures that contribute:
\be \label{termone} \epsilon^{ijp(k}
\chi^{l)}{}_{mp\alpha} (x)
\partial^{\dal\alpha} \frac{1}{(z-x)^2}
,\ee
\be \label{termtwo} \frac{1}{3} \del_m{}^{[i}
\epsilon^{j]pq(k} \chi^{l)}{}_{pq\alpha} (x)
\partial^{\dal\alpha} \frac{1}{(z-x)^2},\ee
and
\be \label{termthree} \epsilon^{ijp(k} \del_m{}^{l)}
\partial_\nu^{-1} J^\nu_{p\alpha} (x)
\partial^{\dal\alpha} \frac{1}{(z-x)^2} .\ee
We will explain the meaning of the formal expression~\C{termthree}\ momentarily.
It is useful to rewrite the $\chi$ term contributions using
the relations
\be \label{simp1}
\epsilon^{ijp(k} \chi^{l)}{}_{mp} =
-\frac{1}{g_{YM}^2} \left(\varphi^{ij} \lambda^{(k} \del_m{}^{l)}
+3 \del_m{}^{[i} \varphi^{j](k} \lambda^{l)}
-\lambda^{[i} \varphi^{j](k} \del_m{}^{l)} \right),
\ee
and
\be \label{simp2}
\frac{1}{3}  \del_m{}^{[i} \epsilon^{j]pq(k}
\chi^{l)}{}_{pq} = \frac{1}{g_{YM}^2} \left(\del_m{}^{[i} \varphi^{j](k}
\lambda^{l)} \right).
\ee
(The possibility
\be \epsilon^{ijpq} \chi^{(k}{}_{pq} \del_m{}^{l)} =
\frac{1}{g_{YM}^2} \left(\varphi^{ij} \lambda^{(k} \del_m{}^{l)}
-\lambda^{[i} \varphi^{j](k} \del_m{}^{l)} \right)
\ee
is not linearly independent.)

We now consider the formal expression~\C{termthree}\
and evaluate it. Using the definition of $J$ in~\C{vardef}, we
see that
\be \label{formaleqn}
\partial_\nu^{-1} J^\nu_p =
\frac{1}{g_{YM}^2} \left(\frac{i}{2} \bar\varphi_{pq}
\lambda^q -4i \partial_\nu^{-1} ( \bar\varphi_{pq}
\partial^\nu \lambda^q) - \partial_\nu^{-1}
(F_{\rho\s} \s^{\rho\s} \s^\nu \bar\lambda_p)\right),
\ee
using $\partial_\mu \partial_\nu^{-1} =\eta_{\mu\nu}$.
Now the second term in this expression, when
inserted in~\C{termthree}, vanishes on integrating by parts
because on-shell
\be \partial^{\dal\alpha} \frac{1}{(z-x)^2}
\partial_\nu^{-1} \left(\bar\varphi_{pq}
\partial^\nu \lambda^q \right) (x) = \frac{1}{(z-x)^2}
\left(\bar\varphi_{pq} \partial^{\dal\alpha}
\lambda^{q}{}_\alpha\right)(x) =0.
\ee
Similarly, the third term
when inserted in~\C{termthree}\ also vanishes on integrating by parts
because
\bea \partial^{\dal\alpha} \frac{1}{(z-x)^2}
\partial_\nu^{-1}
(F_{\rho\s} (\s^{\rho\s} \s^\nu \bar\lambda_p)_\alpha) &=&
\frac{1}{(z-x)^2} {(\bar\s_\mu)}^{\dal\alpha}
(F_{\rho\s} (\s^{\rho\s} \s^\mu \bar\lambda_p)_\alpha)
= (\s^{\rho\s})_\alpha{}^\alpha \times \ldots \non\\ &=& 0.\eea
So we see that~\C{termthree}\ is a well-defined local operator given by
\be \epsilon^{ijp(k} \del_m{}^{l)}
\partial_\nu^{-1} J^\nu_{p\alpha} (x)
\partial^{\dal\alpha} \frac{1}{(z-x)^2} =
\frac{i}{2g_{YM}^2} \epsilon^{ijp(k} \del_m{}^{l)}
\left(\bar\varphi_{pq} \lambda^q{}_\alpha\right) (x)
\partial^{\dal\alpha} \frac{1}{(z-x)^2}.\ee
It is more useful to simplify this using the relation
\be \label{simp3}
\epsilon^{ijp(k} \del_m{}^{l)}
\left(\bar\varphi_{pq} \lambda^q{}_\alpha \right) =
\left(\varphi^{ij} \del_m{}^{(k} \lambda^{l)}
+2 \lambda^{[i} \varphi^{j](k} \del_m{}^{l)} \right).\ee
{}From the expressions~\C{simp1},~\C{simp2} and~\C{simp3}, it
is easy to see that we get all the terms in~\C{sublead}. This provides
constraints on the OPE coefficients from the terms~\C{lesssing4}.
In fact, we can write~\C{sublead}\ as
\bea \label{subleadfinal}
{\mathcal{E}}^{kl} (z) \bar\chi^{ij\dal}{}_m (x) &\sim &
\frac{3i}{8\pi^2} \partial^{\dal\alpha} \frac{1}{(z-x)^2}
\Big(\epsilon^{ijp(k} \chi^{l)}{}_{mp}
+\frac{20}{9} \del_m{}^{[i} \epsilon^{j]pq(k} \chi^{l)}{}_{pq}
\nonumber \\
&& +i \epsilon^{ijp(k} \del_m{}^{l)}
\partial_\nu^{-1} J^\nu_p \Big)_\alpha (x).
\eea
We can now physically see the role of the last term in
\C{subleadfinal}. Without this term, the right hand side of
\C{subleadfinal} would contain a term proportional to
\be \lambda^{[i}{}_\alpha \varphi^{j](k} \del_m{}^{l)}
\partial^{\dal\alpha} \frac{1}{(z-x)^2},
\ee
which does not appear on the left hand side of~\C{subleadfinal}.
Also, $\mathcal{E}$
and $\bar\chi$ are both elements of the current multiplet which
closes on-shell and so their OPE should involve only operators
in the current multiplet, which is seen to be the case.

\subsection{The  $\mathcal{E}(z) T(x)$ OPE}

The prior two checks of the $\del^2 \OT$ OPE involved free field
theory. This final check involves the full interacting theory.
Let us consider the leading terms
in the OPE of the stress tensor with
$\mathcal{E}$, i.e., the $T(z) \mathcal{E} (x)$ OPE. Clearly,
these terms (with $z$
and $x$ interchanged) should appear in the
$\mathcal{E}(z)T(x)$ OPE and we will demonstrate that this is the
case.

{}First consider the $T{\mathcal{E}}$ OPE. From~\C{OPE1}\ we see that,
to leading order,
\be \label{ETzero}
T_{\mu\nu} (z) {\mathcal{E}}^{ij} (x) \sim
{\mathcal{E}}^{ij} (x) \partial_\mu \partial_\nu
\frac{1}{(z-x)^2} +  {\mathcal{E}}^{ij} (x)
\eta_{\mu \nu} \del^4 (z-x). \ee
Now consider the terms in the ${\mathcal{E}}T$ OPE. Since
$$\bar\del^2 T \sim \bar{B}, \qquad \bar\del^3 \del T \sim \bar{B} +
\bar{\mathcal{E}}, \qquad \del^3 \bar\del T \sim B +{\mathcal{E}},$$
we see that only terms from~\C{lesssing4}\ can be non-zero
again. All the coefficients
from~\C{Sans2}, \C{lesssing1},~\C{lesssing2} and~\C{lesssing3}\  must vanish. Among
the terms in~\C{lesssing4}, we consider the contribution of two
terms to the ${\mathcal{E}}T$ OPE. First consider
\bea \label{ETone}
{\mathcal{E}}^{ij} (z) T_{\mu\nu} (x) &\sim & i \epsilon^{klm(i}
(\sigma^{\lambda\rho})_\gamma{}^\delta
\frac{r_\sigma}{r^2}
(\sigma^\sigma)_{\theta\dt} \times \non\\ &&
\{Q_l{}^\gamma, [Q_{k\delta} ,
\{Q_m{}^\theta,[{\bar{Q}}^{j)\dt},
(S_{\lambda\rho} T)_{\mu\nu} (x)]\}]\},
\eea
where
\be (S_{\lambda\rho} T)_{\mu\nu}
=-i (\eta_{\rho\mu} T_{\lambda\nu} -\eta_{\lambda\mu}
T_{\rho\nu} +\eta_{\rho\nu} T_{\mu\lambda} -\eta_{\lambda\nu}
T_{\mu\rho}).\ee
Using the supersymmetry variations of~\C{supervar},
\be
[\bar{Q}^i , T_{\mu\nu}] =\bar\s_{\mu\lambda} \partial^\lambda
\bar{J}^i_\nu +\bar\s_{\nu\lambda} \partial^\lambda
\bar{J}^i_\mu, \ee
\be \{ Q_j{}^\alpha , \bar{J}^{i\dal}_\mu \} =
(\bar\s^\nu)^{\dal\alpha} T_{\mu\nu} \del_j{}^i
+2(\bar\s_\lambda \s_{\mu\nu} -\frac{1}{3} \bar\s_{\mu\nu}
\bar\s_\lambda)^{\dal\alpha} \partial^\nu R^{\lambda i}{}_j,
\ee
\be
[Q_k , R^{\mu i}{}_j] = -\del_k{}^i J^\mu_j +\frac{1}{4}
\del_j{}^i J^\mu_k +\frac{8}{3} i \s^{\mu\nu} \partial_\nu
\chi^i{}_{jk},  \ee
and \C{eqntwo}, we can evaluate~\C{ETone}. To simplify the
computation, let us restrict to on-shell non-vanishing
contributions. Any additional contributions will only change the value of the
coefficients.
On integrating by parts and using $\partial_\mu R^{\mu i}{}_j =0$ (on-shell),
we see that~\C{ETone}\ becomes
\be \label{ETtwo}
{\mathcal{E}}^{ij} (z) T_{\mu\nu} (x) \sim
{\mathcal{E}}^{ij} (x) \left(\partial_\mu \partial_\nu
\frac{1}{(z-x)^2} + \pi^2
\eta_{\mu \nu} \del^4 (z-x) \right). \ee

Next consider another term in the ${\mathcal{E}}T$ OPE given by
\bea \label{ETthree} {\mathcal{E}}^{ij} (z) T_{\mu\nu} (x) &\sim &
\epsilon^{klm(i}
(\sigma^{\s\rho})_\gamma{}^\delta
(\sigma^\omega)_{\theta\dt}
\frac{r^\lambda r_\omega r_\s}{r^4}  \times \non\\ &&
\{Q_l{}^\gamma, [Q_{k\delta} ,
\{Q_m{}^\theta,[{\bar{Q}}^{j)\dt},
(S_{\lambda\rho} T)_{\mu\nu} (x)]\}]\}.
\eea
Again evaluating this term gives
\be \label{ETfour}
{\mathcal{E}}^{ij} (z) T_{\mu\nu} (x) \sim
{\mathcal{E}}^{ij} (x) \left(3\partial_\mu \partial_\nu
\frac{1}{(z-x)^2} -2\pi^2
\eta_{\mu \nu} \del^4 (z-x)\right). \ee
In this way, we recover the $T \mathcal{E}$ OPE structure from the
$\mathcal{E} T$ OPE.

\section{Some Useful Two-Point Functions}
\label{twopoints}
In this Appendix, we list the results for the various two-point
functions needed in section~\ref{evaluate2pt}.
All derivatives act on $x$. For cases with
$\del^2 \bar\del^2$ insertions:
\bea \label{corrtwo} \langle \{Q_i{}^\alpha
,[{\bar{Q}}^i{}_{\dal}, \{{\bar{Q}}^{j\dal}
,[Q_{j\alpha} ,\bar\Om (x)\}]\}] \Om (y) \rangle &=& -48
\partial^2 \langle \bar\Om (x) \Om (y) \rangle, \nonumber\\
\langle \{{\bar{Q}}^i{}_{\dal}, [{\bar{Q}}^{j\dal}
,\{Q_i{}^\alpha ,[Q_{j\alpha} ,\bar\Om (x)\}]\}] \Om (y) \rangle
&=& 32 \partial^2 \langle \bar\Om (x) \Om (y) \rangle, \nonumber\\
(\sigma_{\mu\nu})_\alpha{}^\beta
(\bar\sigma^{\mu\rho})^{\dal}{}_{\dbt}
\langle \{Q_i{}^\alpha ,[Q_{j\beta} ,\{{\bar{Q}}^i{}_{\dal},
[{\bar{Q}}^{j\dbt}
,\bar\Om(x)]\}]\} \Om (y) \rangle
&=& 8 (\eta_{\rho\nu}
\partial^2  -4 \partial_\nu
\partial_\rho) \langle \bar\Om (x) \Om (y) \rangle,
\nonumber\\
(\sigma_{\mu\nu})_\alpha{}^\beta
(\bar\sigma^{\mu\rho})^{\dal}{}_{\dbt}
\langle \{{\bar{Q}}^i{}_{\dal}, [{\bar{Q}}^{j\dbt}
,\{Q_i{}^\alpha ,[Q_{j\beta} ,\bar\Om(x)]\}]\} \Om (y) \rangle
&=& 8 (\eta_{\rho\nu}
\partial^2  -4 \partial_\nu
\partial_\rho) \langle \bar\Om (x) \Om (y) \rangle,
\nonumber\\
(\sigma_{\mu\nu})_\alpha{}^\beta
(\bar\sigma^{\mu\rho})^{\dal}{}_{\dbt}
\langle \{Q_i{}^\alpha ,[{\bar{Q}}^i{}_{\dal}, \{{\bar{Q}}^{j\dbt}
,[Q_{j\beta} ,\bar\Om(x)]\}]\} \Om (y) \rangle
&=& -4 (
\eta_{\rho\nu}
\partial^2 -4 \partial_\nu
\partial_\rho) \langle \bar\Om (x) \Om (y) \rangle. \non\eea
For cases with $\del^3 \bar\del^3$ insertions, we find that
\bea \label{corrthree}
(\sigma_\mu)_{\beta\dbt} \langle \{Q_i{}^\alpha ,[Q_{j\alpha}
,\{{\bar{Q}}^i{}_{\dal}, [{\bar{Q}}^{j\dal}
,\{{\bar{Q}}^{k\dbt},[Q_k{}^\beta,
\bar\Om (x)]\}]\}]\} \Om (y) \rangle &=& 128 i \partial_\mu \partial^2
\langle \bar\Om (x) \Om (y) \rangle,
\nonumber \\
(\sigma_\mu)_{\beta\dbt} \langle \{Q_i{}^\alpha ,
[{\bar{Q}}^i{}_{\dal}, \{{\bar{Q}}^{j\dal}
,[{\bar{Q}}^{k\dbt},\{Q_k{}^\beta,
[Q_{j\alpha} ,\bar\Om (x)]\}]\}]\} \Om (y) \rangle &=&-256 i
\partial_\mu \partial^2 \langle \bar\Om (x) \Om (y) \rangle,
\nonumber \\
(\sigma_\mu)_{\beta\dbt}
\langle \{{\bar{Q}}^i{}_{\dal}, [{\bar{Q}}^{j\dal}
,\{{\bar{Q}}^{k\dbt},[Q_k{}^\beta,
\{Q_i{}^\alpha ,[Q_{j\alpha} ,\bar\Om (x)]\}]\}]\} \Om (y) \rangle &=&
320 i\partial_\mu \partial^2 \langle \bar\Om (x) \Om (y) \rangle, \non
\eea
and,
\bea
 && (\sigma_{\mu\nu})_\alpha{}^\beta
(\sigma^\nu)_{\gamma\dg}
(\bar\sigma^{\mu\lambda})^{\dal}{}_{\dbt}
\langle \{Q_i{}^\alpha ,[Q_{j\beta} ,\{{\bar{Q}}^i{}_{\dal},
[{\bar{Q}}^{j\dbt}
,\{{\bar{Q}}^{k\dg},[Q_k{}^\gamma
,\bar\Om (x)]\}]\}]\}\Om (y) \rangle\nonumber \\
&&= -288 i
\partial_\lambda \partial^2 \langle \bar\Om (x) \Om (y) \rangle,
\nonumber\\ &&
(\sigma_{\mu\nu})_\alpha{}^\beta
(\sigma^\nu)_{\gamma\dg}
(\bar\sigma^{\mu\lambda})^{\dal}{}_{\dbt}
\langle \{Q_i{}^\alpha ,[{\bar{Q}}^i{}_{\dal}, \{{\bar{Q}}^{j\dbt}
,[{\bar{Q}}^{k\dg},\{Q_k{}^\gamma, [Q_{j\beta} ,
,\bar\Om (x)]\}]\}]\} \Om (y) \rangle \nonumber\\ &&= 384 i
\partial_\lambda \partial^2 \langle \bar\Om (x) \Om (y) \rangle,
\nonumber\\ &&
(\sigma_{\mu\nu})_\alpha{}^\beta
(\sigma^\nu)_{\gamma\dg}
(\bar\sigma^{\mu\lambda})^{\dal}{}_{\dbt}
\langle\{{\bar{Q}}^i{}_{\dal}, [{\bar{Q}}^{j\dbt}
,\{{\bar{Q}}^{k\dg}, [Q_k{}^\gamma ,\{Q_i{}^\alpha ,[Q_{j\beta} ,
\bar\Om (x)]\}]\}]\} \Om (y) \rangle \nonumber\\ &&= 48 i
\partial_\lambda \partial^2 \langle \bar\Om (x) \Om (y) \rangle.  \eea




\ifx\undefined\bysame
\newcommand{\bysame}{\leavevmode\hbox to3em{\hrulefill}\,}
\fi

\providecommand{\href}[2]{#2}\begingroup\raggedright  \endgroup

\end{document}